\documentclass[11pt]{article}
\pdfoutput=1
\usepackage{amsmath}
\usepackage{amssymb}
\usepackage{graphicx,bbm,mathrsfs}
\usepackage{nicefrac}
\usepackage{slashed}
\usepackage{bbm}
\usepackage{geometry}
\usepackage{empheq}
\usepackage{stackrel}
\usepackage{ulem}
\usepackage{jheppub}
\usepackage{setspace}
\usepackage{relsize}
\usepackage{empheq}
\usepackage{pifont}
\usepackage{mathtools}
\usepackage{enumitem}
\usepackage{dcolumn}   
\usepackage{bm}        
\usepackage{graphicx,mathrsfs}
\usepackage{nicefrac}
\usepackage{multirow}
\usepackage{color}
\usepackage{mathtools}
\usepackage{makecell}
\usepackage{environ} 
\usepackage{lipsum} 
\usepackage{wasysym}
\usepackage{hhline,colortbl}  
\usepackage{tikz}
\usepackage{graphicx}
\usepackage{tensor}
\usepackage{dsfont}
\usepackage{simpler-wick}

\usepackage{booktabs}
\usepackage{mdframed}
\usepackage{xcolor}
\definecolor{EqFrame}{RGB}{235,245 ,250 }

 \NewEnviron{Smaller11}{
           \scalebox{1.1}{$\BODY$} 
 } 
 \NewEnviron{Smaller08}{
           \scalebox{0.8}{$\BODY$} 
 }

\newcommand{\be}{\begin{equation}}
\newcommand{\ee}{\end{equation}}
\newcommand{\bea}{\begin{eqnarray}}
\newcommand{\eea}{\end{eqnarray}}

\renewcommand{\O}{\mathcal O}

\newcommand{\DeltaF}{\Delta_F}
\newcommand{\TildeDeltaF}{\bar{\Delta}_F}

\newcommand{\Id}{\mathds{1} }

\renewcommand\labelenumi{(\roman{enumi})}
\renewcommand\theenumi\labelenumi

\newcommand{\nn}{\nonumber}

\usepackage{titlesec}

\titleformat*{\section}{\Large\bfseries}
\titleformat*{\subsection}{\large\bfseries}
\titleformat*{\subsubsection}{\large\bfseries}
\titleformat*{\paragraph}{\large\bfseries}
\titleformat*{\subparagraph}{\large\bfseries}

\makeatletter
\newcommand*{\prodsym}{%
  \DOTSB
  \mathop{
    \mathchoice
      {\rlap{\kern.3em\rotatebox[origin=c]{-90}{}}{\prod}}
      {\vcenter{\rlap{\kern.2em\rotatebox[origin=c]{-90}{}}}{\prod}}
      {\sum}{\sum}
  }\slimits@
}
\makeatother

\makeatletter
\DeclareFontFamily{OMX}{MnSymbolE}{}
\DeclareSymbolFont{MnLargeSymbols}{OMX}{MnSymbolE}{m}{n}
\SetSymbolFont{MnLargeSymbols}{bold}{OMX}{MnSymbolE}{b}{n}
\DeclareFontShape{OMX}{MnSymbolE}{m}{n}{
    <-6>  MnSymbolE5
   <6-7>  MnSymbolE6
   <7-8>  MnSymbolE7
   <8-9>  MnSymbolE8
   <9-10> MnSymbolE9
  <10-12> MnSymbolE10
  <12->   MnSymbolE12
}{}
\DeclareFontShape{OMX}{MnSymbolE}{b}{n}{
    <-6>  MnSymbolE-Bold5
   <6-7>  MnSymbolE-Bold6
   <7-8>  MnSymbolE-Bold7
   <8-9>  MnSymbolE-Bold8
   <9-10> MnSymbolE-Bold9
  <10-12> MnSymbolE-Bold10
  <12->   MnSymbolE-Bold12
}{}

\let\llangle\@undefined
\let\rrangle\@undefined
\DeclareMathDelimiter{\llangle}{\mathopen}%
                     {MnLargeSymbols}{'164}{MnLargeSymbols}{'164}
\DeclareMathDelimiter{\rrangle}{\mathclose}%
                     {MnLargeSymbols}{'171}{MnLargeSymbols}{'171}
\makeatother

\begin{document}

\vspace*{4mm}

\thispagestyle{empty}

\begin{center}

\begin{minipage}{20cm}
\begin{center}
\hspace{-5cm }
\huge
\sc
Running Love Numbers      \\   
\hspace{-5cm } of Charged Black Holes
\end{center}
\end{minipage}
\\[30mm]

\renewcommand{\thefootnote}{\fnsymbol{footnote}}

{\large  
Sergio~Barbosa$^{\, a}$ \footnote{sergio.barbosa@aluno.ufabc.edu.br}\,, 
Sylvain~Fichet$^{\, a,b}$ \footnote{sylvain.fichet@gmail.com}\,, 
Lucas~de~Souza$^{\, c,d}$ \footnote{souza.l@ufabc.edu.br}\,, 
}\\[12mm]
\end{center}

\indent \; ${}^a\!$ 
\textit{CCNH, Universidade Federal do ABC,} \textit{Santo Andr\'e, 09210-580 SP, Brazil}

\indent \; ${}^b\!$ 
\textit{IFT - UNESP \&  ICTP - SAIFR,} \textit{ S\~ao Paulo,  2019-032 SP, Brazil}

\indent \; ${}^c\!$ 
\textit{CMCC, Universidade Federal do ABC,} \textit{Santo Andr\'e, 09210-580 SP, Brazil}

\indent\; ${}^d\!$\textit{
ITP, TU Wien, Wiedner Hauptstrasse 8–10/136, A-1040 Vienna, Austria}
\\



\addtocounter{footnote}{-3}

\vspace*{8mm}
 
\begin{center}
{  \bf  Abstract }
\end{center}
\begin{minipage}{15cm}
\setstretch{0.95}

Loops of virtual particles from the vacuum of quantum field theory (QFT) render  black holes tidally deformable. 
We compute the  static tidal response of unspinning charged black holes at arbitrary radius, using the perturbative formalism developed in \cite{Barbosa:2025uau}.  Since the gravitational and electromagnetic tidal responses  mix, we generalize
the notion of Love numbers  to Love matrices. 
We derive the coupled equations of motion for the metric and electromagnetic fluctuations around  purely  electric and magnetic backgrounds.  For large charged black holes,  which are described by the Effective Field Theory (EFT) of gravity,  we compute the full set of  Love matrices induced by an arbitrary tower of  $F^{2n}$ operators. 
We find that, although quantum corrections break electromagnetic duality, the Love matrices in  electric and magnetic backgrounds  are  related by a $Z_2$ symmetry under electric-magnetic exchange. 
Going beyond EFT, we compute the Love matrices of small magnetic black holes. We show that the running of the Love matrices  is  governed by  the running of the $U(1)$ gauge coupling, and we derive the  correspondence between  Love and  $U(1)$ beta functions for arbitrary harmonics. 
The overall picture that emerges is that the QFT-induced tidal response of magnetic black holes saturates in the strong-field regime. 
These results imply that  nearly-extremal magnetic black holes charged under an Abelian dark sector could be probed by  gravitational-wave observations.

    \vspace{0.5cm}
\end{minipage}

\newpage
\setcounter{tocdepth}{2}

\tableofcontents  

\vspace{1cm}
\hrule
\vspace{1cm}

\section{Introduction \label{se:Intro}}

In four-dimensional General Relativity (GR), black hole solutions in the {classical} vacuum exhibit no static response to most  tidal fields. Such black holes behave as perfectly rigid objects\,\cite{1972ApJ...175..243P,Martel:2005ir,Fang:2005qq,Damour:2009va,Binnington:2009bb,Kol:2011vg,Landry:2015cva,Landry:2015zfa,Porto:2016pyg,Poisson:2020mdi,LeTiec:2020spy,LeTiec:2020bos,Chia:2020yla,Goldberger:2020fot,Hui:2020xxx,Charalambous:2021mea,Ivanov:2022qqt,Pereniguez:2021xcj,Rai:2024lho,Landry:2014jka,Charalambous:2021kcz,Hui:2021vcv,Hui:2022vbh,Charalambous:2022rre,
Berens:2025jfs,Sharma:2025xii,
Charalambous:2025ekl,Xia:2025zfp}, even at the nonlinear level \cite{Poisson:2020vap,Poisson:2021yau,DeLuca:2023mio,Riva:2023rcm,Hadad:2024lsf,Iteanu:2024dvx,Combaluzier-Szteinsznaider:2024sgb,Kehagias:2024rtz,
Gounis:2024hcm,Parra-Martinez:2025bcu}.
The absence of tidal response is accidental; it results from a delicate set of conditions that are easily violated.\,\footnote{The vanishing tides
can be understood in terms of accidental symmetries arising in the near-horizon region. See \cite{Charalambous:2021kcz,Hui:2021vcv,Hui:2022vbh,Charalambous:2022rre,Sharma:2024hlz,Combaluzier-Szteinsznaider:2024sgb,Kehagias:2024rtz,Gounis:2024hcm,Parra-Martinez:2025bcu,Berens:2025jfs,Sharma:2025xii} for developments. On the other hand,  dynamical  Love numbers are found to be non-zero \cite{Mandal:2023hqa,Chakraborty:2025wvs}. The static  Love numbers of
asymptotically (anti)-de Sitter black holes do not vanish either, see e.g. \cite{Emparan:2017qxd, Nair:2024mya}. } 
 In particular, a black hole admits non-vanishing tides if \textit{(i)} its environment is non-empty \cite{Pani:2019cyc,Datta:2019epe,DeLuca:2021ite,DeLuca:2022xlz,Coviello:2025pla} or \textit{(ii)} GR is modified \cite{Cardoso:2018ptl,
 Chakravarti:2018vlt,Cai:2019npx,
 DeLuca:2022tkm, Charalambous:2022rre, 
 Katagiri:2024fpn,Chakraborty:2024gcr,
 Barbosa:2025uau, Cano:2025zyk,Bhattacharyya:2025slf,Garcia-Saenz:2025urd}.

In the real world, however, vacuum is not so empty since it exhibits quantum fluctuations.  These fluctuations are described in Quantum Field Theory (QFT)  as bubbles of virtual particles. 
The resulting ensemble of vacuum bubbles surrounding a black hole may be interpreted as a non-empty environment, and can therefore exhibit non-vanishing tidal responses  \cite{Barbosa:2025uau}.

Computing black hole properties  in the presence of vacuum fluctuations is  generally challenging, but  a simplification arises when the black hole horizon is much larger than the Compton wavelengths of the virtual particles. In this regime, the effect of vacuum bubbles can be encapsulated in a long-distance effective field theory (EFT) of gravity \cite{Donoghue:1995cz,Burgess:2003jk}. This gravitational EFT corresponds to GR modified by higher-curvature operators. The effect of heavy particle loops can then be viewed either as a nontrival environment (departing from \textit{(i)}) or as a modification of GR itself (departing from \textit{(ii)}). For example, in the gravitational EFT of the real world (GREFT) \footnote{Our  use of the name  {GREFT} is by analogy with the SMEFT, the Standard Model  EFT that describes the real world of particles.}, the leading tidal effects for neutral black holes originate from neutrino loops, while for charged black holes they come from electron loops \cite{Barbosa:2025uau}.~\footnote{Apart from \cite{Barbosa:2025uau}, computations of EFT-induced tidal responses for neutral black holes can also be found in \,\cite{Cardoso:2018ptl, Cai:2019npx, DeLuca:2022tkm,Charalambous:2022rre, Cano:2025zyk}.
We emphasize that the Riem$^4$ operators motivated by superstring theories \cite{Jack:1988sw,Gross:1986iv,Gross:1986mw,Kikuchi:1986rk,Becker:2006dvp} are  negligible with respect to the particle-induced Riem$^3$. The Riem$^3$ operator vanishes in supersymmetric theories \cite{Goon:2016mil, Barbosa:2025uau}, however this is irrelevant for the real world. 
Other aspects of black hole physics affected by the GREFT   include  gravitational waves \cite{
Endlich:2017tqa,Brandhuber:2019qpg,AccettulliHuber:2020dal,Cayuso:2023xbc,Cardoso:2019mqo,McManus:2019ulj,deRham:2020ejn,Cano:2020cao,Sennett:2019bpc,Silva:2022srr,Maenaut:2024oci}, 
quasinormal modes  \cite{Cano:2023jbk,Miguel:2023rzp,Melville:2024zjq, Cano:2024ezp,Cano:2024wzo} 
and UV conjectures on extremal black holes \cite{Kats:2006xp,
Cheung:2014ega,
Endlich:2017tqa,
Cheung:2018cwt,
Hamada:2018dde,
Loges:2019jzs,Goon:2019faz,Jones:2019nev, Chen:2019qvr, Bellazzini:2019xts,Loges:2020trf,Arkani-Hamed:2021ajd,Cao:2022iqh,DeLuca:2022tkm, Bittar:2024xuc, Knorr:2024yiu}. 
More generally, gravitational EFTs are constrained by infrared consistency conditions based on causality and unitarity, see e.g. \cite{ 
Camanho:2014apa,
 Goon:2016une, 
Bellazzini:2015cra, Cheung:2016wjt,Hamada:2018dde, Chen:2019qvr, 
Bellazzini:2019xts, 
Arkani-Hamed:2020blm,Alberte:2020jsk,Alberte:2020bdz,
Arkani-Hamed:2021ajd,
Bern:2021ppb, 
deRham:2021bll,
 Caron-Huot:2022ugt,
Caron-Huot:2022jli,
Hamada:2023cyt,
Bellazzini:2023nqj,
Bittar:2024xuc, 
Eichhorn:2024wba,
Knorr:2024yiu}.
  }

Naturally, when a charged black hole is deformed by a gravitational tidal field, a response in the  electromagnetic field can be expected, and vice-versa. A complete description of charged black hole tides should therefore take into account the  coupled   gravitational and electromagnetic responses. A computation of this kind has been done for pure Reissner-Nordstr\"om (RN) black holes in arbitrary dimension, in which the results were expressed as a function of the relative intensity of the sources {\cite{Pereniguez:2021xcj}}. In the case of EFT-deformed geometries, our computation of the charged black hole gravitational response made in \cite{Barbosa:2025uau} ignored the mixing to electromagnetism, hence the results should be completed.

The first goal of the present work is thus to extend the computation of the EFT-induced tides of large black holes of \cite{Barbosa:2025uau}, developing a complete treatment of the tidal response of unspinning electric and magnetic black holes in the EFT regime. To this end we introduce the notion of  {\it Love matrices} (see subsection below).

The second goal of this work is to reveal what happens to the tidal response of \textit{small} black holes, that lie beyond the EFT regime \cite{Barbosa:2025smt}.
Our focus is on  charged black holes, both because computations are simpler and because the physical effects are larger than for neutral ones. 
 Beyond the EFT regime, electrically charged black holes efficiently evaporate due to Schwinger effect, making the notion of static tidal response irrelevant. The situation is completely different for magnetically charged black holes, which can evaporate only into magnetic monopoles. Assuming that magnetic monopoles have Planckian mass,  small magnetic black holes do not evaporate, hence we can meaningfully study their tides.

\subsection{Love Numbers and Love Matrices}

In GR, the black hole response to a given tidal field $\psi$ is encoded  into a set of \textit{Love numbers}. These can be computed through the asymptotic behavior of the solution to the tidal equation of motion that is regular on the horizon.
 The asymptotic expansion  of the $\ell$-th harmonic takes the form
\be \psi_{\ell}(r) \Big|_{\rm regular}  \underset{r\gg r_h}{\sim}   r^{\ell+1}\,a_{\ell}+ \ldots + \frac{b_{\ell}}{r^{\ell}}+ \ldots
\equiv  r^{\ell+1} \left( 1+\ldots 
+  k_\ell \left(\frac{r_h}{r}\right)^{2\ell+1}    +\ldots \right) a_{\ell} \,,
\label{eq:TLN_def}
\ee  where
the dimensionless coefficient $k_\ell$ is the   Love number for   harmonic $\ell$.

In the presence of various coupled tidal fields,  the responses can mix. We set up the most general description for $N$ independent fields ${\bm \psi}_\ell=(\psi^{\alpha_1},\ldots,  \psi^{\alpha_N})^t$ as follows. 
The asymptotic expansion  of the  $\ell$-th  harmonic takes the form
\begin{align}
{\bm \psi}_{\ell}(r) \Big|_{\rm regular}  &\underset{r\gg r_h}{\sim}
r^{\ell+1} {\bm a}_{\ell+1}+\ldots 
+  \frac{{\bm b}_\ell }{r^{\ell}}   +\ldots \\ &\equiv r^{\ell+1} \left( \mathds{1}+\ldots 
+  K_\ell \left(\frac{r_h}{r}\right)^{2\ell+1}    +\ldots \right) {\bm a}_{\ell+1} 
,\end{align} where $K_\ell$ is the   \textit{Love matrix} for   harmonic $\ell$. 
Details and properties of the Love matrices are further discussed in section \ref{se:Love_number}.

\subsection{Review: Black Holes as Point Particles}
\label{se:WEFT}

Viewed from far away, a black hole looks just like a point particle living in flat space. Hence black holes can, just like  any other spatially localized object, be described via a long-distance  EFT. At leading order, this EFT simply encodes the worldline  of the point particle. The  subtler information about the object's shape and response to external  fields are encoded order by order into local higher-dimensional operators (see e.g. \cite{Goldberger:2004jt, Goldberger:2005cd,Porto:2016pyg,Hui:2020xxx,Ivanov:2022hlo}).\,\footnote{
The worldine EFT also has application in other fields such as atomic physics \cite{Burgess:2017mhz} or superradiance \cite{Endlich:2016jgc}. The  electromagnetic polarizability of composite objects such as neutral ions \cite{Feinberg:1968zz} or neutral strings \cite{Fichet:2016clq}, which amounts to  deformability  under a vector tidal field,  
is also similarly described via EFTs. 
} 

\paragraph{Worldline EFT.}
Schematically, the worldline EFT describing the black hole response to  $N$   gauge-invariant external fields encoded in a vector ${\bm \phi}$ is
\begin{align}
&S_{\rm WL}[{\bm \phi}]=\int d\tau e \left( {\cal L}_{\rm point} + {\cal L}_{\rm quad}[{\bm \phi}] \right)+S_{\rm kin}[{\bm \phi}] \,, \\
& {\cal L}_{\rm quad}[{\bm \phi}] = \sum_{\ell=1}^\infty 
\frac{1}{2\ell!} {\cal O}_\ell\,,\quad\quad 
{\cal O}_\ell = \left(\partial_{(i_1} \partial_{i_2}\ldots \partial_{i_\ell)_{\rm T}}{\bm \phi} \right)^t \Lambda_\ell \left(\partial_{(i_1} \partial_{i_2}\ldots \partial_{i_\ell)_{\rm T}}{\bm \phi} \right)\,.
\end{align}
The leading term ${\cal L}_{\rm point}$ is simply the Polyakov  Lagrangian for a point particle in Minkowsky space. 
The ${\cal L}_{\rm quad}$ effective Lagrangian 
describes the tidal responses, with $\Lambda_\ell$ a $N\times N$ a symmetric matrix of coefficients.  
The quadratic operators ${\cal O}_\ell$  are defined such that  each of them transform in a different irreducible representation of the spatial rotation group, and thus describes the deformability of the black hole in a given spherical harmonic $\ell$.

\paragraph{Matching.}
The worldline EFT can be used as the infrared EFT of the black hole solutions computed in the full, curved space theory. 
Namely, the matrix $\Lambda_\ell$ can be computed from the Love matrix $K_\ell$ by matching physical observables from both theories. 
Since the ${\cal O}_\ell$ operators are quadratic, a convenient setup   to probe them is to assume that the tidal field is sourced in a given harmonic. The strength of the response, computed perturbatively on both worldline EFT and gravity sides, gives access to $\Lambda_\ell$.
The matching is performed on a sphere with radius $r|_{\rm matching}\equiv L$. This aspect is important when the coefficients feature a renormalization flow, in which case $\Lambda_\ell = \Lambda_\ell(L)$. {In summary, the matching can be viewed as}
\be \hbox{\includegraphics[width=0.7\linewidth,trim={1.5cm 6cm 2cm 5cm},clip]{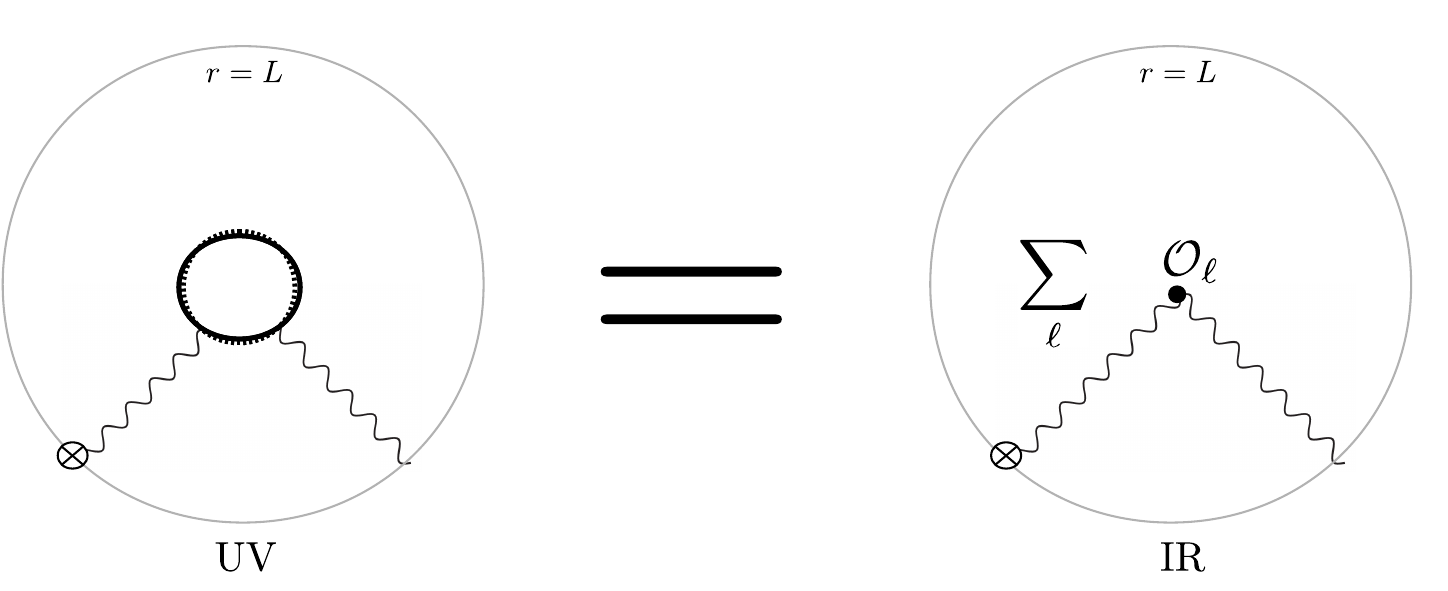}} \nn \ee

\noindent 
The correspondence  between the elements of $K_\ell$ and $\Lambda_\ell$   involves a normalization factor $N_\ell$ such that 
\be
(\Lambda_\ell)_{ij} \equiv (N_\ell)_{ij} (K_\ell)_{ij}\,.  
\label{eq:matching_coef}
\ee

\paragraph{Renormalization Flow.}

Like in any EFT, the coefficients may experience a logarithmic dependence on the matching scale $L$. 
Applying $-L \frac{d}{dL}$  gives the beta function of the wordline EFT coefficients, 
\be
\beta_{\Lambda_\ell} \equiv - \frac{d}{d\log L} \Lambda_\ell(L) \,. 
\ee
The minus sign is introduced to match the usual definition of the beta function from QFT. 
Formally, this beta function describes how the wordline EFT coefficient changes if we look at the theory at different scales $L$. 
It also  describes a physical phenomenon. Starting from an observed value $\Lambda_0$ at a scale $r_0$, the beta function controls how the observable effect changes at a scale $r_1$. At first order, $\beta_\Lambda$ is constant and   the renormalization flow simply is $\Lambda(r_1)=\Lambda(r_0)+ \beta_\Lambda\log\left(\frac{r_1}{r_0}\right)$, where $\Lambda(r_0)=\Lambda_0$.\,\footnote{The renormalization flow of Love numbers was first pointed out in \cite{Kol:2011vg} for non-physical (half-integer) values of $\ell$ and in \cite{Charalambous:2022rre,DeLuca:2022tkm}
for the $R_{\mu\nu}^3$ operator.  A running of the dynamical Love numbers is also discussed in \cite{Mandal:2023hqa}. 
A general perturbative formula for the Love beta functions is derived in \cite{Barbosa:2025uau}, and general considerations along these lines are also presented in 
 \cite{Garcia-Saenz:2025urd}. 
The running of dimensionless quantities at the classical level might seem surprising at first view, but this phenomenon
also occurs in certain holographic quantities, see e.g. \cite{Contino:2002kc,Randall:2001gb,Goldberger:2002hb,Fichet:2019owx,Fichet:2021xfn}. }

\subsection{Outline}

This work can be considered as a continuation of \cite{Barbosa:2025uau}. While we review essential content for self-containedness, a number of details exposed in \cite{Barbosa:2025uau} will not be repeated here.

Here is a roadmap for the paper.
In section  \ref{se:BH} we define the background. We present the charged black hole spacetime corrected by quantum effects and review some of its properties, including the existence of a strong-field regime and the dominance of electromagnetic corrections. 
In section \ref{se:Love_number} we formally introduce the Love numbers and Love matrices. We extend the perturbative formalism of \cite{Barbosa:2025uau} to compute them. 
In section \ref{se:fluctuations} we introduce the fluctuations. We compute by brute force the equations of motion for fluctuations in the purely electric and magnetic backgrounds. 
In section  \ref{se:weak_field} we compute the Love numbers and matrices in the weak-field regime, i.e. for sufficiently large black holes.
In section  \ref{se:strong_field} we compute the Love numbers and matrices in the strong-field regime, i.e. for sufficiently small black holes.
Section \ref{se:conclusion} summarizes. 
The Appendix contains technical details about variational calculations (\ref{se:Formulas}), 
the gravitoelectromagnetic mixing  (\ref{app:mixing}), 
computation of the Love numbers and matrices 
  (\ref{se:Test_Fields}, \ref{app:Computing_Love}),
  an explicit cross-check (\ref{app:Derivation}), 
and details on the results~(\ref{app:Love_beta_gen}).

 \section{Charged Black Hole Geometry and  the Quantum Vacuum } \label{se:BH}

In this section, we compute the spacetime background. We lay down the basic formalism to compute charged black hole solutions beyond pure GR, and compute the corrections to geometry induced by matter loops  in both weak and strong field regimes. The weak field regime is described by a generic EFT, i.e. a power series of higher-dimensional operators. The strong field regime is dominated by the gauge sector corrections, and is thus governed by Euler-Heisenberg-type effective actions \cite{Dunne:2004nc}. 

\subsection{Quantum Effective Action and Effective Field Theory}

Consider a theory of gravity coupled to a $U(1)$ gauge field $A_\mu$ and a generic  charged field $\Phi$ with mass $m$ and charge $q$. By integrating out exactly $\Phi$ in the functional formalism, we obtain a quantum effective action $\Gamma$:
\begin{equation}
    e^{i\Gamma[g_{\mu\nu},A_\mu]} = \int \mathcal{D}\Phi \, \exp{\left[i \int d^4x \sqrt{-g}\left(\frac{R}{2\kappa^2} -\frac{1}{4e^2}F_{\mu\nu}F^{\mu\nu} + \mathcal{L}_\Phi[\Phi, g_{\mu\nu}, A_\mu] \right) \right]}\,,
    \label{eq:Partition_function}
\end{equation}
where $R$ is the Ricci scalar, and $F_{\mu\nu} = \nabla_\mu A_\nu - \nabla_\nu A_\mu$  the $U(1)$ field strength.
$\Gamma[g_{\mu\nu},A_\mu]$ is generally non-local. 
It must respect  the diffeomorphism and $U(1)$ gauge invariances of the parent theory, hence it should be built from covariant derivatives  $\nabla$, from the curvature tensors $R$, $R_{\mu\nu}$, $R_{\mu\nu\rho\sigma}$, and from the field strength $F_{\mu\nu}$ and its dual $\Tilde{F}_{\mu\nu} = \frac{1}{2}\epsilon_{\mu\nu\rho\sigma}F^{\rho\sigma}$. 
It follows that we  can generically write the quantum effective action as 
\begin{align}
\Gamma[R_{\mu\nu\rho\sigma},F_{\mu\nu},\nabla] &= \int d^4x \sqrt{-g} \Bigg[\frac{R}{2\kappa^2} - \frac{1}{4e^2}F_{\mu\nu}F^{\mu\nu} +  
     \Delta {\cal L}[R_{\mu\nu\rho\sigma},F_{\mu\nu},\nabla] \Bigg]\label{eq:effective_action}\,. 
\end{align}

The $\Delta {\cal L}$ piece encodes  all the quantum effects from the $\Phi$ field. Diagrammatically, $\Delta {\cal L}$  contains  all the vacuum bubbles of $\Phi$ with an arbitrary number of insertions of the background fields. More generally, we can say that it  describes the effects of fluctuations of the QFT vacuum. $\Delta {\cal L}$ will be treated perturbatively throughout this work.

Depending on the physical scales involved in the system studied, the quantum effective action can be organized as an expansion in the derivative   $\nabla$,  in  the curvature (denoted Riem), or in the $U(1)$ field strength  (denoted $F$). 
 In the gravitational sector, derivatives can be converted into curvatures; the $\nabla$ that can be independently expanded are those acting on $F$.

Throughout this work, we always use the expansion in $\nabla$ and in curvature. 
By contrast, the expansion in $F$ is not systematically done, because we consider both the weak-field and strong-field regimes of electromagnetism.  In fact, 
the delimitation between these weak and strong regimes is precisely given by the convergence of the expansion in $F$:  such expansion  is valid in the weak-field regime, whereas the strong-field regime requires a specific UV completion ---  which we take to be Euler-Heisenberg electrodynamics. The formal description of these points is provided further below in this section (see in particular subsections \ref{se:BH_properties} to   \ref{se:Deformation_strong_field}).

Based on the above assumptions, the general structure of our $\Delta {\cal L}$ Lagrangian is 
\begin{align}
& \Delta {\cal L}\left[R_{\mu\nu\rho\sigma},F_{\mu\nu},\nabla\right]  =   ~\Delta {\cal L}_F[F_{\mu\nu}] \times  \left(1+{\cal O}\left( \frac{\nabla^2}{m^2} \right)\right) \nonumber  \\ \nonumber
& +  \alpha_1 R^2 + \alpha_2 R_{\mu\nu}R^{\mu\nu} 
 + \beta_1\frac{q^2}{m^2}  R F_{\mu\nu}F^{\mu\nu} + \beta_2 \frac{q^2}{m^2} R_{\mu\nu}F^{\mu\alpha}F\indices{^\nu _\alpha} + \beta_3 \frac{q^2}{m^2} R_{\mu\nu\rho\sigma}F^{\mu\nu}F^{\rho\sigma}  \\ 
& + {\cal O} \left( \frac{{\rm Riem}^3}{m^2}, \frac{q^2{\rm Riem}^2\,F^2}{m^4}, \frac{q^4{\rm Riem}\, F^4}{m^6} \right)\,. \label{eq:DeltaL}
\end{align}
The first line represents a generic function $\Delta {\cal L}_F[F_{\mu\nu}]\equiv \Delta {\cal L}_F[F^2, F\tilde F]$ that constitutes a correction to Maxwell electromagnetism. $\Delta {\cal L}_F$ is assumed to be analytical in $F^2$, $F\tilde F$. 
The second line contains all the  EFT operators at $\partial^4$ order involving curvatures. The $\alpha_i$, $\beta_i$ are dimensionless,  the $m$-dependence is fixed by dimensional analysis, using the assumption that the effective operators originate from integrating out the $\Phi$ field.

\subsection{Field Equations}

Varying \eqref{eq:effective_action} provides the  coupled Maxwell and Einstein equations.  The contribution  of $\Delta {\cal L}$ can be conveniently cast in terms of effective sources. The field equations read
\begin{align}
    \nabla_\mu F^{\mu\nu} &= e^2 \mathcal{J}^{\nu}\,, \quad \quad \nabla_\mu \Tilde{F}^{\mu\nu} = 0\,, \quad \quad \mathcal{J}^{\mu} = -\frac{\partial \Delta {\cal L}}{\partial A^{\mu}}\,,\label{eq:Maxwell} \\
    {{G_{\mu\nu}}}&= \kappa^2 T_{\mu\nu}\,, \quad \quad T_{\mu\nu} = T_{\mu\nu}^{\mathrm{e.m.}} + \mathcal{T}_{\mu\nu}\,, \quad\quad \mathcal{T}_{\mu\nu} = \frac{-2}{\sqrt{-g}}\frac{\partial( \sqrt{-g}\Delta {\cal L})}{\partial g^{\mu\nu}}\,, \label{eq:Einstein}
\end{align}
with {$G_{\mu\nu}=R_{\mu\nu} - \frac{1}{2} R g_{\mu\nu}$ and} $   T_{\mu\nu}^{\mathrm{e.m.}} = \frac{1}{e^2} F_{\mu\rho}F\indices{_\nu ^\rho} - \frac{1}{4 e^2} g_{\mu\nu}F_{\rho\sigma}F^{\rho\sigma}$. 
The effective  current ${\cal J}_\mu$ and effective stress–energy tensor ${\cal T}_{\mu\nu}$ must satisfy the conservation laws $\nabla_\mu \mathcal{J}^\mu = 0$ and $\nabla_\mu \mathcal{T}^{\mu\nu} = 0$,  as a consequence  of the $U(1)$ gauge and diffeomorphism invariances, respectively.

\subsection{Charged Black Hole Solutions}

Throughout this section we compute non-spinning charged black hole solutions of the coupled Maxwell-Einstein field equations \eqref{eq:Maxwell}, \eqref{eq:Einstein}. The general ansatz assumed for the metric is 
\begin{align}
    ds^2 = - A(r) dt^2 + \frac{dr^2}{B(r)} + r^2 d\Omega^2\,.
\end{align}
An analysis of the field equations (see e.g. \cite{Barbosa:2025smt}, from which we follow most conventions) reveals that, if the curvature operators are neglected, which is our working assumption,  then 
\be
A(r)=B(r)\equiv f(r)\,. 
\ee
This is tied to the fact that $T\indices{^t_t}=T\indices{^r_r}$ for an effective stress–energy tensor arising purely from corrections to the Maxwell sector.\footnote{This property follows from Einstein's equation:
\begin{equation}
    G\indices{^t_t} - G\indices{^r_r} = \frac{A(r)}{r}\frac{d}{dr}\left(\frac{B(r)}{A(r)} \right) = \kappa^2 (T\indices{^t_t} - T\indices{^r_r})\,, \label{eq:EE}
\end{equation} 
which implies  $B(r) \propto A(r)$ when $T\indices{^t_t} = T\indices{^r_r}$. Requiring that the metric reduces to the canonical Minkowski metric asymptotically fixes the proportionality constant to $1$. Conversely, when curvature operators such as those shown in \eqref{eq:DeltaL} are included, one has  $T\indices{^t_t} \neq T\indices{^r _r}$, and \eqref{eq:EE} shows that the functions $A(r)$ and $B(r)$ are, in general, different.}

We treat the $\Delta {\cal L}$ Lagrangian perturbatively. Notice  this is a semiclassical calculation since, by definition, $\Delta {\cal L}$ encodes the effects of the quantum vacuum. The solutions to the field equations will take the general form
\begin{align}
    f(r) &= f^{(0)}(r)+f^{(1)}(r) +\ldots\,, \label{eq:metric_expansion}\\
    {\cal E}(r) &= F_{rt}= {\cal E}^{(0)}(r)+{\cal E}^{(1)}(r) +\ldots\,, \quad \quad {\cal B}(r)= -\Tilde{F}_{rt} ={\cal B}^{(0)}(r)+{\cal B}^{(1)}(r) +\ldots\,,
\end{align}
where $\mathcal{E}(r)$ and $\mathcal{B}(r)$ are the radial electric and magnetic fields, respectively. The deformations $f^{(1)}$, ${\cal E}^{(1)}$, ${\cal B}^{(1)}$ are computed  in subsections \ref{se:Deformation_weak_field}, \ref{se:Deformation_strong_field}.

\subsubsection{Geometry at leading order}\label{se:RN_metric}

At zeroth-order, the action in the expansion \eqref{eq:effective_action} is simply GR coupled to Maxwell electromagnetism. Hence, the charged black hole geometry is described by the Reissner–Nordström solution:
\begin{align}
    f^{(0)}(r) = 1 - \frac{\kappa^2 M}{r} + \frac{\kappa^2 Q^2}{2 r^2}\,,
\end{align}
where $M$ is the black hole mass and $Q$ represents either the total electric charge $Q_e$ or the magnetic charge $Q_m$. Spherical symmetry and time independence impose that the electromagnetic field takes a Coulomb form. The  electric or magnetic field take respectively the form\,\footnote{The common $e$ factor comes from the $\frac{1}{e^2}$ normalization of the kinetic term, see  \eqref{eq:Partition_function}. }
\begin{equation}
    \mathcal{E}^{(0)}(r)  = \frac{e Q_{e}}{r^2}\,, \quad \quad \mathcal{B}^{(0)}(r) =  \frac{e Q_{m}}{r^2}\,. \label{eq:zeroth_E_B}
\end{equation}
Throughout this work we consider black holes carrying either a purely electric or magnetic charge, thus we denote the black hole charge as $Q$ without ambiguity.

The inner and outer horizon radii obtained from the roots of $f^{(0)}(r)$ are given by
\begin{equation}
    r_{\pm} = \frac{1}{2}\Big(\kappa^2 M \pm \kappa \sqrt{\kappa^2 M^2 - 2 Q^2} \Big)\,.
\end{equation}
When the two horizons coincide, i.e. for $\kappa M = \sqrt{2} |Q|$, the black hole is extremal and the horizon radius reduces to $r_h = \frac{1}{2}\kappa^2 M$.

\subsubsection{On mass, charge, and horizon parametrization}\label{sec:parametrization}

Once the black hole is perturbed, the parameters $M$ and $Q$ do not necessarily coincide with the physical mass and charge measured by an asymptotic observer. The parameters are corrected as $M \rightarrow M + \delta M$ and $Q \rightarrow Q + \delta Q$, where $\delta M$ and $\delta Q$ are first-order corrections. The horizon radii are shifted accordingly $r_\pm \rightarrow r_\pm + \delta r_\pm$, where the first-order correction is determined by imposing $f(r_\pm + \delta r_\pm) = 0$.

Expanding this condition to first order, using \eqref{eq:metric_expansion} and taking into account the shifts in  mass, charge, and horizon radius, we obtain
\begin{equation}
    \partial_M f^{(0)}(r_\pm) \delta M + \partial_Q f^{(0)}(r_\pm) \delta Q + \frac{d f^{(0)}(r_\pm)}{dr} \delta r_\pm + f^{(1)}(r_\pm) = 0\,,\label{eq:first_law}
\end{equation}
where we have used $f^{(0)}(r_\pm)=0$. As a side note, notice that \eqref{eq:first_law}  resembles the first law of thermodynamics for a non-rotating charged black hole. This implies that EFT corrections may be interpreted as modifications of the black hole thermodynamic quantities.

Equations \eqref{eq:first_law} provide two constraints for the four variables $\delta M$, $\delta Q$, $\delta r_+, \delta r_-$. Therefore, one may solve for any two of them after choosing the remaining two arbitrarily. For example, one may choose $\delta M = \delta Q = 0$, so that the $M$ and $Q$ coincide with the physical mass and charge, while the horizon radii receive corrections. This is the parametrization adopted throughout this work. Alternatively, one may choose $\delta r_\pm = 0$, so that both horizon radii remain unchanged, while the corrections are included into $\delta M$ and $\delta Q$  as in \cite{Noumi:2026shc}.

\subsection{Black Hole  Properties from Dimensional Analysis}

\label{se:BH_properties}

Before computing the deformations to the black hole geometry induced by $\Delta {\cal L}$, it is useful to  identify the scales of the system and the related expansion parameters. As we will see, dimensional analysis alone is highly instructive. 

We estimate the building blocks of $\Delta {\cal L}$, namely $\nabla$, $\rm Riem$ and $F_{\mu\nu}$, in the black hole background.  This section extends the analysis made in \cite{Barbosa:2025smt}. 
Focusing first on these quantities near the outer horizon, we find 
\footnote{Throughout this section we ignore  the numerical factors in the estimates.  }
\be
\nabla \sim \frac{1}{r_+}\,,\quad \quad {\rm Riem} \sim \frac{1}{r_+^2}, \,\quad\quad F_{\mu\nu}\sim \frac{ e Q}{r_+^2}\,. 
\ee
The corresponding dimensionless expansion parameters, chosen to carry two derivatives each, can be written as
 \be
\frac{\square}{m^2 } \sim \frac{\rm Riem}{m^2} \sim \frac{1}{m^2 r^2_+} \equiv \varepsilon \,,\quad  \quad 
\frac{q^2 F^2}{m^4}\sim \frac{ q^2 e^2 Q^2 }{ m^4 r^4_+ } \equiv\varepsilon_F\,.
\label{eq:expansion_parameters}
 \ee

We see that  $\epsilon_F$ fundamentally differs from $\varepsilon$ due   to the dependence on the black hole charge. 
The $\varepsilon$ parameter controls the derivative and curvature expansion. Our focus in this work is on $\varepsilon < 1 $. This corresponds to assume that $r_+>\frac{1}{m}$, i.e. the Compton wavelength of the charged particle is smaller than the black hole radius. The $\varepsilon_F$ parameter controls the electromagnetic field expansion. 

\subsubsection{Weak and strong field regimes } \label{eq:strong_weak_def}
 The case $\varepsilon_F<1$ corresponds to the \textit{weak-field} regime, for which ${\cal L}_F$ can be expanded and truncated, with  schematically ${\cal L}_F \sim F^4+O(F^6) $. Together with the curvature operators of \eqref{eq:DeltaL}, the full set of $\partial^4$ order operators forms the  Einstein-Maxwell EFT.  The case $\varepsilon_F>1$ corresponds to the \textit{strong-field} regime, for which ${\cal L}_F$ is not truncated. This can be generally viewed as the UV completion of the Einstein-Maxwell EFT in the electromagnetic sector, and in our present case, ${\cal L}_F$ is a Euler-Heisenberg type Lagrangian. 
  In terms of radius, the limit between both regimes is at 
  \be
r^c_+ \sim \frac{\sqrt{e |q  Q|}}{m} \,. 
\label{eq:rc_def}
\ee
Black holes with $r_+>r_+^c$ (resp. $r_+<r_+^c$) are in the weak-field (resp. strong-field) regimes. 

\subsubsection{Schwinger effect}
An electrically charged black hole can dissipate its charge through the dissociation of $\Phi\Phi^*$ particles pairs from the vacuum. The decay rate happens to be exponentially suppressed precisely for black holes satisfying $r_+>r_+^c$ (see e.g. \cite{Hiscock_Weems,Brown:2024ajk,Barbosa:2025uau}). Hence electrically charged black holes in the weak-field regime are stable to evaporation. In contrast, electrically charged black holes in the strong-field regime efficiently evaporate. The situation differs for magnetic black holes, that can only discharge into magnetic monopoles which are assumed to be heavy. Therefore we can safely consider magnetic black holes in the strong-field regime.

\subsubsection{The strong-field sphere } 
Our definition of the  strong-field regime is based on the electromagnetic field intensity near the horizon.  For a black hole in the strong-field regime, the electromagnetic field decreases with $r$, and there exists a radius $r_+^s>r_+$ at which the weak-field regime emerges. This radius  is given by the condition $\varepsilon|_{r_+^s}\sim\varepsilon|_{r_+^c}$, and therefore coincides with the critical radius, $r_+^s=r^c_+$, independently of the black hole size.

\subsubsection{Dominance of Maxwell corrections}

 In the generic Lagrangian \eqref{eq:DeltaL}, the  curvature factors   are subleading with respect to the $U(1)$ field strengths if $\varepsilon<\varepsilon_F$. In terms of the physical parameters, this translates as the condition $r_+ < \frac{e |q  Q|}{m}$. {This is more restrictive than the strong-field condition $r_+<r_+^c$}. Therefore, a black hole in the strong-field regime is automatically dominated by the field strength corrections encoded into $\Delta {\cal L}_F$ ---  the electromagnetic field is so strong that gravity corrections are neglected.\,\footnote{See also \cite{Barbosa:2025uau}
for a more detailed demonstration of this point at the level of Love numbers.} 
 This fact  is very useful because, in practice, the Maxwell effective Lagrangian is much easier to compute than the gravitational one. Moreover the computations at the level of the black hole metric are also simplified, as we will see in next section.

\subsubsection{Extremal Black Holes and Weak Gravity Conjecture}

For an extremal black hole, the critical radius becomes
\be
r_h^c \sim \frac{e |q|}{\kappa m^2}\,. 
\label{eq:rhc_def}
\ee
The extremal case maximizes the domain of strong-field regime. Moreover, as  pointed out in \cite{Barbosa:2025smt}, the existence of the strong field domain for extremal black holes  is guaranteed by the particle version of the weak gravity conjecture, which states that  $\frac{e |q|}{\kappa m^2}\gtrsim 1$. (see e.g. \cite{Arkani-Hamed:2006emk} and the reviews \cite{vanBeest:2021lhn,Grana:2021zvf, Agmon:2022thq}). This implies that $r_h^c >\frac{1}{m}$ hence the strong-field domain is non-empty unless the charged particle has Planckian mass. 

\subsubsection{Brief summary}

We have identified a weak-field and a strong-field regime for respectively $r_+>r_+^c$ and $r_+<r_+^c$. We have shown that the electromagnetic sector dominates the corrections to the geometry of any charged black hole in the strong field regime. 

\subsection{Deformed  Geometry in Weak-Field Regime
\label{se:Deformation_weak_field}
}

Given the analysis of the previous subsection,  we focus on the electromagnetic corrections produced by $\Delta {\cal L}_F$.\,\footnote{The leading curvature corrections to neutral and charged black holes have been computed in \cite{Kats:2006xp, DeLuca:2022tkm, Barbosa:2025uau}. } 
We consider purely electric or purely magnetic black holes, so that the invariant $F_{\mu\nu}\Tilde{F}^{\mu\nu}$ vanishes.

In this section, we consider the weak-field regime, for which the EFT expansion of the $\Delta {\cal L}_F$ Lagrangian applies.
We could truncate the series to $O(F^6)$ to keep only the leading $F^4$ operators, however we will be more ambitious and keep the whole series with arbitrary coefficients. We write $\Delta {\cal L}_F$ as 
\begin{equation}
 \Delta{\cal L}_F = \sum_{n=2}^\infty c_n ({F}^{2})^n\,, \quad\quad c_n = \frac{ q^{2n}}{m^{4(n-1)}}\gamma_n\,. \label{eq:EFT_action}
\end{equation}
where ${F}^2 = F_{\mu\nu} F^{\mu\nu}$.
Many of our subsequent weak-field results are presented in terms of the $c_n$ coefficients, and are therefore valid for any EFT of electromagnetism. 
The parametrization in terms of $\gamma_n$ is useful more specifically for charged matter loops.

The  corrections to the background can be obtained from the modified Einstein–Maxwell equations \eqref{eq:Einstein}, \eqref{eq:Maxwell}. Details can be found in {\cite{Barbosa:2025smt, Barbosa:2025uau}}, here we summarize the results.

{For backgrounds with $F_{\mu\nu}\Tilde{F}^{\mu\nu} = 0$, the electric and magnetic solutions to the metric differ only by the sign of the invariant ${F}^2 = 2(\mathcal{B}^2 - \mathcal{E}^2)$. It is thus convenient to treat both cases simultaneously by introducing the parameter}
\begin{equation} \label{eq:sigma}
\sigma \equiv \mathrm{sign}({F}^2)=
\begin{cases}
-1\, & \text{if electric black hole}\,, \\
+1\, & \text{if magnetic black hole}\,.
\end{cases}
\end{equation}
Using this, the first-order corrections to the metric and electromagnetic fields are found to be
\begin{align}
    f^{(1)}(r) &= - \sum_{n=2}^{\infty}  \frac{(2\sigma)^n \kappa^2 e^{2n} Q^{2n} }{(4n-3) r^{4n-2}}c_n\,, \label{eq:EFT_correction_solution0} \\
    \mathcal{E}^{(1)}(r)  &=  \sum_{n=2}^{\infty} \frac{n\,(-2)^{n+1} \, e^{2n+1} Q^{2n-1}}{r^{4n-2}}c_n 
    \,, \quad \quad \mathcal{B}^{(1)}(r) = 0\,.
    \label{eq:EFT_correction_solution}
\end{align}
Notice that the magnetic field of the black hole receives no corrections. This is
because a radial magnetic field identically satisfies the Bianchi identity, $\nabla_\mu \Tilde{F}^{\mu\nu} = 0$, outside the singularity. The asymmetry between electric and magnetic fields can be viewed as a consequence of the
EFT being generated from an electrically charged particle. As a result, vacuum polarization effects modify only electric fields.

\subsubsection{Approximations in the electric case}

In the purely magnetic case, the first-order metric corrections $f^{(1)}(r)$  do not exhibit higher-order nonlinearities, because the magnetic field receives no corrections. {Hence, $f = f^{(0)} + f^{(1)}$ is exact, with $f^{(1)}$ given by \eqref{eq:EFT_correction_solution0}.} In contrast,  nonlinearities do arise in the purely electric case. In that case, we can treat each term in the expansion as an independent first-order correction, given the following  conditions. 

Through the calculation, nonlinearities  produce contributions which are products of coefficients, that can compete with a $c_n$ with same mass dimension. Namely at order $1/m^{4(n-1)}$, the  coefficient $c_n$  has same dimension as a product $c_{n_1} c_{n_2} \cdots c_{n_N}$ if
\begin{equation}
\sum_{i=1}^{N} n_i = n + N - 1\,, \quad N \geq 2\,.
\end{equation}
However, such products carry extra powers of $qe$ relative to the individual term, because $(qe)^{2n} \gg (qe)^{2(n+N-1)}$. Therefore, if $qe \ll 1$, the corrections considered in \eqref{eq:EFT_correction_solution} can be viewed as the leading ones at each order. This fact is also evident from the matching to the  strong-field regime studied in next section. 

In contrast, we may also compare the contribution of non-linearities at fixed $ (q e)^{2n} $. We find that these contributions are negligible if $r_+<\sqrt{|Q|}/m$. On the  other hand, the  weak-field regime condition  $r_+>r_+^c$ is $r_+>\sqrt{e |qQ|}/m$. Hence the non-linearities are negligible in a domain close to the strong-field regime: 
\be \frac{\sqrt{|Q|}}{m}> r_+>\frac{\sqrt{e |qQ|}}{m}\,. \ee 
Instead, for larger $r_+$ i.e. well into the weak-field regime, we should truncate the corrections at leading order --- since  the effect of non-linearities should be taken into account at higher order.

\subsubsection{Charged matter loops}

The $\Delta {\cal L}_F$ is generated by loops of charged particles. For example, for a Dirac fermion (such as the electron) or a complex scalar,
 the Euler-Heisenberg effective action takes the form \eqref{eq:effective_action} and admits an asymptotic expansion with coefficients\,\cite{Dunne:2004nc} 
\begin{equation}
\gamma_{n}^{{\rm Dirac}} = -2 \gamma_{n}^{{\rm scalar}} = -\frac{2^{n}\,B_{2n}}{32\pi^2\, n(n-1)(2n-1)}\,. \label{eq:loops_coef}
\end{equation}
{where $B_{2n}$ are the Bernoulli numbers.}

\subsection{Deformed Geometry in Strong-Field Regime}
\label{se:Deformation_strong_field}

In the strong-field regime, since our focus is on a perturbative treatment of $\Delta{\cal L}$, we focus on the generic effective Lagrangian encoding  the $U(1)$ gauge coupling at large field values: 
\be
\Delta{\cal L}_F= \beta \frac{ F^2}{  8 }\log{\left(\frac{ F^2}{\mu_0^4} \right)}\,. \label{eq:DeltaL_LargeField}
\ee
The $\beta$ parameter can be viewed as the 1-loop beta function coefficient, $\beta_{1/e^2}=\mu \frac{d}{d\mu} \frac{1}{e^2(\mu)}$. The $\mu_0$ scale serves as a reference scale to set the value of $e$ when $F^2=\mu_0^4$. 

The form \eqref{eq:DeltaL_LargeField} can be derived by applying standard renormalization group arguments to  $\Gamma[F, \mu^2 ]$, analogous to the application to the two-point function   $\Pi[p^2, \mu^2 ]$.  It  can also be shown via trace anomaly considerations \cite{Dunne:2004nc}. 
Eq. \eqref{eq:DeltaL_LargeField} emerges from charged matter loops, in which case $\beta_{\rm Dirac}= 4 \beta_{\rm scalar} =\frac{q^2}{12\pi^2}$ \cite{Dunne:2004nc}. It also appears if the $U(1)$ gauge field couples to a CFT sector (see e.g. \cite{ArkaniHamed:2000ds,Gherghetta:2010cj,Barbosa:2025smt}), and from AdS braneworld models \cite{Pomarol:2000hp, ArkaniHamed:2000ds,Contino:2002kc,Randall:2001gb,Goldberger:2002hb,Friedland:2009zg,Fichet:2019owx}.

Following the analysis of section \ref{se:BH_properties}, we focus only on \textit{magnetic} black holes, since the electric ones efficiently evaporate in the strong-field regime. 
The resulting first-order correction to the metric is
\begin{equation}
    f^{(1)}(r) = - \beta\frac{\kappa^2 e^2 Q^2}{2 r^2}\Bigg[\log{\left(\frac{\sqrt{2}e Q}{ r^2 \mu_0^2} \right)} -2\Bigg]\,.
\end{equation}
Finally, as in the weak-field regime, the magnetic field receives no corrections, $ \mathcal{B}^{(1)}(r) = 0$.

\section{Love Numbers from Perturbation Theory}\label{se:Love_number}

This section introduces Love numbers, Love matrices and reviews our  perturbative framework for their computation. We present two simple formulaes extending that of \cite{Barbosa:2025uau} to Love matrices, Eqs.\,\eqref{eq:K_def}, \eqref{eq:B_full}.

\subsection{Tidal Equation of Motion}

Consider a generic tidal field  $\Psi$ living on a non-spinning black hole background. The black hole may be either neutral or charged. In the charged case, the $r_h$ coordinate denotes  the outer horizon.
 We denote the exact wave operator of the tidal field as ${\cal D}$.
 
Since the background is spherically symmetric, the tidal field can be decomposed in spherical harmonics, e.g. ${\Psi}=\sum_{\ell,m} Y_{\ell m} \Psi_{\ell m}$ for a scalar. Similarly, the wave operator can be decomposed in  spherical harmonics, e.g.  ${\cal D}=\sum_{\ell,m} Y_{\ell m} {\cal D}_r^{(\ell)}$. In the following, we focus on the wave operator for a given harmonic $\ell$, ${\cal D}_r^{(\ell)}$, and we omit the $\ell$ indices for simplicity.

In this section, we consider the most general case of $N$ coupled tidal fields $\Psi_{I_1},\Psi_{I_2},\ldots$, that we join in a $N$-vector 
\be
{\bm \Psi } = \begin{pmatrix}  
\Psi_{I_1} \\
\vdots \\ 
\Psi_{I_N}
\end{pmatrix} \,.
\ee

\subsubsection{Structure and Properties}

We review the analysis of \cite{Barbosa:2025uau} and extend it to multiple tides. 
The tidal equation of motion in a given harmonic is 
\begin{align}
{\cal D}_r {\bm \Psi}(r) =0\,, \quad\quad\quad  {\cal D}_r = \mathds{ 1}   \frac{d^2}{d r_\star^2} - \hat{V}(r)\,, \label{eq:EOM}   
\end{align}
where the wave operator ${\cal D}$  is a matrix. $\mathds{1}$ is the $N\times N$ identity, $\hat{V}$ is the matrix potential, and $r_\star$
 is the tortoise coordinate. This equation of motion is a coupled second order linear differential equation, satisfying the following elementary properties: 
\begin{enumerate}[label=(\roman*)]
    \item ${\cal D}_r {\bm \Psi}(r) =0$ reduces to the decoupled flat space equation of motions for  $\frac{r_h}{r}\to 0$. \label{Prop:flat}
    \item ${\cal D}_r {\bm \Psi}(r) =0$ has a regular singularity at the outer horizon $r=r_h$.   \label{Prop:sing}
    \item The coupling  is only  caused by the non-derivative term.
    \label{Prop:coupling}
\end{enumerate}
The  singular point is of the  regular kind, as writing ${\cal D}_r \Psi(r) \propto \Psi''(r)+P(r)\Psi'(r)+Q(r)\Psi(r)$  we have $P(r),Q(r)\sim \frac{1}{r-r_h}$ near the singularity. 
Given that the  equation of motion has a  regular singularity, we know that it admits only two independent solutions and that the Frobenius (i.e. generalized power series) method \cite{gray2008linear} applies.

\subsubsection{Near-horizon Behavior}

The solutions of the tidal equation of motion near the horizon can be understood using the Frobenius method for $r\sim r_h$.  In this region, the non-derivative term  $Q(r)$ is subdominant.  Hence due to Prop.\,\ref{Prop:coupling} there is no mixing. Combined with  Prop.\,\ref{Prop:sing}, one gets that the indicial equation  has degenerate roots. This  implies  that 
in the vicinity of the horizon, there is one regular solution (noted $\bm \psi$) and one that diverges logarithmically (noted $\tilde  { \bm\psi}$).
 This fact  depends only on the singularity structure of the tidal equation of motion, that is not altered in the presence of the EFT corrections.

\subsubsection{Asymptotic Behavior}

The solutions of the tidal equation of motion far from the horizon can be understood using the Frobenius method for $r\gg r_h$. To this end, we change variable to $r\propto \frac{1}{z}$ and use that the point $z=0$ is regular singular. 

Using Prop.\,\ref{Prop:flat},  we know that each of the $\Psi^{\alpha}$ has an associated indicial equation from $\Psi^\alpha\sim z^q$ giving the roots $q_1=\ell$, $q_2=-\ell -1$. It follows from the Frobenius method that whenever $\ell$ takes its physical values $\ell\in \mathbb{N}$, we have $q_1-q_2\in \mathbb{N}^*$ so that the solution translated back to the $r$ coordinates 
takes the general form
\be
{\bm \Psi}_\ell(r) = r^{\ell+1} \, \sum^{\infty}_{n=0}  {\bm a}_{n,\ell}   r^{-n}
+  \frac{1}{r^\ell}\left( { \Id } + B_\ell \log\left(\frac{r_h}{r}    \right) \right)  \sum^{\infty}_{n=0}  {\bm b}_{n,\ell} r^{-n}\,.
\label{eq:Psi_asymptotic}
\ee
The  coefficients  ${\bm a}_{n,\ell}$, ${\bm b}_{n,\ell}$  of the series are written as $N$-vectors, and  $B_\ell$ is an $N\times N$ matrix.\,\footnote{The $2N$ independent integration constants are absorbed into the definitions of the ${\bm a}_{n,\ell}$, ${\bm b}_{n,\ell}$ vectors and the $B_\ell $ matrix.  
In \eqref{eq:Psi_asymptotic}, even though the log term has been factored into the $r^{-\ell}$ term, it forms an independent solution  together with   the $r^{\ell+1}$ term.  This is the  solution with lower indicial root, consistent with the Frobenius method.   }

The presence of the logarithm in  \eqref{eq:Psi_asymptotic} is fundamental. If $B_\ell$ is  nonzero,   it causes the running of the Love matrix (see section \ref{se:WEFT}). 
The fact that only the $r^{-\ell}$ term in \eqref{eq:Psi_asymptotic} can have a log factor is an intrinsic property of the general solution dictated by the Frobenius method. 
The  asymptotic form \eqref{eq:Psi_asymptotic} is valid  in the presence of the EFT corrections, which decrease faster than the leading asymptotic terms and thus do not modify the singularity structure  at $z=0$.

\subsection{Love Matrices}
\label{se:TLM}

The Love matrix generalizes the notion of Love number. It measures the static response of the black hole geometry under a set of $N$ tidal sources with multipole structure $\ell$ and amplitudes $a^{\alpha_1}_{0,\ell}, \ldots , a^{\alpha_N}_{0,\ell}$ at $r\gg r_h$. 

The Love matrix for a given $\ell$ is obtained by considering the asymptotic behavior of the solution that is regular on the horizon. Regularity relates the integration constants. We write the regular solution as
\begin{align} {\bm \Psi}_{\ell}(r) \bigg|_{\rm regular}  &\underset{r\gg r_h} {\sim}    \,r^{\ell+1}    {\bm a}_{0,\ell}    + \frac{1}{r^{\ell}} \left(\Id + B_\ell  \log\left(\frac{r_h}{r} \right)\right)  {\bm b}_{0,\ell}     
\\
 & \,\,\,\,\, {\sim}    \,r^{\ell+1}   \left[  \Id   + \left(\bar K_\ell + \bar B_\ell  \log\left(\frac{r_h}{r} \right)\right)\left(\frac{r_h}{r}\right)^{2\ell+1} \right] {\bm a}_{0,\ell}     
\label{eq:Psireg_expansion}
\end{align}
with 
\be
  \bar K_\ell \, {\bm a}_{0,\ell} \equiv \frac{1}{r_h^{2\ell+1}} \,{\bm b}_{0,\ell} \,,\quad \quad
  \bar B_\ell\, \equiv  B_\ell \bar K_\ell  \,.
  \label{eq:barKB_def}
\ee
The dimensionless matrix $\bar K_\ell$ is the  Love matrix for harmonic $\ell$.

The logarithmic correction in \eqref{eq:Psireg_expansion}   can be viewed as a renormalization flow governed by the  dimensionless matrix $\bar B_\ell$, related to the Love beta function   $\beta_{\Lambda_\ell}$. It is convenient to introduce the $r$-dependent Love matrix
\be
  K_{\ell} \equiv \bar  K_\ell + \bar B_\ell \log\left(\frac{r_h}{r} \right) \,.\label{eq:K_def}
\ee
This matrix is matched with $\Lambda_\ell$ at the scale $r\sim L$ (see section \ref{se:WEFT}), such that the corresponding beta function is 
\be
\beta_{K_\ell}\equiv - L\frac{d}{d L} K_\ell = N_\ell \beta_{\Lambda_\ell} \,.
\ee

In practice, although $ K_{\ell}$  has logarithmic dependence in $r$, it  is often possible to treat it as a constant  from the viewpoint of subsequent calculations.  
Note, however that, as shown in \cite{Barbosa:2025uau}, the  constant terms are not physical whenever the beta function is nonzero, i.e. $ \bar B_\ell \neq 0$. In such case,  it is correct to only report  the Love beta function.

Importantly, unlike for Love numbers (i.e. $N=1$),  Love matrices depends on the choice of the tidal field basis. For a generic transformation $\mathbf{\Psi}_1 = P_{12}\mathbf{\Psi}_2$ between  bases 1 and  2, the Love matrices transform as 
\begin{align}
    {K}_1 = P_{12} {K}_2 P^{-1}_{12}\,. 
    \label{se:K_rotation}
\end{align}
Notice that  Love matrices are invariant  under an overall rescaling of the ${\bm \Psi}$ field, i.e. for $P_{12}\propto \Id$, just like Love numbers.  
Hence the $P_{12}$ matrix can be taken to be unitary without loss of generality.

Finally, let us compare the Love matrices defined here to the tidal response from the worldline EFT, see section~\ref{se:WEFT}. We notice that the matrix of Wilson coefficients $\Lambda_\ell$ is symmetric, and has therefore $\frac{N(N+1)}{2}$ independent elements. On the other hand, we know that each element of the Love matrix $K_\ell$ is related to an element of $\Lambda_\ell$ up to a coefficient. This implies that, for every $\ell$, the Love matrix itself must only depend on $\frac{N(N+1)}{2}$ parameters. 

Furthermore, still considering the matching procedure, one may choose to write  the matched Love matrices using directly the  variables of the worldline EFT. Using these variables,  $K_\ell$  should be symmetric for any $\ell$ since $\Lambda_\ell$ is.  While finding the correct variables on the UV side is a priori nontrivial,  consistency with the worldline EFT guarantees their existence. We thus conclude that 
 \be
 \shortstack[l]{
\textrm{
There is a field basis for which the Love matrices $K_\ell$}  are symmetric for any $\ell$. } 
\label{Prop:basis}
 \ee
This  property will be exemplified via direct calculation    of the charged black hole Love matrices in sections \ref{se:fluctuations} to \ref{se:strong_field}.

\subsection{Tidal Perturbation Theory}
\label{se:tidal_PT}

The deviations to GR+Maxwell theory presented in section \ref{se:BH}  can be treated perturbatively.  In this section and the next we consider a single perturbation with coefficient $\alpha$. The generalization to the various perturbations introduced in section \ref{se:Deformation_weak_field} is trivial. We also omit the $\ell$ index of all the $\Psi_{I_i}$'s and  $\psi_{I_i}$'s for simplicity.

We write the  wave operator and the tidal field from \eqref{eq:EOM} as
\be
{\cal D}= {\cal D}^{(0)}+ \alpha {\cal D}^{(1)}  +\O(\alpha^2) \,,\quad \quad  \Psi=\Psi^{(0)}+\alpha \Psi^{(1)}+ \O(\alpha^2)\,.
\label{eq:PT_WO}
\ee
Plugging these expansions into the equation of motion provides the equation for the tidal field perturbation
\be
 {\cal D}^{(0)}  \Psi^{(1)} =-{\cal D}^{(1)} \Psi^{(0)}  
\label{eq:EOM_Psi1}
\,.
\ee

We are interested in extracting the Love matrix from the regular solution ${\hat \Psi}_{\rm regular}$.
We split the regular solution as 
\be \Psi_{\rm regular}\equiv  \psi^{(0)}+ \alpha  \psi^{(1)} +\O(\alpha^2)  \,.\ee
The properties of the perturbed solutions and  the overall structure of the problem  have been studied in details in \cite{Barbosa:2025uau}.  
While a pedestrian approach would imply to solve \eqref{eq:EOM_Psi1}, 
we have shown that this cumbersome task can be entirely avoided by working  with Green's functions.  In \cite{Barbosa:2025uau} we derived  a pair of formulas --- one for the Love numbers and one for the beta function --- that only require knowledge of ${\cal D}^{(1)} $ and of the unperturbed solution. 
Here we generalize these formulaes to $N$ coupled tidal fields.

\subsubsection{Solving via Green function} \label{sec:Solving_via_Green function}

We write the potential as 
\be
\hat V = \hat V^{(0)}+  \alpha \hat V^{(1)} + \O(\alpha^2)\,. \label{eq:V_expansion}
\ee
We assume with no loss of generality that the 
leading-order potential is diagonal, and that the  wave operator is canonically normalized, such that
\be
{\cal D}_r^{(0)} = \mathds{ 1}   \frac{d^2}{d r_\star^2} - \hat V^{(0)}(r)\,.
\ee
The $\hat V^{(1)} $ term is generally nondiagonal, and thus will causes a mixing in the perturbed solution.

{Since the leading-order equation decouples, it is convenient to put all the zeroth-order $N$-vectors in $N\times N$ diagonal matrices.\,\footnote{Alternatively we would have to work with the elementwise product. }
We denote these matrices with a \textit{hat}, for instance: 
\be
\hat \Psi^{(0)}\equiv {\rm diag}( {\bm \Psi}^{(0)})\,. 
\ee
At any point, the vector notation can be recovered as $\Psi = \hat{\Psi} \,\vec{\mathds{1}} $, where $\vec{\mathds{1}} $ is a $N$-vector of 1's.}

We introduce the Green's function that inverts {the leading-order wave} operator, 
\be
{\cal D}_r^{(0)} \hat G(r,r') = \delta(r_\star-r_\star') \mathds{ 1}\,, \label{eq:Green_def}
\ee
where $\delta(r_\star-r_\star')= f(r)\delta(r-r')$.  The r.h.s. in \eqref{eq:Green_def} can in principle  be determined by the variation of the  action of the tidal field. It can also be deduced directly from consistency with the Wronskian of ${\cal D}_r^{(0)}$ \cite{Fichet:2023dju}. 

For the Green's function  boundary conditions, we require regularity at the horizon and at infinity. We thus define the leading-order solution that is regular  at infinity, $ {\hat \psi}_+^{(0)}$. 
The solving of \eqref{eq:Green_def} follows standard ODE  techniques, see e.g. \cite{Fichet:2019owx} and also \cite{Barura:2024uog}.

We obtain the tidal  Green function 
\be
\hat G(r,r')= {\hat \psi}^{(0)}(r_<)\,\hat N^{-1}_{\psi,\psi_+} \,  {\hat \psi}_+^{(0)}(r_>) \,, 
\label{eq:Green}
\ee
where  $r_>={\rm max}(r,r') $, $r_<={\rm min}(r,r') $,    and  
$\hat N_{\psi,\psi_+} = {\rm diag}(\bm N_{\psi,\psi_+})$ is the matrix of 
 normalization factors computed from the Wronskian via
$W(\psi_1,\psi_2)=\psi_1 \psi'_2- \psi_1' \psi_2 \equiv \frac{N_{\psi_1, \psi_2 }}{ f(r)} $. 
The presence of the normalization factors makes the Green function invariant under rescaling of any of the solutions.

The general  solution to the equation of motion is {given by ${\psi}^{(1)} = {\hat \psi}^{(1)}\,\vec{\mathds{1}} $, where}
\begin{align}
& {\hat \psi}^{(1)}(r)=\int^{\infty}_{r_h} dr_\star' \hat G(r,r') \cdot {\cal D}^{(1)}{\hat \psi}^{(0)}(r') 
\label{eq:psi1solgen}
\\
&= 
\hat N^{-1}_{\psi,\psi_+} \,\left[{\hat \psi}_+^{(0)}(r) \, \int^{r}_{r_h} dr_\star'  {\hat \psi}^{(0)}(r')   {\cal D}^{(1)}{\hat \psi}^{(0)}(r')
+
{\hat\psi}^{(0)}(r)  \int^{\infty}_{r} dr_\star' {\hat \psi}_+^{(0)}(r') {\cal D}^{(1)}{\hat \psi}^{(0)}(r') \right]\,. \nn
\end{align}
In the second line, we have used the explicit expression  \eqref{eq:Green}. {Note that ${\hat \psi}^{(1)}$ is not necessarily a diagonal matrix, since the} perturbation of the wave operator ${\cal D}^{(1)}$  is generally not diagonal.

\subsubsection{Extracting the Love matrices \label{se:extracting_TLMs} }

We take the large $r$ limit of \eqref{eq:psi1solgen}.
 Using the asymptotics from \eqref{eq:Psi_asymptotic}, we know that the condition of regularity at infinity implies that 
 \be
\hat  \psi^{(0)} \underset{r\gg r_h}{\sim}  \hat a^{(0)}_{\ell}\,r^{\ell+1}\,, 
\quad\quad \hat  \psi_+^{(0)} \underset{r\gg r_h}{\sim}  \frac{\hat b^{(0)}_{+,\ell}}{r^{\ell}}\,, 
 \label{eq:psiPM_asymptotics}
 \ee
 where $\hat a^{(0)}_{+,\ell}=0$ due to regularity. Performing the same analysis as in \cite{Barbosa:2025uau},  we find that only the first term of \eqref{eq:psi1solgen} contributes to the Love matrices. The integral may or not produce a logarithm. In either case we are able to extract the relevant coefficient of the Laurent series. 

In the absence of logarithmic term, the Love matrix is extracted as 
\global\mdfdefinestyle{EqFrame}{ linecolor=white,linewidth=3pt,
backgroundcolor=EqFrame,
leftmargin=0cm,rightmargin=0cm }
\begin{mdframed}[style=EqFrame]
\begin{equation}
    \bar {K}_\ell =  \hat N_{\psi \psi_+}^{-1} \hat{b}^{(0)}_{+,\ell}  \Bigg[ \frac{1}{2 \pi i} \oint \frac{dr}{r} \int_{r_h}^r dr_{\star}^\prime \hat{\Psi}^{(0)}(r^\prime) {\mathcal{D}}^{(1)}\hat{\Psi}^{(0)}(r^\prime) \Bigg]\left(\hat{a}_{\ell}^{(0)}\right)^{-1}\,. 
    \label{eq:K_full}
\end{equation}
\end{mdframed}
In the presence of a logarithm $\log(\frac{r_h}{r})$, the matching to the worldline EFT done at matching scale $r\equiv L$ produces the structure of a renormalization flow for the worldline EFT coefficient.  The relevant information is the beta function of the Love matrix $\beta_{K}\equiv - L\frac{d}{d L} K$,  conveniently extracted from the integrand of \eqref{eq:psi1solgen} as
\begin{mdframed}[style=EqFrame]
\begin{equation}
    {\beta}_{K_\ell} = - 
     \hat N_{\psi \psi_+}^{-1} \hat{b}^{(0)}_{+,\ell}  \Bigg[ \frac{1}{2 \pi i} \oint dr_\star  \hat{\Psi}^{(0)}(r) {\mathcal{D}}^{(1)}\hat{\Psi}^{(0)}(r) \Bigg]\left(\hat{a}_{\ell}^{(0)}\right)^{-1}\,.
     \label{eq:B_full}
\end{equation}
\end{mdframed}
We recall that $dr_\star =\frac{dr}{f(r)}$.

The above formulae are manifestly invariant under an overall rescaling of either the $\Psi$ or $\Psi_+$ solutions. On the other hand, as explained in subsection \ref{se:TLM}, the Love matrices change under unitary rotations of the field vector. For a change of basis ${\bm \Psi}_1= P_{12}{\bm \Psi}_2$, we have that $\hat V_1 =  P_{12} \hat V_2  P_{12}^{-1} $,  $\hat G_1 =  P_{12} \hat G_2  P_{12}^{-1} $, and thus  
${ K}_1=  P_{12} { K}_2  P_{12}^{-1}  $,
consistent with \eqref{se:K_rotation}.

We validate the above formulas through numerous examples.  We provide an example of pedestrian solving that reproduces the results given by  \eqref{eq:K_full} in App.\,\ref{app:Derivation}.

\subsubsection{Freedom of parametrization versus Love numbers} 

As explained in section~\ref{sec:parametrization}, two parametrizations of the black hole background are particularly convenient. In the first, $M$ and $Q$ are identified with the physical mass and charge, so that $\delta M =  \delta Q = 0$, while the horizon radii receive corrections. In the second, the corrections are encapsulated into $\delta M$ and $\delta Q$, while the horizon radii are kept fixed, $\delta r_\pm = 0$.

A Love number $k(M, Q)$ changes under a reparameterization according to
\begin{equation}
    k(M + \delta M, Q + \delta Q) = k(M, Q) + \partial_M k(M, Q) \delta M + \partial_Q k(M, Q) \delta Q\,.
\end{equation}
However, since the zeroth-order Love numbers vanish in 4d general relativity,  $k(M, Q)$ is already a first-order quantity in $\alpha$. Since $\delta M$ and $\delta Q$ are also first-order in $\alpha$, it follows that
\begin{equation}
    k(M + \delta M, Q + \delta Q) = k(M, Q) + \mathcal{O}(\alpha^2)\,.
\end{equation}
Therefore, at first order, the Love numbers are independent of the choice of parametrization. In particular, they are unaffected by whether the horizon radii are allowed to shift or are kept fixed.

\section{The Fluctuations of Charged  Black Holes \label{se:fluctuations}}

In this section, we derive the equations of motion for small perturbations around the charged black hole background from Section \ref{se:BH}. The perturbations are  written with lower indices as 
\begin{align}
    \left(g\big|_{\rm full}\right)_{\mu\nu} = g_{\mu\nu} + \delta g_{\mu\nu}\,, \quad \quad \left(A\big|_{\rm full}\right)_{\mu} = A_\mu + \delta A_\mu\,.
\label{eq:bg_perturbation}\end{align}
The background fields $g_{\mu \nu}$ and $A_\mu$ satisfy the field equations \eqref{eq:Maxwell} and  \eqref{eq:Einstein}. 
We define the fluctuation fields
\be
h_{\mu\nu} \equiv \delta g_{\mu\nu}  \,,\quad \quad a_{\mu} \equiv  \delta A_{\mu} \,.
\label{eq:var_def}
\ee
The field strength for the $a_\mu$ fluctuation satisfies $f_{\mu\nu} \equiv \delta F_{\mu \nu} = \nabla_\mu a_\nu - \nabla_\nu a_\mu$, where $\nabla$ is the Levi-Civita covariant derivative defined by the background metric $g_{\mu \nu}$. 
\footnote{
Notice that the variations of quantities with raised indices pick up minus signs, with for instance 
$ \left(g\big|_{\rm full}\right)^{\mu\nu} = g^{\mu\nu} + \delta g^{\mu\nu}$, $\delta g^{\mu\nu}=-h^{\mu\nu}=-g^{\mu\rho}g^{\nu\sigma}h_{\rho\sigma}$ and $\left(F\big|_{\rm full}\right)^{\mu\nu} = F^{\mu\nu}+ \delta F^{\mu\nu}$, 
$\delta F^{\mu\nu} = g^{\mu\rho}g^{\nu\sigma} f_{\rho\sigma} - h^\mu_\rho F^{\rho\nu}- h^\nu_\rho F^{\rho\mu}$. 
The variations in App.\,\ref{se:Formulas} are directly expressed in terms of the $h_{\mu\nu}$, $a_{\mu}$  fields. 
}

\subsection{Spherical Harmonics and Parity}

We decompose the $a_\mu$ and $h_{\mu\nu}$  fluctuations into vector and tensor spherical harmonics on the two-sphere \cite{Hui:2020xxx} (see also \cite{Barbosa:2025uau}). These harmonics can be classified with respect to their parity eigenvalues, hence we can decompose the fields as
\begin{equation}
    a_\mu=a^{\rm even}_\mu+a^{\rm odd}_\mu \,, \quad \quad h_{\mu\nu}=h_{\mu\nu}^{\rm even}+h_{\mu\nu}^{\rm odd}\,.
\end{equation}
Since the background is spherically symmetric, the modes with definite angular momentum and parity eigenvalues are kinetically decoupled and can be treated independently.

Our focus  is on the \textit{parity-odd} components of $h_{\mu\nu}$. Depending on whether the background is electric or magnetic, either the even or odd components of $a_\mu$ mix with $h_{\mu\nu}^{\rm odd}$.\footnote{For a dyonic black hole, the even and odd sectors of gravitational and electromagnetic perturbations are coupled. Here we restrict our analysis to purely electric or purely magnetic black holes.} The relevant components are 
\begin{align}
    h_{a i} &= \sum_{\ell , m} h^{\ell  m}_a(t,r) \,  \mathcal{Y}^{\ell m}_i(\theta, \phi)\,, \quad  \quad h_{ij} = \sum_{\ell , m} h^{\ell  m}_2(t,r) \, \mathcal{Y}^{\ell m}_{ij}(\theta, \phi)\,, \quad \quad \\
    a_b &= \sum_{\ell , m} a^{\ell m}_b(t,r) Y^{\ell m}(\theta, \phi),\,\, a_i = \sum_{\ell , m} \Big[a^{\ell m}_{L}(t,r)\nabla_i Y^{\ell m}(\theta, \phi)+ a^{\ell m}_{T}(t,r)\, \mathcal{Y}^{\ell m}_i(\theta, \phi) \Big] \,. \label{eq:harmonics}
\end{align}
The indices $a,b,\dots$ from the beginning of the Latin alphabet label time and radial components and indices $i,j,\dots$ from the middle of the Latin alphabet label angular components. In some expressions, the $\ell m$ indices will be suppressed for convenience. The vector and tensor spherical harmonics in odd sector are
\begin{equation}
    \mathcal{Y}^{\ell m}_i = \epsilon\indices{_i ^j} \nabla_j Y^{\ell m}\,, \quad \quad \mathcal{Y}^{\ell m}_{ij} =  \epsilon\indices{_( _i ^k}\nabla\indices{_j _) _T}\nabla_k Y^{\ell m}\,.
\end{equation}
Here $(\ldots)^T$ denotes traceless symmetrization. 
In the decomposition \eqref{eq:harmonics}, the $a_T$, $h_{ai}$, $h_{ij}$ components are parity-odd while $a_b$, $a_L$ are parity-even.  

The spherical harmonics satisfy
\begin{align}
    \square_{S^2} Y^{\ell m} &= - j^2 Y^{\ell m}\,, \quad \quad
    \square_{S^2} \mathcal{Y}^{\ell m}_i = - (j^2-1) \mathcal{Y}^{\ell m}_i\,, \quad \quad
    \nabla^i \mathcal{Y}^{\ell m}_i = 0\,,
\end{align}
where $j^2 = \ell(\ell + 1)$ and $\square_{S^2}$ is the two-sphere Laplacian $\square_{S^2} = \gamma^{ij}\nabla_i\nabla_j$, with $\gamma^{ij}$ the  two-sphere metric.
Note that ${\cal Y}_{ij}^{\ell m}$ is  traceless, 
\begin{equation}
    \gamma^{ij}\mathcal{Y}_{ij}^{\ell m} = 0\,.
\end{equation}
This implies $g^{\mu\nu}h^{\rm odd}_{\mu\nu} = 0$.
In the following, we set $h_{\mu\nu}^{\rm even}=0$ and simply use $ h_{\mu\nu}$ for  $h^{\rm odd}_{\mu\nu}$.

\subsection{Gauge Redundancies and Invariants} 

The effective action \eqref{eq:effective_action} is invariant under diffeomorphism and $U(1)$ gauge transformations of the form
\begin{equation}
    h_{\mu\nu} \rightarrow h_{\mu\nu} +\nabla_\mu \xi_\nu + \nabla_\nu \xi_\mu\,, \quad\quad a_{\mu} \rightarrow a_\mu + \nabla_\mu \Lambda\,.
\label{eq:dif_gauge_transf}
\end{equation}
For parity-odd components of $h_{\mu\nu}$, only the angular components of the diffeomorphism contribute. We decompose the corresponding $\xi_i$ parameter into spherical harmonics as
\begin{equation}
    \xi_i = \sum_{\ell, m} \xi^{\ell m}(t,r)\mathcal{Y}^{\ell m}_i(\theta, \phi)\,,
\end{equation}
such that 
\begin{align}
    h_{0} &\rightarrow h_0 + \partial_t \xi\,, \quad\quad
    h_1 \rightarrow h_1 + \partial_r \xi - \frac{2}{r}\xi\,, \quad \quad
    h_2 \rightarrow h_2 + 2 \xi\,. \label{eq:gauge_h2}
\end{align}
We choose a gauge transformation to eliminate $h_2$, which is the Regge-Wheeler gauge. Then, the particular combination
\begin{align}
    \chi &= \dot{h}_1 - h_0^\prime + \frac{2}{r}h_0
\end{align}
is gauge-invariant. {Here $\dot{f} = \partial_t f$ and $f^\prime = \partial_r f$.}

Similarly, the  $U(1)$ gauge parameter can be decomposed in spherical harmonics as
\begin{equation}
    \Lambda = \sum_{\ell , m} \lambda^{\ell m}(t,r)Y^{\ell m}(\theta, \phi) \,.
\end{equation}
The components of $a_\mu$ transform as
\begin{align}
    a_{b} &\rightarrow a_b + \partial_b \lambda\,, \quad\quad\quad
    a_{L} \rightarrow a_{L} + \lambda\,, \quad \quad\quad
    a_{T} \rightarrow a_{T}\,. \label{eq:gauge_a_L}
\end{align}
We choose a gauge transformation to eliminate $a_L$. 
The transverse component $a_{T}$ and the combination $f_{r t}= a_0^\prime -  \Dot{a_1}$ are gauge invariants. As we will see in the next sections, the former is relevant for an electrically charged black hole background, while the latter plays analogous role for a magnetically charged background.

\subsection{Field Equations }

\subsubsection{Background  }

Considering only the correction to Maxwell electromagnetism, such that $\Delta {\cal L} = \Delta {\cal L}_F$, it is convenient to write the quantum effective action \eqref{eq:effective_action} as
\begin{align} 
\Gamma_F[R_{\mu\nu\rho\sigma},F_{\mu\nu}] = \int d^4 x \sqrt{-g}\Bigg[\frac{R}{2\kappa^2} - \frac{1}{4e^2}{F}^2 \Big(1 + \TildeDeltaF \Big) \Bigg]\,, \label{eq:quantum_eff_action_v2}
\end{align}
where we introduced  $\bar{\Delta}_F = -4e^2 \Delta \mathcal{L}_F/F^2$. 
The effective  current and effective stress–energy tensor are given by
\begin{align}
  \mathcal{J}^{\nu} &= -\frac{1}{e^2}\nabla_\mu\left( \DeltaF F^{\mu\nu}\right)\,,\quad \quad \DeltaF = \TildeDeltaF + F^2 \frac{\partial \TildeDeltaF}{\partial F^2}\,,
  \\\mathcal{T}_{\mu\nu}&= \frac{\DeltaF}{e^2}F\indices{_\mu ^\sigma}F_{\nu \sigma}  - \frac{\TildeDeltaF}{4e^2}g_{\mu\nu} F^2 \,.\label{eq:mathcalT}
\end{align}
The background equations of motion take the form 
\begin{align}
     &\nabla_\mu\Big[(1 + \DeltaF)F^{\mu \nu}\Big] = 0\,, \quad \quad \nabla_\mu \Tilde{F}^{\mu \nu} = 0\,, \label{eq:backhround_max}\\
    &G_{\mu \nu} = \frac{\kappa^2}{e^2} \Big[F\indices{_\mu ^\sigma}F_{\nu \sigma}(1+\DeltaF)  - \frac{1}{4}g_{\mu\nu} F^2 (1+\TildeDeltaF) \Big]\label{eq:backhround_einstein} \,.
\end{align}

\subsubsection{Fluctuations }

The quadratic action for the fluctuations, derived from  the second variation of the quantum effective action \eqref{eq:quantum_eff_action_v2}, is 
\begin{align}
    \delta^2 \Gamma_F = \int d^4x \sqrt{-g} \Big[- \frac{(1+\DeltaF)}{4e^2}  f_{\mu\nu}f^{\mu\nu}  
    -\frac{h^{\mu\nu}}{4\kappa^2}  \delta G_{\mu\nu}[h_{\mu\nu}] + {{\frac{1}{4}}}h^{\mu\nu}\delta T_{\mu\nu}[h_{\mu\nu},f_{\mu\nu}]  \Big]\,.\label{eq:perturbation_action}
\end{align}
Notice the gauge-invariant term $\DeltaF$ corrects the kinematic term for $a_\mu$ analogously to  $\bar \Delta_F$ correction to the background kinetic term. 
The variations $\delta G_{\mu \nu}$ and $\delta T_{\mu \nu}$ are given in App.~\ref{se:Formulas}.

The $\delta T_{\mu\nu}$ term mixes the metric and electromagnetic fluctuations, a detailed discussion of the mixing is given in App.~\ref{app:mixing}. The purely electric and magnetic cases correspond respectively to 
\begin{align}
    \mathcal{B}(r) &= 0 \,, \quad  a_\mu = a^{\rm odd}_\mu 
\quad\quad \, {\rm (Electric~background )}\,,\\
    \mathcal{E}(r) &= 0 \,, \quad a_\mu = a^{\rm even}_\mu 
\quad\quad {\rm  (Magnetic~background)}\,.
\end{align}

Focusing on the odd component of $h_{\mu\nu}$, the subsequent coupled equations of motion for both  purely 
 electric and magnetic backgrounds can be written covariantly as~\footnote{The left-hand side of \eqref{eq:perturbation_max} expands as
\begin{align}
    \nabla^\mu \Big[(1+\DeltaF)f_{\mu \nu}  \Big] = \Big[\partial_{r_{\star}} \DeltaF\Big] f_{r \nu}+(1+\DeltaF)\Big[\square a_\nu - R_{\nu \lambda} a^\lambda - \nabla_\nu \nabla^\lambda a_{\lambda} \Big]\,.
\end{align}
We have $\nabla^\lambda a_{\lambda} = 0 $ in the electric case. } 
\begin{align}
    \nabla^\mu \Big[(1+\DeltaF)f_{\mu\nu}  \Big]&= -\nabla^\mu\Big[2 {F^\sigma}_{[\mu}h_{\nu]\sigma}(1+\DeltaF)\Big] \,, \label{eq:perturbation_max} \\
    \delta G_{\mu \nu} &=  \frac{\kappa^2}{e^2}\Big[2f\indices{_(_\mu ^\lambda}F\indices{_\nu_)_\lambda}(1+\DeltaF)-\frac{1}{4}h_{\mu \nu}F^2(1+\TildeDeltaF) \Big] \label{eq:perturbation_einstein}
    \,.
\end{align}
We further use the gauge redundancies to remove components of the fluctuations. Based on  \eqref{eq:gauge_a_L} and \eqref{eq:gauge_h2}
we  set $h_2=0$ and  $a_L=0$.  The gravitational equations \eqref{eq:perturbation_einstein}  reduce to 
\begin{align}
        \delta G_{b i} &=  -\frac{\kappa^2}{e^2}\Big[\left(r^2 \mathcal{B}(r)\epsilon_{i j}\partial^ja_b  -\mathcal{E}(r)\epsilon_{b c}\partial^{c}a_i\right)(1+\DeltaF)+\frac{1}{4}h_{b i}F^2(1+\TildeDeltaF) \Big]\,,\label{eq:geral_deltaG_bi}
        \\
        \delta G_{i j} &= 0\,, \label{eq:geral_deltaG_ij}
\end{align}
while the electromagnetic equations \eqref{eq:perturbation_max} reduce to 
\begin{align}
    \nabla^\mu \Big[(1+\DeltaF)f_{\mu i}  \Big] &= \chi \mathcal{E}(r)\mathcal{Y}_i \nonumber\\ &+ \Big[\partial_r \big((1+\DeltaF)   \mathcal{B}(r) f h_1\big)-\frac{1}{f^2}\partial_t\big((1+\DeltaF)  \mathcal{B}(r) f h_0 \big)\Big]\nabla_i Y\,,\label{eq:max_i}\\
    \nabla^\mu \Big[(1+\DeltaF)f_{\mu b}  \Big] &= h_b \frac{j^2}{r^2} (1+\DeltaF)\mathcal{B}(r) Y\,. \label{eq:max_b}
\end{align}
Integration by parts were used to derive \eqref{eq:max_b}.
We apply $F_{a b} = -\mathcal{E}(r) \epsilon_{a b}$ in the electric case and $F_{i j} =  r^2\mathcal{B}(r)  \epsilon_{i j}$ in the magnetic  case.\footnote{ We use $\epsilon_{t r} = -\epsilon_{r t} = 1$ and  $\epsilon_{\theta \phi} = -\epsilon_{\phi \theta} = \sin{\theta}$. Note that $F^{a b} = \mathcal{E}(r)\epsilon^{a b}$ and $F^{i j} = r^2\mathcal{B}(r)\epsilon^{i j}$.}

\subsection{The Fluctuations of  Electrically-Charged Black Holes}
\label{se:EOM_electric}

We derive the coupled equations of motion for the fluctuations in the purely electric case where $F_{r t} = -F_{tr}= \mathcal{E}(r)$, with all other components vanishing. Note that $\DeltaF$ is a function of $\mathcal{E}^2$. 

From the third term in \eqref{eq:perturbation_action}, which includes the interaction between metric and electromagnetic fluctuations, we can see that  only the parity-odd sector of $a_\mu$ couples with the   $h_{a i}$ components (see App. \ref{app:mixing}). Accordingly, in this subsection we restrict to perturbations of the form 
\begin{equation}
    a_\mu=a^{\rm odd}_\mu=\Big(0,0,\sum_{m,\ell}a^{\ell m}_T {\cal Y}_i^{\ell m} \Big)\,.
\end{equation}
Hence the gauge-invariant variable $a_{T}$ is the only  degree of freedom for the electromagnetic fluctuation. 
In the  following, we work at the level of the $a^{\ell m}_T$ harmonics and omit the $\ell m$ indices. 

Given these assumptions, the equations of motion for the perturbations take the form
\begin{align}
    &\delta G_{a i} = - \frac{\kappa^2}{e^2} \Big[ (1+\DeltaF) F\indices{_a _b}\partial^b a_i + \frac{1}{4} h_{ai} (1+\TildeDeltaF) F^2 \Big]\,, \quad \quad \delta G_{ij} = 0\,, \label{eq:grav_eq}\\
    &\nabla^\mu\Big[(1+\DeltaF)f_{\mu i} \Big] =  \chi (1+\DeltaF) \mathcal{E}\mathcal{Y}_i\,\label{eq:max_eq_ai}\,.
\end{align}
The gravitational sector \eqref{eq:grav_eq} yields three independent equations. Two of them act as constraints, allowing us to eliminate the variables $h_0$ and $h_1$ in favor of the gauge-invariant combination $\chi$. We begin with rearranging the equation carrying a radial index,  
\be
\delta G_{r i} - \frac{\kappa^2}{e^2} \Big[ (1+\DeltaF) \frac{\mathcal{E}}{f}\Dot{a}_i + \frac{1}{2} h_{r i} (1+\TildeDeltaF) \mathcal{E}^2 \Big] =0\,,
\ee
or more explicitly,
\be
\frac{\partial}{\partial t}\Big( \chi - 2 \frac{\kappa^2}{e^2} \mathcal{E} (1+\DeltaF)a_T\Big) + \frac{h_1}{r^2}\left(j^2 -2  + \frac{d}{dr}(r^2 f^\prime) - \frac{\kappa^2}{e^2} r^2 (1+\TildeDeltaF)\mathcal{E}^2\right)f  = 0\,. \label{eq:equation_h1}
\ee
The $\TildeDeltaF$-dependence is exactly canceled by the background field equation $G_{\theta\theta} = \frac{\kappa^2}{e^2}T_{\theta\theta}$, which implies
\begin{equation}
 \frac{d}{dr}(r^2 f^\prime) = \frac{\kappa^2}{e^2} r^2 (1+\TildeDeltaF) \mathcal{E}^2\,. \label{eq:ele_background_simplification}
\end{equation}
From \eqref{eq:equation_h1} and similarly from the angular index equations $\delta G_{ij} = 0$, we solve for $h_0, h_1$ to find 
\begin{align}
   h_1= -\frac{\kappa}{e}\frac{2 r}{f\sqrt{2(j^2-2)}}\frac{\partial}{\partial t}\Psi_h  \,, \quad \quad h_0=-\frac{\kappa}{e}\frac{2 f}{\sqrt{2(j^2-2)}}\frac{\partial}{\partial r} \left( r \Psi_h \right)\,,
   \label{eq:h0_h1}
\end{align}
where we have defined a new gauge-invariant variable
\begin{equation}
    \Psi_h = \frac{e}{\kappa}\frac{r}{\sqrt{2(j^2-2)}}\left(\chi - 2 \frac{\kappa^2}{e^2} \mathcal{E} (1+\DeltaF)a_T\right)\,.\label{eq:def_Psi_h}
\end{equation}
Note that $\ell=1$ is ill-defined. We therefore restrict to $\ell \geq 2$ in this work.\footnote{The corresponding gravitational perturbation for this harmonic is non-dynamical. See \cite{Pereniguez:2021xcj} for a discussion.} The remaining gravitational equation of motion reads
\be
    \delta G_{t i}-  \frac{\kappa^2}{e^2} \left( f \mathcal{E}a_i^\prime - \frac{1}{2}h_{t i}\mathcal{E}^2 \right) = 0 
    \ee
    or, more explicitly, 
    \be 
    \Big(2 f + j^2 - 2 \Big)h_{0} + r^2 f \left(\frac{2}{r} \Dot{h}_1 + \frac{\partial}{\partial r}\big(\Dot{h}_1 - h_0^{\prime}\big) \right) = 2 \frac{\kappa^2}{e^2} r^2 f \mathcal{E} (1+\DeltaF) a_T^\prime\,. \label{eq:Gti_ele}
\ee%
To derive a second order partial differential equation involving only gauge-invariant variables, we isolate the combination
\begin{equation}
    \Dot{h}_1 - h_0^{\prime} = \frac{\kappa}{e}\frac{\sqrt{2(j^2-2)}}{r}\Psi_h - \frac{2}{r}h_0 + 2 \frac{\kappa^2}{e^2} \mathcal{E}(1+\DeltaF)a_T\,\label{eq:psih_v1}
\end{equation}
and substitute the expressions for $h_0, h_1$ from \eqref{eq:h0_h1},  leading to the wave equation\footnote{
Note that for $q = \mathcal{E} = 0$, the  RN background reduces to Schwarzschild. The metric and electromagnetic perturbations decouple, and the variable $\Psi_h$ becomes the Regge–Wheeler variable (except for a $\kappa$ factor  introduced here for convenience), which describes purely metric perturbations (see  (6.51) of \cite{Barbosa:2025uau}). Consistently, the equation for $\Psi_h$ in the neutral limit reproduces the one from the Regge–Wheeler variable,  (6.52) of \cite{Barbosa:2025uau}.}
\begin{align}
    \Bigg[-\partial_t^2 + \partial^2_{r_\star} - \frac{f}{r^2}\Big(j^2 - 2 + 2 f - r f^\prime \Big)\Bigg]\Psi_h &= -\frac{\kappa}{2 e}\sqrt{2(j^2-2)}  \partial_{r_\star}\Big[ (1+\DeltaF)\mathcal{E}\Big] a_T \,, 
    \label{eq:Psi_h}
\end{align}
where
\begin{equation}
    \partial^2_{r_\star} = f^2 \partial_r^2 + f f^\prime \partial_r\,.
\end{equation}
The r.h.s can be further simplified using the background field equation $\nabla_\mu \big[(1+\DeltaF)F^{\mu t}\big]=0$, which implies
\begin{align}
    \partial_{r_{\star}} \Big[ (1+\DeltaF)\mathcal{E}\Big] =-\frac{2}{r}(1+\DeltaF) f \mathcal{E}\,. \label{eq:background_max_eq_ele}
\end{align}

We now consider Maxwell's equations \eqref{eq:max_eq_ai}. It turn out that only one equation is independent,  
\begin{align}
    \partial_r \DeltaF\, \partial^r a_i
+
\left(1 + \DeltaF \right)\Big[
\Box a_i
- R_{i j} a^j 
\Big] = \chi (1+\DeltaF) \mathcal{E}\mathcal{Y}_i\,, \label{eq:a_i}
\end{align}
where $\nabla_\mu a^\mu = 0$, which is valid specifically in the case of a purely electric background. Substituting $\chi$ in terms of $\Psi_h$ from \eqref{eq:def_Psi_h} and rearranging, we obtain the equation of motion
\begin{align} 
\Bigg[-\partial_t^2 + \partial_{r_\star}^2 -\frac{f}{r^2}  \bigg(j^2+2\frac{\kappa^2}{e^2}r^2\mathcal{E}^2(1+\DeltaF)\bigg) -
&
\frac{\partial_{r_\star}^2\left(\sqrt{1+\DeltaF}\right)}{ \sqrt{1+\DeltaF}}
\Bigg]  \Psi_a =  \label{eq:Psi_a}\\ & ~~~~~~~~~~~~~~~~~~~\frac{\kappa}{e}  \sqrt{2(j^2-2)(1+\DeltaF)}  f \mathcal{E} \frac{\Psi_h}{r}  \,, \nonumber
\end{align}
where we have defined the gauge-invariant variable
\begin{align}
\Psi_a = \sqrt{1+\DeltaF}\,a_T\,. 
\label{eq:defPsi_a}\end{align} 

The coupled equations of motion \eqref{eq:Psi_h} and \eqref{eq:Psi_a} can be written in  matrix form. 
Our final result for the electrically-charged black hole case (denoted $|_{\cal E}$) is
\be
\mathcal{D} \mathbf{\Psi}\big|_{\cal E} = 0\,, \label{eq:matrix_form_E}
\ee
where
\begin{align}
\mathcal{D}\big|_{\cal E} = \mathds{1} \left(-\frac{\partial^2}{\partial t^2} + \frac{\partial^2}{\partial r^2_\star}\right)\bigg|_{\cal E}- \hat V\big|_{\cal E}\,, \quad \quad      \mathbf{\Psi}\big|_{\cal E} =  \begin{pmatrix}
        \Psi_h \\ \Psi_a\end{pmatrix}\,,\label{eq:matricial_eq}
\end{align}
with the potential matrix given by
\begin{mdframed}[style=EqFrame] 
\begin{align}
    {\hat V}\big|_{\cal E} &= \begin{pmatrix}
        \frac{f}{r^2}\Big(j^2 - 2 + 2 f - r f^\prime \Big) & \frac{\kappa}{e}\sqrt{2(j^2-2)(1+\DeltaF)}  \frac{f \mathcal{E}}{r} \\
         \frac{\kappa}{e}\sqrt{2(j^2-2)(1+\DeltaF)}  \frac{f \mathcal{E}}{r}& \quad \frac{f}{r^2}\left(j^2+2\frac{\kappa^2}{e^2}r^2\mathcal{E}^2(1+\DeltaF)\right)  + \frac{\partial_{r_\star}^2\left(\sqrt{1+\DeltaF}\right)}{ \sqrt{1+\DeltaF}}
    \end{pmatrix}\,.
    \label{eq:potential_full}
\end{align}
\end{mdframed}
We see that the potential is symmetric to all orders.~\footnote{Property \eqref{Prop:basis}, obtained from analyzing the worldline EFT, is a statement about symmetry of the Love matrices. Since the Love matrices directly follow from solving the equations of motion,  Prop.~\eqref{Prop:basis} also guarantees that the same set of variables  makes the potential symmetric. Here we establish that these variables are the $\Psi_h$, $\Psi_a$ fields.}

\subsection{The Fluctuations of Magnetically-Charged Black Holes }\label{se:Magnetically_Charged_Black_Hole}

We derive the coupled equations of motion for the fluctuations in the purely magnetic case where  $F_{\theta \phi} = -F_{\phi \theta } =  r^2\mathcal{B}(r)  \sin{\theta}$,  with all other components vanishing. Note that $\DeltaF$ is a function of $\mathcal{B}^2$. 

We mention that in the absence of corrections to the Maxwell sector, and  since the purely magnetic black hole is a monopole, an elegant  derivation could be done using electromagnetic duality.~\footnote{It is  possible to perform a change of variable in the path integral such that we integrate over the dual potential $B_\mu$ \cite{Lozano:1995aq,Kehagias:1995ic} defined as
$
\tilde F_{\mu\nu}=\partial_\mu B_\nu - \partial_\nu B_\mu\,.
$
We then expand the dual field as background plus fluctuation,  $(B|_{\rm full})_\mu=B_\mu+b_\mu$. We have verified that this derivation correctly reproduces at leading order the  potential derived in this section. } 
Here, however, electromagnetic duality does not hold, hence we perform a brute force calculation. 

For a magnetically charged black hole background, the parity-even sector $a_b, a_L$ mix with the $h_{ai}$ metric perturbation. 
Since we can set $a_L=0$,  in this section we restrict to perturbations of the form
\begin{equation}
    a_\mu=a^{\rm even}_\mu= \Big(\sum_{m,\ell}a^{\ell m}_b  Y^{\ell m},0,0 \Big)\,.
\end{equation}
We omit  the $\ell m$ indices in the following. 

Our goal is to derive wave equations in terms of gauge-invariant variables. Compared to the electric case, the challenge is that the electromagnetic variables $a_t$, $a_r$ are not gauge-invariant. Instead, we should make appear the {$f_{rt}=\partial_r a_0-\partial_t a_1$} combination.

It is convenient to start with the electromagnetic equations \eqref{eq:max_b}. From components $\mu = 0, 1$ we obtain
\begin{align}
     \partial_{r_\star}\operatorname{log}(1+\DeltaF)\left(\frac{r^2}{j^2}f_{rt}\right)+ 
    \partial_{r_\star} \left( \frac{r^2}{j^2}f_{rt} \right)- \mathcal{B} h_0  = a_0\,,  \label{eq:a_mag_0}\\
    \frac{1}{f}\partial_t \left( \frac{r^2}{j^2}f_{rt}\right)-\mathcal{B}  h_1= a_1\,.\label{eq:a_mag_1}
\end{align}
To derive an equation of motion in terms of gauge-invariant variables, we differentiate Eqs. \eqref{eq:a_mag_0} and \eqref{eq:a_mag_1} with respect to $r$
and $t$, respectively, and subtract. We obtain
\begin{align} 
\Bigg[-\partial_t^2 + \partial_{r_\star}^2-f\frac{j^2}{r^2}   -
&
\frac{\partial_{r_\star}^2\left(\frac{1}{\sqrt{1+\DeltaF}}\right)}{ \frac{1}{\sqrt{1+\DeltaF}}}
\Bigg]  \Tilde{\Psi}_a = -\sqrt{1+\DeltaF}f \mathcal{B}\left(\Dot{h_1}-h_0^\prime - \frac{\mathcal{B}^\prime}{\mathcal{B}}h_0 \right) \,, \label{eq:mag_psi_a_chi}
\end{align}
where we have defined the gauge-invariant variable 
\begin{equation}
    \Tilde{\Psi}_a  
    = \sqrt{1+\DeltaF}\dfrac{r^2}{j^2}f_{r t}  \,.
\end{equation}
Using the background equation $\nabla_\mu\Tilde{F}^{\mu t} =0$, which implies~\footnote{Note that the magnetic background Bianchi identities play a role analogous to the electric background Maxwell's equations \eqref{eq:background_max_eq_ele}.}
\begin{align}
    r \mathcal{B}^\prime + 2\mathcal{B} = 0\,,
\end{align} 
we can identify the gravitational gauge-invariant variable $\chi$ on the right hand side of \eqref{eq:mag_psi_a_chi}.

We then consider  the gravitational equations \eqref{eq:geral_deltaG_bi}. From components $ri$  we obtain
\begin{align}
 \Dot{\chi}+\frac{h_1}{r^2} \left(j^2-2+\frac{d}{dr}(r^2 f^\prime)+ \frac{\kappa ^2}{e^2}r^2  (1+\DeltaF)\mathcal{B}\right)f  =2\frac{\kappa ^2}{e^2}  (1+\DeltaF)\mathcal{B} f a_1 \,.
\end{align}
Substituting for $a_1$ via \eqref{eq:a_mag_1}, we obtain a gauge-independent equation without introducing higher-order derivatives,
\begin{align}
   \frac{\partial}{\partial t} \left(
   \chi
   +2 \frac{\kappa^2}{e^2} \mathcal{B}\sqrt{1+\DeltaF} 
   \Tilde{\Psi}_a \right)
   +\frac{h_1}{r^2}\left( j^2-2 +\frac{d}{dr}(r^2 f^\prime) -\frac{\kappa ^2}{e^2} r^2 (1-\TildeDeltaF+2\DeltaF)\mathcal{B}^2 \right)f&=0\,. \label{eq:delta_G_rphi_mag}
\end{align}
Analogously to the electric case \eqref{eq:ele_background_simplification}, the magnetic background field equation $G_{\theta\theta} = \frac{\kappa^2}{e^2} T_{\theta \theta}$ results in the simplification
\begin{equation}
    \frac{d}{dr}(r^2 f^\prime) = \frac{\kappa ^2}{e^2} r^2 (1-\TildeDeltaF+2\DeltaF)\mathcal{B}^2\,. \label{Eq:magG22}
\end{equation}
Solving \eqref{eq:delta_G_rphi_mag} for $h_1$ and solving $\delta G_{ij} = 0$ for $h_0$ yields
\begin{equation}
    h_1 = -\frac{2 \kappa r}{f\sqrt{2\left(j^2 -2 \right)}}\frac{\partial}{\partial t}\Tilde{\Psi}_h \,, \quad \quad h_0 = -\frac{2\kappa f}{\sqrt{2(j^2-2)}}\frac{\partial}{\partial r}\big(r \Tilde{\Psi}_h\big)\,, \label{eq:mag_h0_h1}
\end{equation}
where we have defined a new gauge-invariant variable\footnote{As in the electric case, we restrict to $\ell \geq 2$.}
\begin{equation}
    \Tilde{\Psi}_h = \frac{e}{\kappa}\frac{r}{ \sqrt{2(j^2-2)}}\left(\chi + 2 \frac{\kappa^2}{e^2} \mathcal{B} \sqrt{1+\DeltaF}\Tilde{\Psi}_a  \right)\,. \label{eq:Z_ele}
\end{equation}
Similarly, we substitute for $a_0$ from \eqref{eq:a_mag_0} in the $\delta G_{t i}$ components of the gravitational equation \eqref{eq:geral_deltaG_bi}. Upon simplifying with the background equation \eqref{Eq:magG22}, we obtain 
\begin{align}
\Big(2 f+j^2-2 \Big)h_0 +r^2 f \left(\frac{2 }{r}\Dot{h_1}+\frac{\partial}{\partial r} (\Dot{h_1}-h_0^{\prime})\right)=2 \frac{\kappa^2}{e^2} r^2  \mathcal{B} f  \frac{\partial}{\partial r}\left( \sqrt{1+\DeltaF}\Tilde{\Psi}_a\right)\,.
\end{align} 
To obtain an equation of motion, we use \eqref{eq:Z_ele}, which is equivalent to
\begin{equation}
    \Dot{h}_1 - h_0^{\prime} = \frac{\kappa}{e}\frac{\sqrt{2(j^2-2)}}{r}\Tilde{\Psi}_h - \frac{2}{r}h_0 - 2 \frac{\kappa^2}{e^2} \mathcal{B}\sqrt{1+\DeltaF}\Tilde{\Psi}_a\,
\end{equation}
and substitute the expressions for $h_0, h_1$ from \eqref{eq:mag_h0_h1},  leading to
\begin{align}
    \Bigg[-\partial_t^2 + \partial^2_{r^*} - \frac{f}{r^2}\Big(j^2 - 2 + 2 f - r f^\prime \Big)\Bigg]\Tilde{\Psi}_h &=-\frac{\kappa}{e}\sqrt{2(j^2-2)(1+\DeltaF)}  f \mathcal{B}\frac{\Tilde{\Psi}_a}{r}\,.\label{eq:TildePsi_h}
\end{align}
Finally, using \eqref{eq:Z_ele}  again to substitute $\chi$ in terms of $\Tilde{\Psi}_h$ and $\Tilde{\Psi}_a$ in \eqref{eq:mag_psi_a_chi}, we obtain~\footnote{The remaining electromagnetic equations \eqref{eq:max_i} are redundant.}
\begin{align} 
\Bigg[-\partial_t^2 + \partial_{r_\star}^2 -\frac{f}{r^2}  \bigg(j^2+2\frac{\kappa^2}{e^2}r^2\mathcal{B}^2(1+\DeltaF)\bigg) -
&
\frac{\partial_{r_\star}^2\left(\frac{1}{\sqrt{1+\DeltaF}}\right)}{\frac{1}{\sqrt{1+\DeltaF}}}
\Bigg]  \Tilde{\Psi}_a =  \label{eq:TildePsi_a} \\& ~~~~~~~~~~~~~~~~~~~-\frac{\kappa}{e}  \sqrt{2(j^2-2)(1+\DeltaF)}  f \mathcal{B} \frac{\Tilde{\Psi}_h}{r}  \,.\nonumber
\end{align}

The coupled equations of motion \eqref{eq:TildePsi_h} and \eqref{eq:TildePsi_a} for the magnetically-charged black hole case (denoted $|_{\cal B}$) can be written in  the matrix form $\mathcal{D} \Tilde{\mathbf{\Psi}}\big|_{\cal B} = 0$, analogously to \eqref{eq:matrix_form_E}, by defining
\begin{align}
\mathcal{D}\big|_{\cal B} = \mathds{1} \left(-\frac{\partial^2}{\partial t^2} + \frac{\partial^2}{\partial r^2_\star}\right)\bigg|_{\cal B}- {\hat V}\big|_{\cal B}\,, \quad \quad      \Tilde{\mathbf{\Psi}}\big|_{\cal B} =  \begin{pmatrix}
        \Tilde{\Psi}_h \\ \Tilde{\Psi}_a\end{pmatrix}\,,\label{eq:mag_matricial_eq}
\end{align}
with the potential matrix given by
\begin{mdframed}[style=EqFrame] 
\begin{align}
    {\hat V}\big|_{\cal B} &= \begin{pmatrix}
        \frac{f}{r^2}\Big(j^2 - 2 + 2 f - r f^\prime \Big) & -\frac{\kappa}{e}\sqrt{2(j^2-2)(1+\DeltaF)}  \frac{f \mathcal{B}}{r} \\
         -\frac{\kappa}{e}\sqrt{2(j^2-2)(1+\DeltaF)}  \frac{f \mathcal{B}}{r}& \quad \frac{f}{r^2}\left(j^2+2\frac{\kappa^2}{e^2}r^2\mathcal{B}^2(1+\DeltaF)\right)  + \frac{\partial_{r_\star}^2\left(\frac{1}{\sqrt{1+\DeltaF}}\right)}{\frac{1}{\sqrt{1+\DeltaF}}}
    \end{pmatrix}\,.
    \label{eq:potential_full_mag}
\end{align}
\end{mdframed}
%

Again the potential is symmetric to all orders. 
The ${\hat V}\big|_{\cal B}$ potential is remarkably similar to ${\hat V}\big|_{\cal E}$ 
\eqref{eq:potential_full_mag}
 despite  the intermediate steps of the derivation being very different. Apart from the sign of the off-diagonal terms (that could  be removed by redefinition of $\tilde \Psi _a $), the key difference is the last term of the $(2,2)$ component of the potential. 

Let us also emphasize that ${\cal E}$ receives corrections while ${\cal B}$ does not.  We will later see that the difference of  the $(2,2)$ terms ends up compensating the corrections to the electric field.

\section{Love Numbers of Weak-Field Charged Black Holes}
\label{se:weak_field}

In this section, we compute the tidal responses of a charged black hole  in the weak-field regime. As explained in section \ref{se:BH}, this is the regime for which the corrections to the geometry are encoded into a long-distance EFT with operators $ \Delta{\cal L}_F = \sum_{n=2}^\infty c_n ({F}^{2})^n$,  see section~\ref{se:Deformation_weak_field}.

For our tidal computations we use the perturbative framework developped in \ref{se:tidal_PT}, that leads to the  formulas \eqref{eq:K_full} and \eqref{eq:B_full}.  We first compute the tidal response to probe fields (both scalar and vector)
 living on the curved background. 
 We then derive the Love matrices encoding the coupled response to metric and electromagnetic tidal perturbations.

As reviewed in section \ref{se:WEFT}, a tidal response with given $\ell$ may feature a renormalization flow, i.e. have a nonzero beta function. In our perturbative formalism the beta function is directly given by  \eqref{eq:B_full}. 
Using \eqref{eq:B_full} together with dimensional analysis of the EFT operators
of \eqref{eq:EFT_action}, we find the following property:  

\be
 \shortstack[l]{
\textrm{Scalar and vector Love number with harmonic $\ell$ induced by  an  operator } \\
 $F^{2n}$ {\textrm runs if}
    $ 2 \leq n \leq \left \lfloor \frac{\ell+1}{2}\right \rfloor $\,, while the Love matrix runs if $ 2 \leq n \leq \left \lfloor \frac{\ell}{2}\right \rfloor + 1$.  } 
\label{Prop:running}
 \ee
Here $\lfloor x\rfloor$ is the floor function. 

Property \eqref{Prop:running} implies that, for fixed $\ell$, only a \textit{finite} number of operators contribute to the Love beta function. This behavior is common to all tidal fields.

\subsection{Scalar and Vector Love Numbers}

The equations of motion and static solutions for scalar and vector probe fields are presented in App. \ref{se:Test_Fields}. The effect of the EFT operators on the metric and to the electromagnetic field  is given in \eqref{eq:EFT_correction_solution}. We treat these corrections as perturbations and apply our perturbative formulas \eqref{eq:K_full},\eqref{eq:B_full}, that reduce here to the case of  single tidal field.

\subsubsection*{Scalar} 

First consider the cases $\ell=1$ and $\ell=2$, for which all operators only contribute to a  constant (i.e. non-running) Love number. We find
\begin{align}
    k_{\ell=1} &= \frac{2 \kappa^2 (e Q)^4}{945 r_+^{11}}\Big[21 r_+^4 (8 r_+ - 3 \kappa^2 M)c_2 + 10 \sigma e^2 Q^2 (16 r_+ - 7 \kappa^2 M) c_3 + \cdots \Big]\,,\\
    k_{\ell=2} &= \frac{2 \kappa^2 (e Q)^4}{4725 r_+^{13}}\Big[63 r_+^4 \Big(36 r_+^3 + 2 \kappa^2 r_+ (Q^2 - 15 M r_+) - \kappa^4 M (Q^2 - 10 M r_+) - \kappa^6 M^3 \Big)c_2 \nonumber \\ & \quad \quad \quad \quad \quad + 2 \sigma e^2 Q^2 \Big(588 r_+^3 + 70 \kappa^2 r_+ (Q^2 - 11 M r_+) \nonumber \\ & \quad \quad \quad \quad \quad - 5 \kappa^4 M (7 Q^2 - 62 M r_+) - 35 \kappa^6 M^3 \Big)c_3 +  \cdots \Big]\,
\end{align}
The contributions of operators with odd $n$ have opposite sign
in the purely electric  ($\sigma = -1$) and magnetic ($\sigma = 1$) cases.

In contrast, for $\ell=3$, there is a running induced by the operator $n=2$, characterized by the beta function~\footnote{The contributions from operators with $n>2$ remain constant, 
\begin{align}
    k_{\ell=3} &= \frac{2 \sigma \kappa^2 (e Q)^6}{525 r_+^{15}}\Big[200 r_+^5 - 400 \kappa^2 M r_+^4 - \frac{5}{3} \kappa^4 M r_+^2(42 Q^2 + 69 \kappa^2 M^2) + 2 \kappa^2 r_+^3 (26 Q^2 + 157 \kappa^2 M^2) \nonumber \\ & -\frac{1}{2} \kappa^6 M (3 Q^4 + 7 \kappa^2 M^2 Q^2 + 2 \kappa^4 M^4) + \frac{1}{7} \kappa^4 r_+ (24 Q^4 + 201 \kappa^2 M^2 Q^2 + 131 \kappa^4 M^4)  
     \Big]c_3 + \cdots
\end{align}
}
\begin{equation}
    \beta_{\ell=3} = \frac{72 \kappa^4 M (e Q)^4}{35 r_+^7}c_2\,.
\end{equation}
Here and in the following we use $\beta_\ell \equiv \beta_{k_\ell}$.

Finally, let us focus on extremal black holes to conveniently present a general formula for the beta functions. 
For the extremal  background, we know the zeroth-order solution in closed form  for arbitrary $\ell$, given by \eqref{eq:Scalar_Extremal}.  Using this solution, together with the fact that a finite number of operators contribute at given $\ell$, we obtain the beta functions for any $\ell$ as the following finite sum,
\begin{equation}
    \beta_\ell = \frac{\ell^2 ( 2\ell -2)!}{2\ell +1}\sum_{n=2}^{\left \lfloor \frac{\ell+1}{2}\right \rfloor} \frac{\sigma^{n} 4^{n+1}(n-1)}{(4n-3)!(2\ell-4n+3)!}\frac{e^{2n}}{(\kappa r_h)^{2(n-1)}} c_n \,.\label{eq:Beta_scalar}
\end{equation}

\subsubsection*{Vector} 

As in the scalar field case, the $\ell=1,2$ Love numbers receive constant contributions from all operators. We find 
\begin{align}
    k_{\ell=1} &= \frac{4 \kappa^2 (e Q)^4}{945 r_+^{12}}\Big[21 r_+^4 (10 r_+^2 - 3 \kappa^2 Q^2)c_2 + 10 \sigma e^2 Q^2 (18 r_+^2 - 7 \kappa^2 Q^2) c_3 + \cdots \Big]\\  
    k_{\ell=2} &= \frac{\kappa^2 (e Q)^4}{2100 r_+^{14}}\Big[63 r_+^4 \Big(64 r_+^4 - 44 \kappa^2 M r_+^3 + 2 \kappa^4 M r_+ (Q^2 + 5 M r_+) - \kappa^6 M^2 Q^2 \Big)c_2  \\ & + 2 \sigma e^2 Q^2 \Big(896 r_+^4 - 980 \kappa^2 M r_+^3 + 10 \kappa^4 M r_+ (7 Q^2 + 27 M r_+) - 35 \kappa^6 M^2 Q^2 \Big)c_3 +  \cdots \Big]\,.\nonumber
\end{align}
In contrast, for $\ell=3$, the $n=2$  generates a running Love number, with beta function\,\footnote{
The $n>2$ operators yield the constant contribution
\begin{align}
    k_{\ell= 3} &= \frac{4 \sigma \kappa^2 (e Q)^6}{4725 r_+^{16}}\Big[1400 r_+^6 - 2600 \kappa^2 M r_+^5 - 280 \kappa^4 M r_+^3 (Q^2 + 2 \kappa^2 M^2) - \kappa^6 (Q^3 + 2 \kappa^2 M^2 Q)^2 \nonumber \\ & \quad \quad \quad \quad \quad \quad+ 10 \kappa^6 M Q^2 r_+ (Q^2 + 2 \kappa^2 M^2) + \frac{2}{7} \kappa^4 r_+^2 (Q^2 + 2 \kappa^2 M^2)(19 Q^2 + 108 \kappa^2 M^2) \nonumber \\ & \quad \quad \quad \quad \quad \quad + 4 r_+^4 (69 \kappa^2 Q^2 + 458 \kappa^4 M^2)  
     \Big]c_3 + \cdots
\end{align}
}
\begin{equation}
    \beta_{\ell=3} = \frac{128 \kappa^4 M (e Q)^4}{35 r_+^7}c_2\,.
\end{equation}
For extremal black holes,  using the leading-order solution \eqref{eq:Vector_Extremal},  we find the beta function for arbitrary $\ell$ to be
\begin{equation}
    \beta_\ell = \frac{(\ell+1)^2 (2\ell-2)!}{2\ell +1}\sum_{n=2}^{\left \lfloor \frac{\ell+1}{2}\right \rfloor} \frac{\sigma^{n} 4^{n+1}(n-1)}{(4n-3)!(2\ell-4n+3)!}\frac{e^{2n}}{(\kappa r_h)^{2(n-1)}} c_n\,.\label{eq:Beta_Vector}
\end{equation}

Comparing the  beta functions of the vector and scalar cases, i.e. \eqref{eq:Beta_Vector}  and  \eqref{eq:Beta_scalar}, we find the remarkably simple relation for any $\ell$, 
\begin{equation}
    \frac{\beta_{\ell}^{\mathrm{scalar}}}{\beta_{\ell}^{\mathrm{vector}}} = \frac{\ell^2}{(\ell+1)^2}\,. \label{eq:ratio_weak}
\end{equation}

\subsection{The Love Matrices}

We turn to the coupled tidal response to metric and electromagnetic tidal fields. 
We use the formulas for the computation of Love matrices derived in Section \ref{se:Love_number}, applied to a pair of coupled tidal fields. Here we express the results in the  basis $\mathbf{\Psi} = (\Psi_h, \Psi_a)^t$, in which the Love matrices are symmetric.~\footnote{
Using the conventions for the worldline EFT from section \ref{se:WEFT},  $\bm \phi$  is here a 2-vector. Defining the electric field $E_i\equiv F_{0i}$, the magnetic tensor field $B_{ij}\equiv F_{ij}$, 
and the  gravitomagnetic field
$B^{(C)}_{ij|k}\equiv C_{0ijk}$ with $C_{\mu\nu\rho\sigma}$ the Weyl tensor, the  $\bm \phi$ in the electric background is  
$ \bm \phi|_{\rm el} = ( B_{ij}  ,   B^{(C)}_{ij|k} )^t $, and $\bm \phi$  in the magnetic background is  $ \bm \phi|_{\rm mag} = ( E_{i}  ,   B^{(C)}_{ij|k} )^t $. The $N_\ell$  coefficients relating the diagonal of $\Lambda_\ell$ to $K_\ell$  can be easily inferred from \cite{Hui:2020xxx}, while the nondiagonal coefficient remains as a pleasant exercise to be done. 
}  

The ingredients of the computation are collected in App.~\ref{app:Computing_Love}. 
Namely, following the method of section \ref{se:tidal_PT}, 
we diagonalize the $2\times2$ potentials \eqref{eq:potential_full},  \eqref{eq:potential_full_mag} at leading order and compute their solutions $\mathbf{\Psi}^{(0)}$ in App.~\ref{app:Computing_Love_LO}. We then compute the 
 perturbation of the wave operator acting on the leading-order solution, i.e. $ \mathcal{D}^{(1)}\mathbf{\Psi}^{(0)}$, in  App.~\ref{app:Computing_Love_NLO}.

We first focus on the effect of the $n=2$ EFT operator (i.e. $F^4$). Based on Prop.~\eqref{Prop:running}, the $\ell=2,3$ Love matrices may run. We find the following beta functions:
\begin{align}
    \beta_{\ell=2} &= \frac{\kappa e^4 Q^2}{5 r_+^5} \left(
    \begin{array}{cc}
         0 &  96 \sigma \sqrt{2}  Q \\
        96 \sigma \sqrt{2}  Q & -432 \kappa M \\
    \end{array}
    \right) c_2\,, \\ \label{eq:beta_l2_n2}
    \beta_{\ell=3} &= \frac{\kappa^3 e^4 Q^2}{21 \sqrt{5} r_+^7} \left(
    \begin{array}{cc}
         -928 \sqrt{5} \kappa M Q^2 & 64 \sigma Q (13 Q^2 + 56 \kappa^2 M^2) \\
        64 \sigma Q (13 Q^2 + 56 \kappa^2 M^2) & -\frac{128}{\sqrt{5}}  \kappa M(67 Q^2 + 78 \kappa^2 M^2)\\
    \end{array}
    \right) c_2\,.
\end{align}

As a consistency check of the whole formalism, we present a pedestrian calculation of the $n=2,\ell=2$ case in App.~\ref{app:Derivation}.

The contributions from EFT operators with $n>2$ are constant entries in the $\ell=2,3$ Love matrices. For simplicity, we present the results for extremal black holes, considering the first $n$ of the series:
\begin{align}
    K_{\ell=2} &= \frac{e^6}{25 \kappa^4 r_h^4} \left(
    \begin{array}{cc}
         -7424 \sigma &  17184 \\
        17184 & -\frac{108928}{3} \sigma \\
    \end{array}
    \right) c_3 + \frac{e^8}{5 \kappa^6 r_h^6} \left(
    \begin{array}{cc}
         -9216 &  22528 \sigma \\
       22528 \sigma & -\frac{4 315 904}{91} \\
    \end{array}
    \right) c_4 + \cdots\\
    K_{\ell=3 }&= \frac{e^6}{7 \kappa^4 r_h^4} \left(
    \begin{array}{cc}
         -\frac{78 080}{27} \sigma &  20064\sqrt{\frac{2}{5}} \\
        20064\sqrt{\frac{2}{5}} & -\frac{5 110 528}{135} \sigma \\
    \end{array}
    \right) c_3 + \frac{e^8}{21 \kappa^6 r_h^6} \left(
    \begin{array}{cc}
         -\frac{3681280}{91} &  306688 \sqrt{\frac{2}{5}} \sigma \\
       306688 \sqrt{\frac{2}{5}} \sigma & -\frac{264367104}{455} \\
    \end{array}
    \right) c_4 + \cdots
\end{align}

In the extremal black hole background, we use the zeroth-order solutions \eqref{eq:Psi_extremal_1} and \eqref{eq:Psi_extremal_2}, valid for arbitrary $\ell \geq 2$. The beta function then takes the form
\begin{align}
        \beta_{\ell} &=   \sum_{n=2}^{\left \lfloor \frac{\ell}{2}\right \rfloor + 1}c_n\beta^{(n)}_{\ell} \,,
\end{align}
where $\beta^{(n)}_{\ell}$ denotes the contribution of the operator $ (F^2)^n \subset \Delta \mathcal{L}$ to the beta function, and is given by
\begin{align}
   \beta^{(n)}_{\ell}  &=   -  \frac{(\ell +1)^2 (2 \ell -4)!}{(2 \ell-1 )^2 (2 \ell +1)}    
     \frac{\sigma^n 4^{n+2} (n-1) }{ ( 4 n-1)! (
    2\ell- 4 n +5 )!}\begin{pmatrix}
        \beta^{(n,\ell)}_{h h} & \beta^{(n,\ell)}_{h a} \\
        \beta^{(n,\ell)}_{a h} & \beta^{(n,\ell)}_{a a}\\
    \end{pmatrix}\frac{e^{2 n}}{(\kappa r_h)^{2(n-1)}}\,, \label{eq:extremal_beta_n}
\end{align}
with components given in App.~\ref{app:Love_beta_gen}. 

At leading order, where the expansion of $\Delta \mathcal{L}$ is truncated to $O(F^6)$, keeping only the leading operator $F^4$, the Love beta function reduces to $\beta_{\ell}\sim c_2 \beta^{(2)}_{\ell}$. In the large-$\ell$ limit, the $n=2$ Love beta function behaves as
\be
   \beta^{(2)}_{\ell} \xrightarrow[\ell \to \infty]{}    
     -   \frac{16 e^4 \ell^6}{63 r_h^2 \kappa^2} \begin{pmatrix}
       ~~1 & -\sigma  \\ -\sigma  & ~~1
   \end{pmatrix} \,. \label{eq:Weak_large_L}
\ee
The leading-order beta function matrix therefore has a vanishing eigenvalue. The combination $\Psi_a+\sigma \Psi_h$  has  Love numbers suppressed as $\sim \frac{1}{\ell^6}$ at large $\ell$, whereas the $\Psi_a-\sigma \Psi_h$ combination has Love numbers growing as $\sim \ell^6$.

\subsubsection{An electromagnetic duality}

In all our results, we find that  the entries of the Love matrices in the electric and magnetic backgrounds are related by 
\begin{align}
\left(K_\ell^{(n)}|_{\cal B}\right)_{hh} &= \sigma^n \left(K_\ell^{(n)}|_{\cal E}\right)_{hh} \,,\quad 
\left(K_\ell^{(n)}|_{\cal B}\right)_{aa} = \sigma^n \left(K_\ell^{(n)}|_{\cal E}\right)_{aa} \,, \nonumber \\ 
\left(K_\ell^{(n)}|_{\cal B}\right)_{ah} &= \sigma^{n+1} \left(K_\ell^{(n)}|_{\cal E}\right)_{ah} \label{eq:duality} \,. 
\end{align}
There is therefore a $Z_2$ symmetry of the Love matrices under ${\cal E} \leftrightarrow {\cal B}$ exchange.  This fact  is highly nontrivial, because the electric and magnetic potentials are not just related by  sign flips. The electric field receives a correction while the magnetic field does not, and furthermore the $\partial^2_{r_\star}$ term differs between both potentials. In fact one can check that any change in these terms would break the $Z_2$ symmetry. It would be very interesting to find a formulation of the problem that makes this  symmetry manifest.

\section{Love Numbers of Strong-Field Magnetic Black Holes}
\label{se:strong_field}

In this section, we focus on a magnetic black hole in the strong-field regime, i.e. a black hole with horizon $r_h<r_h^c$ (see section \ref{se:BH_properties}).  
Depending on the distance $L$ of the source to the center of the magnetic black hole, the source may lie within the strong-field region ($L<r_h^c$) or the weak-field region ($L>r_h^c$). 
Here we focus on $L<r_h^c$, which is pictured  as:
\be \hbox{\includegraphics[width=0.6\linewidth,trim={1.5cm 4.5cm 2cm 3.5cm},clip]{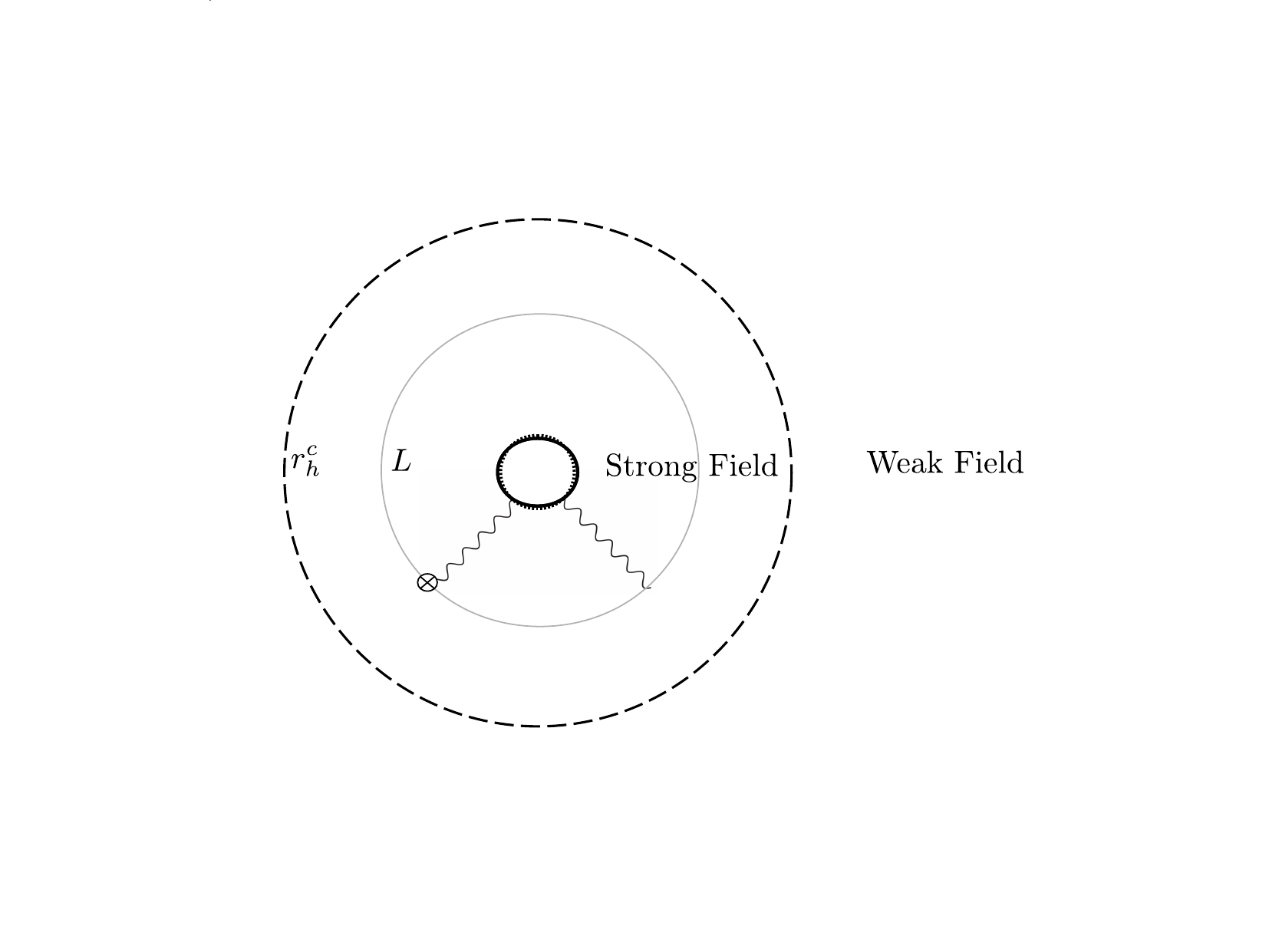}} \nn \ee
In this case we can derive the tidal response using $\Delta{\cal L}_F= \beta \frac{ F^2}{  8 }\log{\left(\frac{ F^2}{\mu_0^4} \right)}$. See section \ref{se:Deformation_strong_field} for details.  
The $L>r_h^c$ case is briefly discussed in  section \ref{se:matching}.

 As in the previous section, we begin by deriving the Love numbers from probe fields. This provides a simple example of the relation between the $U(1)$ running and the Love beta functions. We then derive the running of the Love matrices. 
 The computional framework is the same as in section \ref{se:weak_field}. Here the perturbative expansion is made relative to powers of the gauge coupling. 

\subsection{Scalar and Vector Love Numbers}

We find that for scalar fields the Love numbers run for all $\ell$. For vector fields, the running occurs for all $\ell \geq 2$. In all cases, the Love number beta function is proportional to the $U(1)$ beta function, 
\be \beta_{k_\ell} \propto \beta_{\frac{1}{e^2}}\,\,.  \ee

\subsubsection*{Scalar} 

The beta functions $\beta_{k_\ell}\equiv \beta_\ell$ for the lowest values of $\ell$ are found to be
\begin{align}
    \beta_1 &= \beta \frac{\kappa^4 M (eQ)^2}{6 r_+^3}\,, \quad \quad \quad \beta_2 = \beta \frac{\kappa^6 M (e Q)^2}{30 r_+^5} (Q^2 + \kappa^2 M^2)\,, \\ \beta_3 &= \beta\frac{3 \kappa^8 M (e Q)^2}{1400 r_+^7}(3 Q^4 + 7 \kappa^2 M^2 Q^2 + 2 \kappa^4 M^4)\,, \quad \cdots\,
\end{align}
For a scalar tidal field in an extremal black hole background, the beta function for arbitrary $\ell$ takes the simple form
\begin{equation}
    \beta_{\ell} = \beta \frac{2 \ell e^2}{2 \ell + 1}\,.\label{eq:Scalar_EH}
\end{equation}

\subsubsection*{Vector} 

For a vector field,  the running starts at $\ell=2$. The results for the lowest values of $\ell$ are found to be
\begin{align}
    k_{1} &= - \beta \frac{\kappa^2 (e Q)^2 (\kappa^2 Q^2 + 2 r_+^2)}{6 r_+^4}\log{\left(\frac{\sqrt{2} e Q}{\mu_0^2 r_+^2} \right)}\,, \quad \quad \beta_2 = -\beta \frac{3 \kappa^6 M e^2 Q^4}{40 r_+^5}\,,
\end{align}
\begin{align}
    \beta_3 &= -\beta \frac{\kappa^8 M e^2 Q^4 (Q^2 + 2 \kappa^2 M^2)}{105 r_+^7}\,, \quad \cdots\,
\end{align}
For  arbitrary $\ell \geq 2$ in an extremal black hole background, the general beta function formula reads
\begin{equation}
    \beta_{\ell} = - \beta \frac{2 (\ell^2 - 1) e^2}{\ell(2 \ell + 1)}\,.\label{eq:Vetor_EH}
\end{equation}

We find that the vector and scalar beta functions \eqref{eq:Vetor_EH} and  \eqref{eq:Scalar_EH} are again related by a simple factor,
\begin{equation}
    \frac{\beta_{\ell}^{\mathrm{scalar}}}{\beta_{\ell}^{\mathrm{vector}}} = - \frac{\ell^2}{\ell^2 - 1} \,, \quad \quad \ell\neq1\,.
\end{equation}
This factor differs from the one found in the weak-coupling regime, see \eqref{eq:ratio_weak}. 

\subsection{The Love Matrices}

The coupled tidal responses to metric and electromagnetic perturbations exhibit a running for all $\ell \geq 2$. For the first $\ell$, we find
\begin{align}
    \beta_2 &= \beta \frac{(e Q)^2 \kappa^5}{5 r_+^5}
    \begin{pmatrix}
        \kappa M Q^2 & -\frac{3 Q}{2\sqrt{2}}(Q^2 + \kappa^2 M^2) \\
         -\frac{3 Q}{2\sqrt{2}}(Q^2 + \kappa^2 M^2) & \frac{\kappa M}{8}(11 Q^2 + 9 \kappa^2 M^2) \\
    \end{pmatrix},\\
    \beta_3 &= \beta \frac{(e Q)^2 \kappa^7}{1260 r_+^7}
    \begin{pmatrix}
        \kappa M Q^2(53 Q^2 + 64 \kappa^2 M^2) & \frac{Q}{\sqrt{5}}(45 Q^4 - 307 \kappa^2 M^2 Q^2 - 96 \kappa^4 M^4) \\
         \frac{Q}{\sqrt{5}}(45 Q^4 - 307 \kappa^2 M^2 Q^2 - 96 \kappa^4 M^4) & \frac{8 \kappa M}{5}(5 Q^4 + 76 \kappa^2 M^2 Q^2 + 18 \kappa^4 M^4) \\
    \end{pmatrix}.
\end{align}

In the extremal black hole background, we use the leading-order solutions \eqref{eq:Psi_extremal_1} and \eqref{eq:Psi_extremal_2}, valid for arbitrary $\ell \geq 2$. The general beta function  takes the form
\begin{equation}
    \beta_{\ell} =  -\beta \frac{e^2}{3 \ell (\ell-1) (2\ell-1)^2 (2\ell+1)}     
    \begin{pmatrix}
        \beta_{hh} & \beta_{ha} \\
        \beta_{ah} & \beta_{aa}\\
    \end{pmatrix},
\end{equation}
with components
\begin{align}
    \beta_{hh} &= (\ell+2)(\ell+1)\Big[2 \ell^5-25 \ell^4+51 \ell^3-24 \ell^2+28 \ell - 68 \Big]\,,\\
    \beta_{ha} &= \beta_{ah} = -(\ell+1)\sqrt{\ell(\ell+1)-2} \Big[2 \ell^5-19 \ell^4-6 \ell^3+36 \ell^2-44 \ell +70 \Big]\,,\\
    \beta_{aa} &= (\ell-1)\Big[2 \ell^6-11 \ell^5-58 \ell^4-66 \ell^3+52 \ell^2-7 \ell -80 \Big]\,.
\end{align}

Remarkably, in the large-$\ell$ limit, the Love beta function behaves as
\be
   \beta_{\ell} \xrightarrow[\ell \to \infty]{} -\frac{\beta e^2 \ell^2}{12} \begin{pmatrix}
       ~~1 & -1  \\ -1  & ~~1
   \end{pmatrix} \,. \label{eq:EH_large_L}
\ee
The beta function matrix has thus a vanishing eigenvalue. It turns out that  the combination $\Psi_a+\Psi_h$  has  Love numbers suppressed as $\sim \frac{1}{\ell^2}$ at large $\ell$, while the $\Psi_a-\Psi_h$ combination has Love numbers going as $\sim \ell^2$. {Note that this is analogous  to the large-$\ell$ limit of the Love beta function in the weak-field regime \eqref{eq:Weak_large_L}, with $\sigma = 1$ for the magnetic background.}

\subsection{Scale-Independence of the Beta Functions}

The fact that the Love  beta functions depend solely on the $U(1)$ beta function --- and not on $\log(L)$ --- is beautiful, but  intriguing. Why does $\beta_{\ell}\propto \beta_{U(1)}$ hold?

Inspecting the explicit  computations, one may observe that this fact occurs due to intricate cancellations between the coefficients of the zero-order solution.  These magic cancellations are in fact a manifestation of  the vanishing of the Love numbers at classical level. The property is  shown as follows. 

The strong-field metric deviates from the classical metric only via the effective  $\beta\log(r/L)$ dependence of the black hole charge. 
This metric, when  introduced in the tidal equation of motion, would produce an exactly vanishing Love numbers in the absence of the $\log(r/L)$, i.e. if the black hole charge was constant. This implies that the only terms that contribute to the Love number beta function are those where a $r$-derivative acts on the black hole charge --- because all other terms must cancel exactly. It follows that the Love  beta functions must be proportional to the derivative  of $\beta \log(r/L) $, and are thus independent of  $\log(L)$.

\subsection{On the Computation in the Weak-Field Region\label{se:matching}}

If the worldine EFT is defined in the weak-field region, the knowledge of the solutions in the strong-field region is still necessary to compute the Love numbers, because  the regularity condition on the horizon needs to be used. 
The computation in this case could in principle be  done exactly if the  solutions to the equation of motion perturbed by the exact Euler-Heisenberg Lagrangian were known.

The exact solutions are however not known analytically. Instead, one may use the simpler expressions for the Lagrangian at strong and weak field given in 
\eqref{eq:DeltaL_LargeField}, \eqref{eq:EFT_action}, \eqref{eq:loops_coef}.
This approximate approach should be performed in two steps, with an intermediate matching of the solutions at the transition radius $L\sim r_h^c$. 
The resulting Love number  will then contain a constant logarithmic contribution as well as the contribution from the weak-field region, we schematically 
\be k_\ell(L) =  \beta^{\rm strong}_{\ell}\log\left(\frac{r_h^c}{r_h}\right)+ \beta^{\rm weak}_{\ell}\log\left(\frac{L}{r_h^c}\right) \,.
\label{eq:TLN_weak}
\ee

\section{Summary and Outlook
\label{se:conclusion}
}

\subsection{Summary}

Throughout this work, we have studied the tidal properties of four-dimensional, unspinning charged black holes in the vacuum of quantum field theory. Unlike black holes in the classical vacuum, 
black holes living in the  QFT vacuum  are not completely isolated. Instead, they are surrounded by vacuum bubbles that induce  non-vanishing tidal responses.

Black holes with charge are especially interesting, as they acquire corrections that are much larger than those of their neutral counterparts. Moreover, they present both a weak and a strong-field regime and tend to be dominated by corrections to the Maxwell sector. These features are particularly pronounced for near-extremal black holes. For these reasons, our analysis focuses on the quantum corrections to the  Maxwell sector.

Charged black holes exhibit mixed responses to gravitational and electromagnetic tidal fields. To properly describe these effects, we introduce the  {\it Love matrices}, which generalize the concept of Love numbers to  mixed tidal responses. 
The Love matrices are  independent of an overall rescaling of the tidal fields, but depend on unitary rotations of the field basis.

Another important property of the Love matrices is the existence of a field basis in which they are symmetric. While this property is not  manifest at the level of the curved space description, it follows 
from matching to the worldline EFT description.
As a result, the  Love matrices for $N$ tidal fields  contain $\frac{N(N+1)}{2}$ independent parameters. 
In this  work we focus on purely electric and magnetic black holes, for which  fluctuations  of different parity decouple. In this case,  the Love matrices  are $2\times2$, and we identify  the field basis for which they are symmetric. For a dyonic black hole, carrying both electric and magnetic charges, the Love matrices will be $4 \times 4$, since both even and odd sectors of electromagnetic and gravitational perturbations mix.

As for Love numbers, Love matrices can exhibit  a renormalization flow, in which case the remaining constant terms are unphysical. While this property was demonstrated in \cite{Barbosa:2025uau} using an argument based on field redefinitions,  our analysis provides another argument:  the constant terms are not necessarily symmetric when there is running --- while the Love beta function is always symmetric.

Extending the perturbative method of \cite{Barbosa:2025uau}, we show 
that both constants and running Love matrices can be computed for any $\ell$ by a couple of compact formulas, \eqref{eq:K_full}\, \eqref{eq:B_full}. 
Their derivation involves introducing a Green's function for the mixed tidal fields. 
To provide an explicit consistency check,  we present  a pedestrian derivation in a simple example (see App.\,\ref{app:Derivation}), which reproduces the result given  by our method.

We compute the coupled equation of motion for the metric and electromagnetic fluctuations in both electric and magnetic backgrounds, in the presence of Maxwell corrections.
Due to the quantum corrections, electromagnetic duality does not apply. Hence we separately compute both electric and magnetic equations of motion by brute force.

We compute the tidal response of electric and magnetic black holes
in the weak-field regime, defined by radii larger than a critical radius $r_+^c$ given in \eqref{eq:rc_def}. We consider a full series of EFT operators $F^{2n}$ in these computations. In addition to the Love matrix encoding the mixed tidal response  to metric and electromagnetic perturbations, we also 
 consider the tidal response to scalar and vector probe fields. We show that, for all tidal fields,   a finite number of operators contributes to the Love beta function for a given harmonic.
 
We report the Love matrices for the lowest harmonics and operators, and we present  the Love beta functions for arbitrary harmonics and  operators in the  case of extremal black holes.  
We find that, although the electric and magnetic equations of motions are not related by duality, all the Love numbers and elements of the Love matrices are $Z_2$-symmetric under exchange of electric and magnetic backgrounds. It would be interesting to find a formulation that makes this  symmetry manifest at the level of the equations of motion.

Beyond the EFT regime, we compute the tidal response of magnetic black holes that are sufficiently small to lie in the  strong-field regime, $r<r_+^c$. 
We assume an Euler-Heisenberg UV completion of the Maxwell sector. We find that,  for all kind of tidal fields, the Love numbers and Love matrices exhibit running. The corresponding Love  beta functions are governed by the $U(1)$ gauge-coupling beta function:
\be
\beta_{\rm Love} \propto \beta_{U(1)}\,. 
\ee

While the Love matrices in the weak-field regime are suppressed by powers of the black hole radius $r_+$,  the Love matrices in the strong-field regime depend only logarithmically on $r_+$. The overall picture is that the tidal response increases as  $r_+$ decreases, and saturates when $r_+$ reaches $r^c_+$.  
We also find that, for both weak and strong-field regimes 
at large $\ell$, one combination of the canonical tidal fields experiences vanishing running, due to the   mixing between the gravitational and electromagnetic sectors becoming dominant.

\subsection{Outlook: Probing the Dark  Sector Through Magnetic Black Holes}

We conclude by briefly discussing how our results on charged black holes could be used 
in probes of fundamental physics.  
 There is overwhelming evidence for dark matter and dark energy,  which suggests the existence of a light dark sector. The dark sector may feature a dark Abelian gauge symmetry with gauge coupling $\tilde e$. For concreteness we assume a dark electron with mass $\tilde m$ and unit dark charge.

As pointed out in \cite{Barbosa:2025uau}, black holes may be charged, either electrically or magnetically, under the dark $U(1)$ group, and thus develop large Love numbers.\,\footnote{Primordial dark-charged black holes have been considered as dark matter candidates due to their long-lived properties, see e.g.\,\cite{jess_BN}.}  The magnetic  charge may result from capturing heavy magnetic monopoles in the early Universe,  see e.g. \cite{Stojkovic:2004hz,Zhang:2023tfv}.
Such a magnetic black hole naturally evolves to near-extremality because it cannot dissipate its magnetic charge.

For a magnetically dark-charged black hole in the strong-field regime, i.e. with radius $r_+<r_+^c$  the Love numbers and matrices are governed by the running of the dark $U(1)$, as shown in section \ref{se:strong_field}.
 For black hole masses of roughly 10–100 solar masses probed by LIGO/VIRGO, the dark electron has to satisfy   \be \tilde m \lesssim 100  \sqrt{ \tilde e}~{\rm MeV}  \ee
for the black hole to be in strong-field regime.   

The $\ell=2$ Love matrix that runs in the strong-field regime is at least of order $\beta_{k_2}$, therefore we can use the estimate 
\be k_2 \sim \frac{ \beta \tilde e^2}{5}\begin{pmatrix}
  8 &  -18 \\ 18 & 29 
\end{pmatrix} \approx   \begin{pmatrix}
  0.01 &  -0.03 \\ -0.03 & 0.05 
\end{pmatrix} e^2 \,. \ee
Taking  $\tilde e \sim 4\pi$, the $\ell=2$ Love matrix reaches   $k_2\approx \begin{pmatrix}
  2.1 &  -4.8 \\ -4.8 & 7.7 
\end{pmatrix}$. Such values for $k_2$ could realistically be probed by future gravitational waves experiments, see e.g.  \cite{Chia:2023tle}. 
\\~
\\~
\\
{\textbf{NOTE}:}
In the final days prior to submission of this work, Ref.~\cite{Noumi:2026shc} appeared, which has partial overlap with our results.

\begin{acknowledgments}

 The work of SB is supported by grant 2025/05571-3 of the São Paulo Research Foundation (FAPESP). SF is supported in part by FAPESP under grants 2025/27083-0 and 2021/10128-0. LS is supported by grants 2023/11293-0 and 2025/27243-8 of FAPESP.
 SF thanks the IPhT/CEA-Saclay for funding  a visit during which this work was initiated. 
 
\end{acknowledgments}

\appendix

\section{Variational Formulas}\label{se:Formulas}

\subsection{Einstein Tensor}

 The metric perturbation of \eqref{eq:var_def} induces the first-order variation of the  Riemann tensor,
\begin{equation}
    \delta R\indices{^\rho _\mu _\lambda _\nu} = \nabla_\lambda \delta \Gamma^\rho_{\nu \mu} - \nabla_\nu \delta \Gamma^\rho_{\lambda \mu}\,,\label{eq:R_1}
\end{equation}
where the Christoffel symbols varies as
\begin{equation}
    \delta \Gamma^\rho_{\mu\nu} = \frac{1}{2} g^{\rho \sigma}\Big(\nabla_\mu h_{\nu \sigma} + \nabla_{\nu} h_{\sigma \mu} - \nabla_\sigma h_{\mu\nu} \Big)\,.
\end{equation}
Consequently, the variation of the Ricci tensor is
\begin{align}
    \delta R_{\mu\nu} &= \frac{1}{2}\Big(\nabla_\mu \nabla_\rho h\indices{_\nu ^\rho} + \nabla_\nu \nabla_\rho h\indices{^\rho _\mu} - \square h_{\mu\nu} - \nabla_\mu \nabla_\nu h\Big) + R\indices{_( _\mu ^\rho} h\indices{_\nu _) _\rho} - R_{\mu\rho\nu\lambda}h^{\rho\lambda} \label{eq:R_2}\,.
\end{align}
The Einstein tensor variation is 
\begin{align}
    \delta G_{\mu\nu} &= \delta R_{\mu\nu} - \frac{1}{2}(\delta R) g_{\mu\nu} - \frac{1}{2} R h_{\mu\nu}\\
     &= \delta R_{\mu\nu} - \frac{1}{2}\Big(g^{\rho \sigma}\delta R_{\rho \sigma} - h^{\rho \sigma} R_{\rho \sigma} \Big)g_{\mu\nu} - \frac{1}{2} R h_{\mu\nu}\,,
\end{align}
and combining with  \eqref{eq:R_1}-\eqref{eq:R_2} we obtain
\begin{align}
    &\delta G_{\mu\nu} =-\frac{1}{2}\Big[\square h_{\mu\nu} + \nabla_\mu \nabla_\nu h - \nabla_\mu \nabla_\rho h\indices{^\rho _\nu} - \nabla_\nu \nabla_\rho h\indices{^\rho _\mu} - 2\Big(R\indices{_( _\mu ^\rho}h\indices{_\nu _) _\rho} - R_{\mu \rho \nu \lambda}h^{\rho \lambda} \Big)  \nonumber\\ & \quad \quad \quad \quad \quad - g_{\mu\nu}\Big(h^{\rho \sigma}R_{\rho \sigma} + \square h - \nabla_\rho \nabla_\sigma h^{\rho \sigma} \Big) + R h_{\mu\nu} \Big]\,. \label{eq:deltaGmunu}
\end{align}
This variation appears in the quadratic action \eqref{eq:perturbation_action}. 

\subsection{Stress-Energy Tensor}\label{App:stress_energy_tensor}

The stress-energy tensor corresponding to the quantum effective action \eqref{eq:quantum_eff_action_v2} is given by
\begin{equation}
    T_{\mu\nu} = \frac{1}{e^2} (1 + \Delta_F) F_{\mu \lambda} F\indices{_\nu ^\lambda} - \frac{1}{4 e^2} (1 + \bar \Delta_F) g_{\mu\nu}F^2\,.
\end{equation}
We find its first variation to be
\begin{align}
    \delta  T_{\mu\nu} &= \frac{2}{e^2}(1+\DeltaF) f\indices{_( _\mu ^\lambda} F\indices{_\nu _) _\lambda} + \frac{1}{e^2}F_{\mu \lambda} F\indices{_\nu ^\lambda} \delta \DeltaF -\frac{1}{4e^2} g_{\mu\nu}F^2 \delta \TildeDeltaF \nonumber\\&\quad+ \frac{(1+\TildeDeltaF)}{e^2}\Bigg[ - \frac{1}{2}g_{\mu\nu}F^{\rho \sigma}f_{\rho \sigma} - \frac{1}{4}h_{\mu\nu}F^2 - h^{\sigma \lambda}F_{\mu \lambda}F_{\nu \sigma} + \frac{1}{2}g_{\mu\nu}h^{\rho \lambda}F_{\rho \sigma}F\indices{_\lambda ^\sigma}\Bigg]\,,\label{eq:delta_T}\\
    \delta \DeltaF &= \frac{\partial \DeltaF}{\partial F^2} \delta F^2\,, \quad \quad \delta \TildeDeltaF = \frac{\partial \TildeDeltaF}{\partial F^2} \delta F^2\,, \quad \quad \delta F^2 = 2 F^{\rho \lambda}F\indices{^\sigma _\lambda}h_{\rho \sigma} + 2 F^{\rho \sigma}f_{\rho\sigma}\,. &
\end{align}
This variation appears in \eqref{eq:perturbation_action}. 
The correction terms for the EFT and Euler-Heisenberg backgrounds are given by
\begin{align}
    \TildeDeltaF|_{\rm EFT} &= -\sum_{n=2}^{\infty} 4 e^2 c_n (F^2)^{n-1}\,, \quad \quad \DeltaF|_{\rm EFT} = -\sum_{n=2}^{\infty} 4 n e^2 c_n (F^2)^{n-1}\,,\\
    \TildeDeltaF|_{\rm EH} &= - \frac{\beta e^2}{2}\log{\left(\frac{F^2}{\mu_0^4} \right)}\,, \quad \quad \quad \,\,\,\,\, \DeltaF|_{\rm EH} =  - \frac{\beta e^2}{2}\Bigg[1 + \log{\left(\frac{F^2}{\mu_0^4} \right)}  \Bigg]\,.
\end{align}
\section{Gravitoelectromagnetic Mixing}\label{app:mixing}
Due to absence of trace for the {parity-odd} component of $h_{\mu\nu}$ and using the Regge-Wheeler gauge where $h_{ij}$ is absent, the third term in \eqref{eq:perturbation_action} reduces to
\begin{align}
    h^{a i}\delta  T_{a i} = \frac{2}{e^2}(1+\DeltaF) f\indices{_( _a ^\lambda} F\indices{_i _) _\lambda}h^{a i} &+ \frac{1}{e^2}F_{a \lambda} F\indices{_i ^\lambda} h^{a i}\delta \DeltaF \nonumber\\&- \frac{(1+\TildeDeltaF)}{e^2}\Bigg[  \frac{1}{4}h_{a i}h^{a i}F^2 + h^{\sigma \lambda} h^{a i} F_{a \lambda}F_{i \sigma}\Bigg]\,. \label{eq:a_h_interaction}
\end{align}

The gravitoelectromagnetic mixing term is given by the first line of \eqref{eq:a_h_interaction}. By specifying  a  purely electric or purely magnetic background, the second mixing term, proportional to $\delta \DeltaF$, vanishes exactly due to the factor $F_{a \lambda} F\indices{_i ^\lambda}$. Indeed, the only nonvanishing components of the field-strength tensor are $F_{a b} = -\mathcal{E}(r) \epsilon_{a b}$ in the electric case and $F_{i j} =  r^2\mathcal{B}(r)  \epsilon_{i j}$ in the magnetic  case.  

\subsection*{Electrically-charged background}

The gravitomagnetic mixing term is proportional to \begin{equation}
    h^{a i} \epsilon_{a b}\nabla^b a_i\,. \label{eq:ele_interaction}
\end{equation} 
Hence $a_{0}$ and $a_1$ do not couple with the  odd-parity sector of the metric fluctuation. Furthermore, this contraction selects only the transversal component of $a_i$, namely $a_T$, since $h^{a}\mathcal{Y}^i$ is orthogonal to $ \epsilon_{a b}\nabla^b a_L \nabla_i Y$. 

\subsection*{Magnetically-charged background}

 In this case the situation is reversed.   The gravitoelectric mixing term is proportional to
\begin{align}
\left(\nabla_b (\epsilon_{ij} a^j) - \epsilon_{ij} \nabla^j a_b\right)h^{b i} \,,
\label{eq:mag_interaction}\end{align}
hence the odd- and even-parity sectors are interchanged with respect to the electric case. As a result the  time, radial, and longitudinal components of $a_\mu$ mix with   $h_{\mu\nu}$.

{Given the mixing structure described above, and adopting the $a_L=0$ electromagnetic gauge (see \eqref{eq:gauge_a_L}), we can set $a_\mu=a^{\rm odd}_\mu=(0,0,\sum_{m,\ell}a^{\ell m}_T {\cal Y}_i^{\ell m})$ for an electric background and $a_\mu=a^{\rm even}_\mu = (\sum_{\ell m}a^{\ell m}_b Y^{\ell m },0,0)$ for a magnetic background.}

\section{Probe Fields in the Black Hole Background}\label{se:Test_Fields}

We present the equations of motion for  probe fields propagating on a charged black hole background, and then derive the static solutions that are regular at the horizon in the leading-order (i.e. RN) geometry (see subsection \ref{se:RN_metric}).

\subsection*{Scalar Field}

Consider a free real massless scalar field $\phi$ propagating in the charged black hole geometry. The field equation is:
\begin{equation}
    \square \phi = 0\,.
\end{equation}
Decomposing the field into spherical harmonics,
\begin{equation}
    \phi = \sum_{l, m}\frac{\Psi_{\ell m}(t,r)}{r} Y^{\ell m}(\theta,\phi)\,,
\end{equation}
the different angular momentum modes decouple, and each mode satisfies the radial wave equation
\begin{equation}
    \Big[-\partial_t^2 + \partial_{r_{\star}}^2 - V_\ell(r) \Big]\Psi_{\ell,m}(t,r) = 0\,, \quad \quad V_\ell(r) =  \frac{f}{r^2}(j^2 + r f^\prime)\,.
\end{equation}
In the leading-order RN background, the general static solution that is regular  at the horizon behaves as $\Psi_{\ell}^{(0)} \sim \, r^{\ell+1}$ at infinity. For the first $\ell$ we have
\begin{align}
    \Psi_{\ell=1}^{(0)} &= r^2 - \frac{1}{2}\kappa^2 M r\\
    \Psi_{\ell=2}^{(0)} &= r^3 -\kappa^2 M r^2 + \frac{\kappa^2}{6} \left(Q^2 + \frac{1}{6} \kappa^2 M^2 \right) r\\
    \Psi_{\ell=3}^{(0)} &= r^4 -\frac{3}{2} \kappa^2 M r^3 + \frac{3\kappa^2}{5}\left(\kappa^2 M^2+\frac{Q^2}{2}\right)r^2 -\frac{\kappa^4}{20}
    \left(\kappa^2 M^3+ 3 M Q^2\right) r\,. 
\end{align}
 For an extremal RN black hole, the general solution that is regular at the horizon takes the closed form
\begin{equation}
    \Psi_{\ell}^{(0)} = r^{\ell+1}\left(1 - \frac{r_h}{r} \right)^\ell\,.\label{eq:Scalar_Extremal}
\end{equation}

\subsection*{Vector Field}

Consider a free massless vector field $A_\mu$ propagating in the charged black hole background. The dynamics is governed by the Maxwell equation in curved spacetime,
\begin{align}
    \square A_\mu - R_{\mu \nu}A^\nu -\nabla_\mu \nabla_\nu A^\nu &= 0\,.
\end{align}
Since the even and odd parity sectors decouple, we focus on the odd sector. In this sector we have
\begin{equation}
    A_{i} = A_T(t,r) \epsilon\indices{_i ^j}\nabla_j Y^{\ell m}(\theta,\phi)\,,
\end{equation}
where $\Psi_\ell = A_T$ is a gauge-invariant variable. The radial wave equation is
\begin{equation}
    \Big[-\partial_t^2 + \partial_{r^*}^2 -V_\ell(r) \Big]\Psi_{\ell}(t,r) =  0\,, \quad \quad V_\ell(r) = f\frac{j^2}{r^2}\,.
\end{equation}
In the leading-order RN background, for the first $\ell$, the static solutions that are regular at horizon are
\begin{align}
    \Psi_{\ell=1}^{(0)} &= r^2 - \frac{\kappa^2 Q^2}{2}\\
    \Psi_{\ell=2}^{(0)} &= r^3 - \frac{3}{4}\kappa^2 M r^2 + \frac{1}{8}\kappa^4 M Q^2\\
     \Psi_{\ell=3}^{(0)} &= r^4 -\frac{4}{3}\kappa^2 M r^3 + \frac{1}{5} \kappa^2 (Q^2 + 2 \kappa^2 M^2)r^2 - \frac{1}{60} \kappa^4 Q^2(Q^2 + 2 \kappa^2 M^2)\,. 
\end{align}
For an extremal RN black hole, the general static regular solution is
\begin{equation}
    \Psi_\ell^{(0)} = r^{\ell+1}\left(1 + \frac{r_h}{\ell r} \right)\left(1 - \frac{r_h}{r} \right)^{\ell}\,.\label{eq:Vector_Extremal}
\end{equation}

\section{Computing the Love matrices}\label{app:Computing_Love}

We collect the elements required to apply the perturbative method for computing the Love matrices as described in Section~\ref{se:Love_number}. This procedure requires expanding the equation of motion perturbatively and solving it  in the leading-order   background (i.e. Reissner-Nordstr\"om).

\subsection{Leading-Order \label{app:Computing_Love_LO}}

The zeroth-order tidal equation of motion is
\begin{equation}
    \mathcal{D}^{(0)}\mathbf{\Psi}^{(0)} = \mathds{1} \left[-\frac{\partial^2}{\partial t^2} + (f^{(0)})^2\frac{\partial^2}{\partial r^2} + f^{(0)}\frac{d}{dr}( f^{(0)}) \frac{\partial}{\partial r}\right]\mathbf{\Psi}^{(0)} - \hat V^{(0)}\mathbf{\Psi}^{(0)} = 0\,.\label{eq:Order_Zero}
\end{equation}
The zeroth-order potential for both electric and magnetic backgrounds can be written in unified form as 
\begin{equation}
    \Hat{V}^{(0)} = \begin{pmatrix}
    V_{hh}^{(0)} & V_{ha}^{(0)} \\
    V_{ah}^{(0)} & V_{aa}^{(0)} \\
\end{pmatrix},
\end{equation}
with elements
\begin{align}
    V^{(0)}_{hh} &= \frac{1}{2r^6}(j^2 r^2 + \kappa^2(2 Q^2 - 3 M r))(2 r^2 + \kappa^2(Q^2 - 2 M r))\\
    V^{(0)}_{ha} &= V^{(0)}_{ah} = -\frac{\sigma}{2 r^5} \kappa Q \sqrt{2(j^2-2)}  (2 r^2 + \kappa^2(Q^2 - 2 M r))\\
    V^{(0)}_{aa} &= \frac{1}{2 r^6}(j^2 r^2 + 2 \kappa^2 Q^2)(2 r^2 + \kappa^2(Q^2 - 2 M r))\,,
\end{align}
where $\sigma=-1 (+1)$ for the electric (magnetic) black hole, as defined in \eqref{eq:sigma}. 

We diagonalize the zeroth-order potential matrix, following the perturbative method described in subsection~\ref{sec:Solving_via_Green function}, in order to decouple the equations. Diagonalizing $\hat V^{(0)}$, we obtain the matrix of eigenvalues
\begin{equation}
    \Hat{U}^{(0)} = \begin{pmatrix}
        U_+^{(0)} & 0 \\
        0 & U_-^{(0)} \\
    \end{pmatrix},
\end{equation}
where
\begin{align}
    U_{\pm}^{(0)} = \frac{(2 r^2 + \kappa^2(Q^2 - 2 M r))}{4 r^6}\Big[2 j^2 r^2 + \kappa(4 \kappa Q^2 - 3 \kappa M r \pm \sqrt{9 \kappa^2 M^2 + 8(j^2-2)Q^2})\Big]\,.
\end{align}
The corresponding eigenvectors are
\begin{align}
    \Psi_{\pm} &= N_{\pm}\Bigg[-\sigma Q \sqrt{8(j^2-2)}\Psi_h + \Big(3 \kappa M \pm \sqrt{9 \kappa^2 M^2 + 8(j^2-2)Q^2} \Big)\Psi_a \Bigg]\,,
\end{align}
for two non-zero normalization constants $N_\pm$. 

{Let $P$ be the corresponding basis transformation matrix, such that $\Hat{U}^{(0)} = P\Hat{V}^{(0)}P^{-1}$ and $\mathbf{\Psi}_{\pm} = P \mathbf{\Psi}$. Since the differential part of the wave operator is proportional to identity, it commutes with $P$.} Therefore, the transformed form of Eq.~\eqref{eq:Order_Zero} reads
\begin{equation}
    \left[-\frac{\partial^2}{\partial t^2} + (f^{(0)})^2\frac{\partial^2}{\partial r^2} + f^{(0)}\frac{d}{dr}( f^{(0)}) \frac{\partial}{\partial r}\right]\Psi^{(0)}_{\pm} - U^{(0)}_{\pm} \Psi^{(0)}_{\pm} = 0\,.\label{eq:Order_Zero_1}
\end{equation}
This result agrees with \cite{Kodama:2003kk, Pereniguez:2021xcj} upon  translating   the conventions.

\subsubsection*{Leading-order  Solutions} In the  RN background, the first $\ell$ static solutions that are regular on the  horizon are
\begin{align}
    &\Psi_{\pm,\, \ell=2}^{(0)} = \sigma_{\pm, 2}\left[r^3 - \frac{1}{8}r^2 \left( 3 \kappa^2 M \pm\kappa\sqrt{9 \kappa^2 M^2 + 32 Q^2}\right) \nonumber \right.
    \\ 
    &  \quad \quad \quad \quad \quad \quad \left.+  \frac{1}{16} \kappa^2 Q^2 \left( 3 \kappa^2 M \pm\kappa\sqrt{9 \kappa^2 M^2 + 32 Q^2}\right) - \frac{\kappa^4 Q^4}{4 r} \right]\,,\\
    &\Psi_{\pm,\, \ell=3}^{(0)} = \sigma_{\pm,3}\left[ r^4 -\frac{1}{12} r^3 \left(13 \kappa^2 M\pm\kappa\sqrt{9 \kappa^2 M^2+80  Q^2}\right) \right. \nn 
    \\
    &  \quad \quad \quad \quad \quad \quad  \left. + \frac{1}{15} \kappa^2 r^2 \left(5 Q^2 + 3 \kappa^2 M^2 \pm\kappa M \sqrt{9 \kappa^2 M^2+80 Q^2}\right)\nonumber \right. 
    \\ 
    &  \quad \quad \quad \quad \quad \quad \left. -\frac{1}{60} \kappa^4 Q^2 \left(5 Q^2 + 3 \kappa^2 M^2 \pm\kappa M \sqrt{9 \kappa^2 M^2+80  Q^2}\right) \right. \nn 
    \\
    &  \quad \quad \quad \quad \quad \quad \left.+\frac{\kappa^4 Q^4}{240 r} \left(13 \kappa^2 M \pm \kappa \sqrt{9 \kappa^2 M^2+80  Q^2}\right)\right]\,,
\end{align}
{where $\sigma_{\pm, \ell}$ are integration constants, which can be interpreted as the amplitudes of the sources.} Note that these solutions have vanishing Love Numbers, as expected. 

For an extremal RN  black hole, the general static regular solution is
\begin{align}
    \Psi_{+,\ell}^{(0)} &= \sigma_{+,\ell}\left[ \frac{r^\ell}{\ell}\left(1-\frac{r_h}{r} \right)^{\ell+1}(2 r_h + \ell r) \right]\,,\label{eq:Psi_extremal_1}\\
    \Psi_{-,\ell}^{(0)} &= \sigma_{-,\ell}\left[\frac{r^{\ell-2}}{\ell (2 \ell^2 - 3 \ell +1)}\left(1 - \frac{r_h}{r} \right)^{\ell-1}\left(6 r_h^3 + 9 r_h^2 r(\ell-1) + r^2(2 \ell^2 - 3 \ell +1)(3 r_h  + \ell r) \right)\right].\label{eq:Psi_extremal_2}
\end{align}

\subsection{First-Order \label{app:Computing_Love_NLO}}

The first-order $\mathcal{D}^{(1)}$ operator in the basis $\mathbf{\Psi}^t = (\Psi_h, \Psi_a)$ reads
\begin{equation}
    \mathcal{D}^{(1)}\mathbf{\Psi}^{(0)} = \mathds{1} \left[-\frac{\partial^2}{\partial t^2} + 2 f^{(1)}f^{(0)}\frac{\partial^2}{\partial r^2} + \Bigg(f^{(1)}\frac{d}{dr} f^{(0)}+f^{(0)}\frac{d }{dr} f^{(1)} \Bigg)\frac{\partial}{\partial r}\right]\mathbf{\Psi}^{(0)} - \hat V^{(1)}\mathbf{\Psi}^{(0)}\,.\label{eq:first_order}
\end{equation}
The first-order potential matrix is
\begin{equation}
    \Hat{V}^{(1)} = \begin{pmatrix}
    V_{hh}^{(1)} & V_{ha}^{(1)} \\
    V_{ah}^{(1)} & V_{aa}^{(1)} \\
\end{pmatrix},
\end{equation}
with components, for electric black hole, given by
\begin{align}
    V^{(1)}_{hh} &= \frac{1}{2r^4}\Big[2(r^2(j^2+2) + \kappa^2(3 Q^2 - 5 M r))f^{(1)} - r(2 r^2 + \kappa^2(Q^2 - 2 M r))f^{(1)\prime} \Big]\\
    V^{(1)}_{ha} &= V^{(1)}_{ah} = -\frac{\sigma}{4 r^5} \frac{\kappa}{e} \sqrt{2(j^2-2)}  \Big[2 r^2( (2 r^2 + \kappa^2(Q^2 - 2 Mr)) \mathcal{E}^{(1)} + 2 e Q f^{(1)}) \nonumber \\ & \quad \quad \quad \quad \quad \quad \quad \quad \quad \quad \quad \quad + e Q (2 r^2 + \kappa^2(Q^2 - 2 M r))\Delta_F \Big]\\
    V^{(1)}_{aa} &= \frac{1}{r^6}\Big[r^2(j^2 r^2 + 2 \kappa^2 e^2 Q^2)f^{(1)} + 2 \frac{\kappa^2}{e} Q r^2 (2 r^2 + \kappa^2(Q^2 - 2 M r))\mathcal{E}^{(1)} \nonumber\\
    &+ \kappa^2 Q^2(2 r^2 + \kappa^2(Q^2 - 2 M r))\Delta_F\Big]  \\ &+ \frac{\sigma}{8 r^5}(2 r^2 + \kappa^2(Q^2 - 2 Mr)) \Big[ 2(Q^2 - M r) \Delta_F^\prime - r^7(2 r^2 + \kappa^2(Q^2 - 2 M r))\Delta_F^{\prime \prime}\Big]\,.\nonumber
\end{align}
For the magnetic case, the potential has a similar form, with $\mathcal{E}^{(1)} \rightarrow \mathcal{B}^{(1)} = 0$.

\section{Example:  The $\ell=2$, $n=2$  Love Matrix}\label{app:Derivation}

To validate our general perturbative approach presented in section \ref{se:tidal_PT},  we provide a pedestrian derivation of the $\ell=2$ Love matrix  produced by the $n=2$ EFT operator, namely $c_2 F^4$. Here the Love matrix is obtained by solving the equation of motion perturbatively, requiring regularity, then taking the large $r$ limit. We consider an electrically-charged black hole and work in the original basis $\mathbf{\Psi}^t = (\Psi_h, \Psi_a)$.

The solution to the zeroth-order equation of motion \eqref{eq:Order_Zero} which is regular at the horizon  is given by
\begin{align}
    \mathbf{\Psi}^{(0)}_{\ell=2} &=  \sigma_+ \begin{pmatrix}
        r^3+2 r^2 r_h -2 r_h^3 -\frac{r_h^4}{r} \\ -2 r^3+2 r^2 r_h -2 r_h^3 + \frac{2 r_h^4}{r}
    \end{pmatrix} + \sigma_-\begin{pmatrix}
       -\frac{r^2 r_h^3}{5} + \frac{r_h^5}{5} \\ \frac{r^3 r_h^2}{5}-\frac{3 r^2 r_h^3}{10}+\frac{3 r_h^5}{10}-\frac{r_h^6}{5 r}
    \end{pmatrix}.
\end{align}
This solution is related to the solutions \eqref{eq:Psi_extremal_1} and \eqref{eq:Psi_extremal_2} by setting $N_{\pm} = 1$ and redefining the integration constants as
\begin{align}
 \sigma_+ \to  -\frac{8r_h}{5\kappa}\left(15 \sigma_{+} -2  r_h^2\sigma_{-}\right)\,, \quad \quad 
 \sigma_- \to \frac{r_h}{5\kappa} \left(80 \sigma_{+}-4  r_h^2\sigma_{-}\right)   \,.
\end{align}

Using the notations of \eqref{eq:Psireg_expansion}, we have
\begin{equation}
\mathbf{\Psi}^{(0)}_{\ell=2} =  r^3 {\bm a}_{0,2}+\ldots\,, \quad\quad \quad {\bm a}_{0,2} = a_{0, 2} \cdot {\bm \sigma} = 
\begin{pmatrix}
\sigma_+ \\
-2 \sigma_+ + \frac{r_h^2}{5}\sigma_-
\end{pmatrix},
\end{equation}
where ${\bm a}_{0,2}$ is defined as the product of a $2\times2$ matrix $a_{0,2}$ and a constant vector ${\bm \sigma}$, 
\begin{equation}
    a_{0,2} = \begin{pmatrix}
1 & 0
\\ -2 &  \frac{r_h^2}{5}
\end{pmatrix}, \quad \quad {\bm \sigma} = \begin{pmatrix}
\sigma_+ \\
\sigma_-
\end{pmatrix} \,. 
\end{equation}
Notice that ${\bm b}_{0,2}={\bm 0}$, implying that  the Love matrix vanishes at leading order.

For the $n = 2$ EFT operator, the solution to the equation of motion for the tidal field perturbation  \eqref{eq:first_order} that is regular at $r=r_h$ and at $r\to \infty$ is 
\begin{align}
    \mathbf{\Psi}^{(1)}_{\ell=2} &=  \sigma_+ \begin{pmatrix}
        \frac{32 e^4 r_h^3}{25 \kappa ^2 r^2} \left(281 + 120 \log \left(\frac{r_h}{r}\right)\right) \\ \frac{160 e^4 r_h^2}{\kappa ^2 r} + \frac{512 e^4 r_h^3}{175 \kappa ^2 r^2} \left(311 + 210 \log \left(\frac{r_h}{r}\right)\right)
    \end{pmatrix}c_2 \nonumber \\ & \quad \quad \quad \quad \quad \quad \quad \quad \quad \quad + \sigma_- \begin{pmatrix}
      - \frac{16 e^4 r_h^5}{25 \kappa ^2 r^2} \left(59 + 24 \log \left(\frac{r_h}{r}\right) \right) \\ -\frac{16 e^4 r_h^4}{\kappa ^2 r}-\frac{32 e^4 r_h^5}{875 \kappa ^2 r^2} \left(2873-1890 \log \left(\frac{r_h}{r}\right)\right)
    \end{pmatrix}c_2\,.
\end{align}
The presence of logarithmic terms in the first-order correction indicates the running of the Love numbers. Using the form \eqref{eq:Psireg_expansion}, the tidal response is expressed as
\begin{equation}
     \Psi_{\ell=2}^{(1)} = \frac{1}{r^2}\left(\mathds{1} + B_2 \log{\left(\frac{r_h}{r} \right)} \right)\mathbf{b}_{0, 2}\,,
\end{equation}
where
\begin{align}
    {\bm b}_{0,2} = b_{0,2}\cdot {\bm \sigma} =\frac{e^4 r_h^3}{\kappa^2}\begin{pmatrix}
    \frac{8992}{25}\sigma_+ -\frac{944}{25}r_h^2 \sigma_-
    \\ \frac{159232}{175}\sigma_+  -\frac{91936}{875} r_h^2 \sigma_-
    \end{pmatrix}
    c_2\,, \quad \quad B_2 = \begin{pmatrix}
    \frac{2200}{3493} & -\frac{40}{499}
    \\ \frac{11840}{24451} &  \frac{1690}{3493}
    \end{pmatrix},
\end{align}
with
\begin{equation}
    b_{0,2} = \frac{e^4 r_h^3}{\kappa^2}\begin{pmatrix}
    \frac{8992}{25} & -\frac{944}{25}r_h^2
    \\ \frac{159232}{175} &  -\frac{91936}{875}r_h^2
    \end{pmatrix}c_2\,.
\end{equation}
Using the definitions in \eqref{eq:barKB_def}, the linear transformation mapping the source to the tidal response is 
\begin{equation}
    \Bar{K}_2 = \frac{1}{r_h^{5}} b_{0, 2}\cdot (a_{0, 2})^{-1} = -\begin{pmatrix}
    \frac{448}{25} & \frac{944}{5}
    \\ \frac{704}{5} &  \frac{91936}{175}
    \end{pmatrix} \frac{e^4 c_2}{\kappa^2 r_h^2}\,,
\end{equation}
and the corresponding  Love beta matrix is 
\begin{align}
    \Bar{B}_2 = B_2 \Bar{K}_2 = 
    -\begin{pmatrix}
        0 & 384 \\
        384 & 1728 \\
    \end{pmatrix}\frac{e^4 c_2}{5 \kappa^2 r_h^2}\,.
\end{align}
This result coincides with \eqref{eq:beta_l2_n2} in the extremal limit. 
One can notice that, as expected, the $\Bar{B}_2 $ matrix is symmetric while   $\Bar{K}_2$  is not symmetric.  Hence only the former is a physical observable.

\section{Weak-Field Love Beta Matrices\label{app:Love_beta_gen}}

The components of the $n$-th contribution to the beta function in the weak-field regime for the extremal black hole \eqref{eq:extremal_beta_n} are given by
\begin{align}
    \beta^{(n,\ell)}_{h h} &=(32 n (4 n-7)+16) \ell ^8+16 \left(8 n^2-6 n+1\right) \ell ^7 \nn \\
    &-16 (4 n (n (8 n (2 n-7)+77)-46)+9) \ell ^6+4 (2 n-1) (4 n-1) (4 n (4 n-11)-3) \ell ^5 \nn \\
    &+(2 n (4 n (4 n (4 n (4 n (4 n-25)+281)-1599)+4689)-6221)+267) \ell ^4 \nn \\
    &-2 (2 n-1) (4 n-1) (2 n (284 n-721)+795) \ell ^3 \nn \\
    &-2 (n (4 n (n (4 n (4 n (60 n-431)+4889)-26277)+17294)-19919)+1307) \ell ^2 \nn \\
    &-4 (2 n-1) (4 n-1) (n (n (24 n (2 n-21)+935)-547)+76) \ell \nn \\
    &-8 (n-1) (4 n-5) (n (2 n (16 n (5 n-7)+37)+5)+9)\,,\\
    \beta^{(n,\ell)}_{h a} &= \beta^{(n,\ell)}_{a h} = -\sigma \sqrt{\ell(\ell+1)-2} \Big[(32 n (4 n-7)+16) \ell ^7+(64 n (4 n-5)+32) \ell ^6 \nn \\
    &-16 \left(56 n^2-94 n+7\right) \ell ^5 -4 (2 n (32 n (3 n (2 n-7)+26)-335)+31) \ell ^4 \nn \\
    &+(551-2 n (4 n (4 n (4 n (4 n (4 n-25)+217)-751)+527)+1809)) \ell ^3 \nn \\
    &+(2 n (64 n (n (n (8 n (4 n-25)+515)-650)+400)-6537)+797) \ell ^2 \nn \\
    &+(265-2 n (2 n (2 n (4 n (4 n (4 n+31)-539)+2881)-3213)+1399)) \ell \nn \\ 
    &+2 (4 n-5) (2 n (n (2 n (16 n (5 n-12)+137)-51)+13)-15) \Big]\,,\\
    \beta^{(n,\ell)}_{a a} &= (32 n (4 n-7)+16) \ell ^8+64 (4 n (2 n-3)+1) \ell ^7 \nn \\
    &+(64 n (n (8 n (2 n-7)+53)-4)-48) \ell ^6-4 (2 n (8 n (4 n (2 n-7)+61)-451)+87) \ell ^5 \nn \\
    &+(2 n (4 n (4 n (4 n (4 n (4 n-25)+189)-407)-535)+1239)-105) \ell ^4 \nn \\
    &+(2 n (1465-16 n (n (8 n (6 n (4 n-25)+335)-2651)+1058))+63) \ell^3 \nn \\
    &+(4 n (2 n (n (4 n (4 n (36 n-169)+1295)-5171)+2541)-721)-223) \ell ^2 \nn \\
    &+(4 n (n (4 n (16 n (3 n (7-4 n)+31)-1445)+4227)-1168)+491) \ell \nn \\
    &+2 (4 n-5) (2 (n-1) n (2 n (16 n (5 n-7)+13)-7)-9)\,.
\end{align}

\bibliographystyle{JHEP}
\bibliography{biblio}

@article{Fichet:2019owx,
    author = "Fichet, Sylvain",
    title = "{Braneworld effective field theories --- holography, consistency and conformal effects}",
    eprint = "1912.12316",
    archivePrefix = "arXiv",
    primaryClass = "hep-th",
    doi = "10.1007/JHEP04(2020)016",
    journal = "JHEP",
    volume = "04",
    pages = "016",
    year = "2020"
}

@article{Fichet:2016clq,
      author         = "Fichet, Sylvain",
      title          = "{Shining Light on Polarizable Dark Particles}",
      journal        = "JHEP",
      volume         = "04",
      year           = "2017",
      pages          = "088",
      doi            = "10.1007/JHEP04(2017)088",
      eprint         = "1609.01762",
      archivePrefix  = "arXiv",
      primaryClass   = "hep-ph",
      SLACcitation   = "%%CITATION = ARXIV:1609.01762;%%"
}

@article{Feinberg:1968zz,
      author         = "Feinberg, G. and Sucher, J.",
      title          = "{Long-Range Forces from Neutrino-Pair Exchange}",
      journal        = "Phys. Rev.",
      volume         = "166",
      year           = "1968",
      pages          = "1638-1644",
      doi            = "10.1103/PhysRev.166.1638",
      SLACcitation   = "%%CITATION = PHRVA,166,1638;%%"
}

@article{ArkaniHamed:2000ds,
      author         = "Arkani-Hamed, Nima and Porrati, Massimo and Randall,
                        Lisa",
      title          = "{Holography and phenomenology}",
      journal        = "JHEP",
      volume         = "08",
      year           = "2001",
      pages          = "017",
      doi            = "10.1088/1126-6708/2001/08/017",
      eprint         = "hep-th/0012148",
      archivePrefix  = "arXiv",
      primaryClass   = "hep-th",
      reportNumber   = "LBL-46886, UCB-PTH-00-32, NYU-TH-00-09-01",
      SLACcitation   = "%%CITATION = HEP-TH/0012148;%%"
}

@article{Goldberger:2002hb,
      author         = "Goldberger, Walter D. and Rothstein, Ira Z.",
      title          = "{Effective field theory and unification in AdS
                        backgrounds}",
      journal        = "Phys. Rev.",
      volume         = "D68",
      year           = "2003",
      pages          = "125011",
      doi            = "10.1103/PhysRevD.68.125011",
      eprint         = "hep-th/0208060",
      archivePrefix  = "arXiv",
      primaryClass   = "hep-th",
      SLACcitation   = "%%CITATION = HEP-TH/0208060;%%"
}

@article{Randall:2001gb,
      author         = "Randall, Lisa and Schwartz, Matthew D.",
      title          = "{Quantum field theory and unification in AdS5}",
      journal        = "JHEP",
      volume         = "11",
      year           = "2001",
      pages          = "003",
      doi            = "10.1088/1126-6708/2001/11/003",
      eprint         = "hep-th/0108114",
      archivePrefix  = "arXiv",
      primaryClass   = "hep-th",
      reportNumber   = "MIT-CTP-3176",
      SLACcitation   = "%%CITATION = HEP-TH/0108114;%%"
}

@article{Contino:2002kc,
      author         = "Contino, R. and Creminelli, P. and Trincherini, Enrico",
      title          = "{Holographic evolution of gauge couplings}",
      journal        = "JHEP",
      volume         = "10",
      year           = "2002",
      pages          = "029",
      doi            = "10.1088/1126-6708/2002/10/029",
      eprint         = "hep-th/0208002",
      archivePrefix  = "arXiv",
      primaryClass   = "hep-th",
      reportNumber   = "BICOCCA-FT-02-16",
      SLACcitation   = "%%CITATION = HEP-TH/0208002;%%"
}

@article{Fichet:2021xfn,
    author = "Fichet, Sylvain",
    title = "{On holography in general background and the boundary effective action from AdS to dS}",
    eprint = "2112.00746",
    archivePrefix = "arXiv",
    primaryClass = "hep-th",
    doi = "10.1007/JHEP07(2022)113",
    journal = "JHEP",
    volume = "07",
    pages = "113",
    year = "2022"
}

@article{Bellazzini:2019xts,
    author = "Bellazzini, Brando and Lewandowski, Matthew and Serra, Javi",
    title = "{Positivity of Amplitudes, Weak Gravity Conjecture, and Modified Gravity}",
    eprint = "1902.03250",
    archivePrefix = "arXiv",
    primaryClass = "hep-th",
    doi = "10.1103/PhysRevLett.123.251103",
    journal = "Phys. Rev. Lett.",
    volume = "123",
    number = "25",
    pages = "251103",
    year = "2019"
}

@article{Arkani-Hamed:2021ajd,
    author = "Arkani-Hamed, Nima and Huang, Yu-tin and Liu, Jin-Yu and Remmen, Grant N.",
    title = "{Causality, unitarity, and the weak gravity conjecture}",
    eprint = "2109.13937",
    archivePrefix = "arXiv",
    primaryClass = "hep-th",
    doi = "10.1007/JHEP03(2022)083",
    journal = "JHEP",
    volume = "03",
    pages = "083",
    year = "2022"
}

@article{Cheung:2014ega,
    author = "Cheung, Clifford and Remmen, Grant N.",
    title = "{Infrared Consistency and the Weak Gravity Conjecture}",
    eprint = "1407.7865",
    archivePrefix = "arXiv",
    primaryClass = "hep-th",
    reportNumber = "CALT-TH-2014-146",
    doi = "10.1007/JHEP12(2014)087",
    journal = "JHEP",
    volume = "12",
    pages = "087",
    year = "2014"
}

@article{Hamada:2018dde,
    author = "Hamada, Yuta and Noumi, Toshifumi and Shiu, Gary",
    title = "{Weak Gravity Conjecture from Unitarity and Causality}",
    eprint = "1810.03637",
    archivePrefix = "arXiv",
    primaryClass = "hep-th",
    reportNumber = "CCTP-2018-12, ITCP-IPP 2018/9, KOBE-COSMO-18-08, MAD-TH-18-05",
    doi = "10.1103/PhysRevLett.123.051601",
    journal = "Phys. Rev. Lett.",
    volume = "123",
    number = "5",
    pages = "051601",
    year = "2019"
}

@article{Bellazzini:2015cra,
    author = "Bellazzini, Brando and Cheung, Clifford and Remmen, Grant N.",
    title = "{Quantum Gravity Constraints from Unitarity and Analyticity}",
    eprint = "1509.00851",
    archivePrefix = "arXiv",
    primaryClass = "hep-th",
    reportNumber = "CALT-TH-2015-044, SACLAY-T15-161",
    doi = "10.1103/PhysRevD.93.064076",
    journal = "Phys. Rev. D",
    volume = "93",
    number = "6",
    pages = "064076",
    year = "2016"
}

@article{Cheung:2016wjt,
    author = "Cheung, Clifford and Remmen, Grant N.",
    title = "{Positivity of Curvature-Squared Corrections in Gravity}",
    eprint = "1608.02942",
    archivePrefix = "arXiv",
    primaryClass = "hep-th",
    reportNumber = "CALT-TH-2016-018",
    doi = "10.1103/PhysRevLett.118.051601",
    journal = "Phys. Rev. Lett.",
    volume = "118",
    number = "5",
    pages = "051601",
    year = "2017"
}

@article{DeLuca:2022tkm,
    author = "De Luca, Valerio and Khoury, Justin and Wong, Sam S. C.",
    title = "{Implications of the Weak Gravity Conjecture for Tidal Love Numbers of Black Holes}",
    eprint = "2211.14325",
    archivePrefix = "arXiv",
    primaryClass = "hep-th",
    month = "11",
    year = "2022"
}

@article{Hamada:2023cyt,
    author = "Hamada, Yuta and Kuramochi, Rinto and Loges, Gregory J. and Nakajima, Sota",
    title = "{On (Scalar QED) Gravitational Positivity Bounds}",
    eprint = "2301.01999",
    archivePrefix = "arXiv",
    primaryClass = "hep-th",
    reportNumber = "KEK-TH-2492",
    month = "1",
    year = "2023"
}

@article{Arkani-Hamed:2020blm,
    author = "Arkani-Hamed, Nima and Huang, Tzu-Chen and Huang, Yu-tin",
    title = "{The EFT-Hedron}",
    eprint = "2012.15849",
    archivePrefix = "arXiv",
    primaryClass = "hep-th",
    reportNumber = "NCTS-TH/2014, CALT-TH 2020-061",
    doi = "10.1007/JHEP05(2021)259",
    journal = "JHEP",
    volume = "05",
    pages = "259",
    year = "2021"
}

@article{Arkani-Hamed:2006emk,
    author = "Arkani-Hamed, Nima and Motl, Lubos and Nicolis, Alberto and Vafa, Cumrun",
    title = "{The String landscape, black holes and gravity as the weakest force}",
    eprint = "hep-th/0601001",
    archivePrefix = "arXiv",
    reportNumber = "HUTP-05-A0057",
    doi = "10.1088/1126-6708/2007/06/060",
    journal = "JHEP",
    volume = "06",
    pages = "060",
    year = "2007"
}

@article{vanBeest:2021lhn,
    author = "van Beest, Marieke and Calder\'on-Infante, Jos\'e and Mirfendereski, Delaram and Valenzuela, Irene",
    title = "{Lectures on the Swampland Program in String Compactifications}",
    eprint = "2102.01111",
    archivePrefix = "arXiv",
    primaryClass = "hep-th",
    doi = "10.1016/j.physrep.2022.09.002",
    journal = "Phys. Rept.",
    volume = "989",
    pages = "1--50",
    year = "2022"
}

@article{Grana:2021zvf,
    author = "Gra\~na, Mariana and Herr\'aez, Alvaro",
    title = "{The Swampland Conjectures: A Bridge from Quantum Gravity to Particle Physics}",
    eprint = "2107.00087",
    archivePrefix = "arXiv",
    primaryClass = "hep-th",
    doi = "10.3390/universe7080273",
    journal = "Universe",
    volume = "7",
    number = "8",
    pages = "273",
    year = "2021"
}

@article{Agmon:2022thq,
    author = "Agmon, Nathan Benjamin and Bedroya, Alek and Kang, Monica Jinwoo and Vafa, Cumrun",
    title = "{Lectures on the string landscape and the Swampland}",
    eprint = "2212.06187",
    archivePrefix = "arXiv",
    primaryClass = "hep-th",
    month = "12",
    year = "2022"
}

@article{Kats:2006xp,
    author = "Kats, Yevgeny and Motl, Lubos and Padi, Megha",
    title = "{Higher-order corrections to mass-charge relation of extremal black holes}",
    eprint = "hep-th/0606100",
    archivePrefix = "arXiv",
    reportNumber = "HUTP-06-A0023",
    doi = "10.1088/1126-6708/2007/12/068",
    journal = "JHEP",
    volume = "12",
    pages = "068",
    year = "2007"
}

@article{Cheung:2018cwt,
    author = "Cheung, Clifford and Liu, Junyu and Remmen, Grant N.",
    title = "{Proof of the Weak Gravity Conjecture from Black Hole Entropy}",
    eprint = "1801.08546",
    archivePrefix = "arXiv",
    primaryClass = "hep-th",
    reportNumber = "CALT-TH-2018-007",
    doi = "10.1007/JHEP10(2018)004",
    journal = "JHEP",
    volume = "10",
    pages = "004",
    year = "2018"
}

@article{Loges:2019jzs,
    author = "Loges, Gregory J. and Noumi, Toshifumi and Shiu, Gary",
    title = "{Thermodynamics of 4D Dilatonic Black Holes and the Weak Gravity Conjecture}",
    eprint = "1909.01352",
    archivePrefix = "arXiv",
    primaryClass = "hep-th",
    reportNumber = "KOBE-COSMO-19-15, MAD-TH-19-07",
    doi = "10.1103/PhysRevD.102.046010",
    journal = "Phys. Rev. D",
    volume = "102",
    number = "4",
    pages = "046010",
    year = "2020"
}

@article{Loges:2020trf,
    author = "Loges, Gregory J. and Noumi, Toshifumi and Shiu, Gary",
    title = "{Duality and Supersymmetry Constraints on the Weak Gravity Conjecture}",
    eprint = "2006.06696",
    archivePrefix = "arXiv",
    primaryClass = "hep-th",
    reportNumber = "KOBE-COSMO-20-10, MAD-TH-20-03",
    doi = "10.1007/JHEP11(2020)008",
    journal = "JHEP",
    volume = "11",
    pages = "008",
    year = "2020"
}

@article{Goon:2019faz,
    author = "Goon, Garrett and Penco, Riccardo",
    title = "{Universal Relation between Corrections to Entropy and Extremality}",
    eprint = "1909.05254",
    archivePrefix = "arXiv",
    primaryClass = "hep-th",
    doi = "10.1103/PhysRevLett.124.101103",
    journal = "Phys. Rev. Lett.",
    volume = "124",
    number = "10",
    pages = "101103",
    year = "2020"
}

@article{Jones:2019nev,
    author = "Jones, Callum R. T. and McPeak, Brian",
    title = "{The Black Hole Weak Gravity Conjecture with Multiple Charges}",
    eprint = "1908.10452",
    archivePrefix = "arXiv",
    primaryClass = "hep-th",
    reportNumber = "LCTP-20-01, LCTP-19-20",
    doi = "10.1007/JHEP06(2020)140",
    journal = "JHEP",
    volume = "06",
    pages = "140",
    year = "2020"
}

@article{Cao:2022iqh,
    author = "Cao, Qing-Hong and Ueda, Daiki",
    title = "{Entropy Constraint on Effective Field Theory}",
    eprint = "2201.00931",
    archivePrefix = "arXiv",
    primaryClass = "hep-th",
    month = "1",
    year = "2022"
}

@article{Alberte:2020bdz,
    author = "Alberte, Lasma and de Rham, Claudia and Jaitly, Sumer and Tolley, Andrew J.",
    title = "{QED positivity bounds}",
    eprint = "2012.05798",
    archivePrefix = "arXiv",
    primaryClass = "hep-th",
    reportNumber = "Imperial/TP/2020/LA/03",
    doi = "10.1103/PhysRevD.103.125020",
    journal = "Phys. Rev. D",
    volume = "103",
    number = "12",
    pages = "125020",
    year = "2021"
}

@article{Alberte:2020jsk,
    author = "Alberte, Lasma and de Rham, Claudia and Jaitly, Sumer and Tolley, Andrew J.",
    title = "{Positivity Bounds and the Massless Spin-2 Pole}",
    eprint = "2007.12667",
    archivePrefix = "arXiv",
    primaryClass = "hep-th",
    reportNumber = "Imperial/TP/2020/LA/02",
    doi = "10.1103/PhysRevD.102.125023",
    journal = "Phys. Rev. D",
    volume = "102",
    number = "12",
    pages = "125023",
    year = "2020"
}

@article{Caron-Huot:2022ugt,
    author = "Caron-Huot, Simon and Li, Yue-Zhou and Parra-Martinez, Julio and Simmons-Duffin, David",
    title = "{Causality constraints on corrections to Einstein gravity}",
    eprint = "2201.06602",
    archivePrefix = "arXiv",
    primaryClass = "hep-th",
    reportNumber = "CALT-TH 2021-003",
    doi = "10.1007/JHEP05(2023)122",
    journal = "JHEP",
    volume = "05",
    pages = "122",
    year = "2023"
}

@article{Caron-Huot:2022jli,
    author = "Caron-Huot, Simon and Li, Yue-Zhou and Parra-Martinez, Julio and Simmons-Duffin, David",
    title = "{Graviton partial waves and causality in higher dimensions}",
    eprint = "2205.01495",
    archivePrefix = "arXiv",
    primaryClass = "hep-th",
    reportNumber = "CALT-TH 2022-17",
    doi = "10.1103/PhysRevD.108.026007",
    journal = "Phys. Rev. D",
    volume = "108",
    number = "2",
    pages = "026007",
    year = "2023"
}

@article{Bellazzini:2023nqj,
    author = "Bellazzini, Brando and Isabella, Giulia and Ricossa, Sergio and Riva, Francesco",
    title = "{Massive gravity is not positive}",
    eprint = "2304.02550",
    archivePrefix = "arXiv",
    primaryClass = "hep-th",
    doi = "10.1103/PhysRevD.109.024051",
    journal = "Phys. Rev. D",
    volume = "109",
    number = "2",
    pages = "024051",
    year = "2024"
}

@inbook{Dunne:2004nc,
    author = "Dunne, Gerald V.",
    editor = "Shifman, M. and Vainshtein, A. and Wheater, J.",
    title = "{Heisenberg-Euler effective Lagrangians: Basics and extensions}",
    booktitle = "{From fields to strings: Circumnavigating theoretical physics. Ian Kogan memorial collection (3 volume set)}",
    eprint = "hep-th/0406216",
    archivePrefix = "arXiv",
    doi = "10.1142/9789812775344_0014",
    pages = "445--522",
    month = "6",
    year = "2004"
}

@article{Chen:2019qvr,
    author = "Chen, Wei-Ming and Huang, Yu-Tin and Noumi, Toshifumi and Wen, Congkao",
    title = "{Unitarity bounds on charged/neutral state mass ratios}",
    eprint = "1901.11480",
    archivePrefix = "arXiv",
    primaryClass = "hep-th",
    reportNumber = "KOBE-COSMO-19-03, MAD-TH-19-01, QMUL-PH-19-03, NCTS-TH/1901",
    doi = "10.1103/PhysRevD.100.025016",
    journal = "Phys. Rev. D",
    volume = "100",
    number = "2",
    pages = "025016",
    year = "2019"
}

@article{Eichhorn:2024wba,
    author = "Eichhorn, Astrid and Pedersen, Andreas Odgaard and Schiffer, Marc",
    title = "{Application of positivity bounds in asymptotically safe gravity}",
    eprint = "2405.08862",
    archivePrefix = "arXiv",
    primaryClass = "hep-th",
    month = "5",
    year = "2024"
}

@article{Bittar:2024xuc,
    author = "Bittar, Pedro and Fichet, Sylvain and de Souza, Lucas",
    title = "{Gravity-Induced Photon Interactions and Infrared Consistency in any Dimensions}",
    eprint = "2404.07254",
    archivePrefix = "arXiv",
    primaryClass = "hep-th",
    month = "4",
    year = "2024"
}

@article{Hui:2020xxx,
    author = "Hui, Lam and Joyce, Austin and Penco, Riccardo and Santoni, Luca and Solomon, Adam R.",
    title = "{Static response and Love numbers of Schwarzschild black holes}",
    eprint = "2010.00593",
    archivePrefix = "arXiv",
    primaryClass = "hep-th",
    doi = "10.1088/1475-7516/2021/04/052",
    journal = "JCAP",
    volume = "04",
    pages = "052",
    year = "2021"
}

@article{Bern:2021ppb,
    author = "Bern, Zvi and Kosmopoulos, Dimitrios and Zhiboedov, Alexander",
    title = "{Gravitational effective field theory islands, low-spin dominance, and the four-graviton amplitude}",
    eprint = "2103.12728",
    archivePrefix = "arXiv",
    primaryClass = "hep-th",
    reportNumber = "CERN-TH-2021-035",
    doi = "10.1088/1751-8121/ac0e51",
    journal = "J. Phys. A",
    volume = "54",
    number = "34",
    pages = "344002",
    year = "2021"
}

@article{Jack:1988sw,
    author = "Jack, I. and Jones, D. R. T. and Mohammedi, N.",
    title = "{The Four Loop Metric Beta Function for the Bosonic $\sigma$ Model}",
    reportNumber = "LTH-225, SHEP-88-89-3",
    doi = "10.1016/0370-2693(89)90031-2",
    journal = "Phys. Lett. B",
    volume = "220",
    pages = "171--175",
    year = "1989"
}

@article{Gross:1986iv,
    author = "Gross, David J. and Witten, Edward",
    title = "{Superstring Modifications of Einstein's Equations}",
    reportNumber = "Print-86-0250 (PRINCETON)",
    doi = "10.1016/0550-3213(86)90429-3",
    journal = "Nucl. Phys. B",
    volume = "277",
    pages = "1",
    year = "1986"
}

@article{Gross:1986mw,
    author = "Gross, David J. and Sloan, John H.",
    title = "{The Quartic Effective Action for the Heterotic String}",
    reportNumber = "NSF-ITP-87-02",
    doi = "10.1016/0550-3213(87)90465-2",
    journal = "Nucl. Phys. B",
    volume = "291",
    pages = "41--89",
    year = "1987"
}

@article{Kikuchi:1986rk,
    author = "Kikuchi, Y. and Marzban, C. and Ng, Y. J.",
    title = "{Heterotic String Modifications of Einstein's and {Yang-Mills}' Actions}",
    reportNumber = "IFP-272-UNC",
    doi = "10.1016/0370-2693(86)90924-X",
    journal = "Phys. Lett. B",
    volume = "176",
    pages = "57--60",
    year = "1986"
}

@book{Becker:2006dvp,
    author = "Becker, K. and Becker, M. and Schwarz, J. H.",
    title = "{String theory and M-theory: A modern introduction}",
    doi = "10.1017/CBO9780511816086",
    isbn = "978-0-511-25486-4, 978-0-521-86069-7, 978-0-511-81608-6",
    publisher = "Cambridge University Press",
    month = "12",
    year = "2006"
}

@article{Chia:2023tle,
    author = "Chia, Horng Sheng and Edwards, Thomas D. P. and Wadekar, Digvijay and Zimmerman, Aaron and Olsen, Seth and Roulet, Javier and Venumadhav, Tejaswi and Zackay, Barak and Zaldarriaga, Matias",
    title = "{In pursuit of Love numbers: First templated search for compact objects with large tidal deformabilities in the LIGO-Virgo data}",
    eprint = "2306.00050",
    archivePrefix = "arXiv",
    primaryClass = "gr-qc",
    doi = "10.1103/PhysRevD.110.063007",
    journal = "Phys. Rev. D",
    volume = "110",
    number = "6",
    pages = "063007",
    year = "2024"
}

@article{Melville:2024zjq,
    author = "Melville, Scott",
    title = "{Causality and quasi-normal modes in the GREFT}",
    eprint = "2401.05524",
    archivePrefix = "arXiv",
    primaryClass = "gr-qc",
    doi = "10.1140/epjp/s13360-024-05520-5",
    journal = "Eur. Phys. J. Plus",
    volume = "139",
    number = "8",
    pages = "725",
    year = "2024"
}

@article{Cano:2023jbk,
    author = "Cano, Pablo A. and Fransen, Kwinten and Hertog, Thomas and Maenaut, Simon",
    title = "{Quasinormal modes of rotating black holes in higher-derivative gravity}",
    eprint = "2307.07431",
    archivePrefix = "arXiv",
    primaryClass = "gr-qc",
    doi = "10.1103/PhysRevD.108.124032",
    journal = "Phys. Rev. D",
    volume = "108",
    number = "12",
    pages = "124032",
    year = "2023"
}

@article{Cano:2024ezp,
    author = {Cano, Pablo A. and Capuano, Lodovico and Franchini, Nicola and Maenaut, Simon and V\"olkel, Sebastian H.},
    title = "{Higher-derivative corrections to the Kerr quasinormal mode spectrum}",
    eprint = "2409.04517",
    archivePrefix = "arXiv",
    primaryClass = "gr-qc",
    doi = "10.1103/PhysRevD.110.124057",
    journal = "Phys. Rev. D",
    volume = "110",
    number = "12",
    pages = "124057",
    year = "2024"
}

@article{Cano:2024wzo,
    author = "Cano, Pablo A. and David, Marina",
    title = "{Isospectrality in Effective Field Theory Extensions of General Relativity}",
    eprint = "2407.12080",
    archivePrefix = "arXiv",
    primaryClass = "hep-th",
    month = "7",
    year = "2024"
}

@article{Miguel:2023rzp,
    author = "Miguel, Filipe S.",
    title = "{EFT corrections to scalar and vector quasinormal modes of rapidly rotating black holes}",
    eprint = "2308.03832",
    archivePrefix = "arXiv",
    primaryClass = "gr-qc",
    doi = "10.1103/PhysRevD.109.104016",
    journal = "Phys. Rev. D",
    volume = "109",
    number = "10",
    pages = "104016",
    year = "2024"
}

@inproceedings{Donoghue:1995cz,
    author = "Donoghue, John F.",
    title = "{Introduction to the effective field theory description of gravity}",
    booktitle = "{Advanced School on Effective Theories}",
    eprint = "gr-qc/9512024",
    archivePrefix = "arXiv",
    reportNumber = "UMHEP-424",
    month = "6",
    year = "1995"
}

@article{Burgess:2003jk,
    author = "Burgess, C. P.",
    title = "{Quantum gravity in everyday life: General relativity as an effective field theory}",
    eprint = "gr-qc/0311082",
    archivePrefix = "arXiv",
    doi = "10.12942/lrr-2004-5",
    journal = "Living Rev. Rel.",
    volume = "7",
    pages = "5--56",
    year = "2004"
}

@article{Goon:2016mil,
    author = "Goon, Garrett",
    title = "{Heavy Fields and Gravity}",
    eprint = "1611.02705",
    archivePrefix = "arXiv",
    primaryClass = "hep-th",
    doi = "10.1007/JHEP01(2017)045",
    journal = "JHEP",
    volume = "01",
    pages = "045",
    year = "2017",
    note = "[Erratum: JHEP 03, 161 (2017)]"
}

@article{Cardoso:2018ptl,
    author = "Cardoso, Vitor and Kimura, Masashi and Maselli, Andrea and Senatore, Leonardo",
    title = "{Black Holes in an Effective Field Theory Extension of General Relativity}",
    eprint = "1808.08962",
    archivePrefix = "arXiv",
    primaryClass = "gr-qc",
    doi = "10.1103/PhysRevLett.121.251105",
    journal = "Phys. Rev. Lett.",
    volume = "121",
    number = "25",
    pages = "251105",
    year = "2018",
    note = "[Erratum: Phys.Rev.Lett. 131, 109903 (2023)]"
}

@article{Charalambous:2021kcz,
    author = "Charalambous, Panagiotis and Dubovsky, Sergei and Ivanov, Mikhail M.",
    title = "{Hidden Symmetry of Vanishing Love Numbers}",
    eprint = "2103.01234",
    archivePrefix = "arXiv",
    primaryClass = "hep-th",
    reportNumber = "INR-TH-2021-003",
    doi = "10.1103/PhysRevLett.127.101101",
    journal = "Phys. Rev. Lett.",
    volume = "127",
    number = "10",
    pages = "101101",
    year = "2021"
}

@article{Hui:2021vcv,
    author = "Hui, Lam and Joyce, Austin and Penco, Riccardo and Santoni, Luca and Solomon, Adam R.",
    title = "{Ladder symmetries of black holes. Implications for love numbers and no-hair theorems}",
    eprint = "2105.01069",
    archivePrefix = "arXiv",
    primaryClass = "hep-th",
    doi = "10.1088/1475-7516/2022/01/032",
    journal = "JCAP",
    volume = "01",
    number = "01",
    pages = "032",
    year = "2022"
}

@article{Hui:2022vbh,
    author = "Hui, Lam and Joyce, Austin and Penco, Riccardo and Santoni, Luca and Solomon, Adam R.",
    title = "{Near-zone symmetries of Kerr black holes}",
    eprint = "2203.08832",
    archivePrefix = "arXiv",
    primaryClass = "hep-th",
    doi = "10.1007/JHEP09(2022)049",
    journal = "JHEP",
    volume = "09",
    pages = "049",
    year = "2022"
}

@article{Charalambous:2022rre,
    author = "Charalambous, Panagiotis and Dubovsky, Sergei and Ivanov, Mikhail M.",
    title = "{Love symmetry}",
    eprint = "2209.02091",
    archivePrefix = "arXiv",
    primaryClass = "hep-th",
    doi = "10.1007/JHEP10(2022)175",
    journal = "JHEP",
    volume = "10",
    pages = "175",
    year = "2022"
}

@article{Combaluzier-Szteinsznaider:2024sgb,
    author = "Combaluzier-Szteinsznaider, Oscar and Hui, Lam and Santoni, Luca and Solomon, Adam R. and Wong, Sam S. C.",
    title = "{Symmetries of Vanishing Nonlinear Love Numbers of Schwarzschild Black Holes}",
    eprint = "2410.10952",
    archivePrefix = "arXiv",
    primaryClass = "gr-qc",
    month = "10",
    year = "2024"
}

@article{Kehagias:2024rtz,
    author = "Kehagias, Alex and Riotto, Antonio",
    title = "{Black Holes in a Gravitational Field: The Non-linear Static Love Number of Schwarzschild Black Holes Vanishes}",
    eprint = "2410.11014",
    archivePrefix = "arXiv",
    primaryClass = "gr-qc",
    month = "10",
    year = "2024"
}

@article{Gounis:2024hcm,
    author = "Gounis, L. -R. and Kehagias, A. and Riotto, A.",
    title = "{The Vanishing of the Non-linear Static Love Number of Kerr Black Holes and the Role of Symmetries}",
    eprint = "2412.08249",
    archivePrefix = "arXiv",
    primaryClass = "gr-qc",
    month = "12",
    year = "2024"
}

@ARTICLE{1972ApJ...175..243P,
       author = {{Press}, William H.},
        title = "{Time Evolution of a Rotating Black Hole Immersed in a Static Scalar Field}",
      journal = {The Astrophysical Journal},
         year = 1972,
        month = jul,
       volume = {175},
        pages = {243},
          doi = {10.1086/151551},
       adsurl = {https://ui.adsabs.harvard.edu/abs/1972ApJ...175..243P}
}

@article{Martel:2005ir,
    author = "Martel, Karl and Poisson, Eric",
    title = "{Gravitational perturbations of the Schwarzschild spacetime: A Practical covariant and gauge-invariant formalism}",
    eprint = "gr-qc/0502028",
    archivePrefix = "arXiv",
    doi = "10.1103/PhysRevD.71.104003",
    journal = "Phys. Rev. D",
    volume = "71",
    pages = "104003",
    year = "2005"
}

@article{Fang:2005qq,
    author = "Fang, Hua and Lovelace, Geoffrey",
    title = "{Tidal coupling of a Schwarzschild black hole and circularly orbiting moon}",
    eprint = "gr-qc/0505156",
    archivePrefix = "arXiv",
    doi = "10.1103/PhysRevD.72.124016",
    journal = "Phys. Rev. D",
    volume = "72",
    pages = "124016",
    year = "2005"
}

@article{Damour:2009va,
    author = "Damour, Thibault and Lecian, Orchidea Maria",
    title = "{On the gravitational polarizability of black holes}",
    eprint = "0906.3003",
    archivePrefix = "arXiv",
    primaryClass = "gr-qc",
    reportNumber = "IHES-P-09-28",
    doi = "10.1103/PhysRevD.80.044017",
    journal = "Phys. Rev. D",
    volume = "80",
    pages = "044017",
    year = "2009"
}

@article{Binnington:2009bb,
    author = "Binnington, Taylor and Poisson, Eric",
    title = "{Relativistic theory of tidal Love numbers}",
    eprint = "0906.1366",
    archivePrefix = "arXiv",
    primaryClass = "gr-qc",
    doi = "10.1103/PhysRevD.80.084018",
    journal = "Phys. Rev. D",
    volume = "80",
    pages = "084018",
    year = "2009"
}

@article{Kol:2011vg,
    author = "Kol, Barak and Smolkin, Michael",
    title = "{Black hole stereotyping: Induced gravito-static polarization}",
    eprint = "1110.3764",
    archivePrefix = "arXiv",
    primaryClass = "hep-th",
    doi = "10.1007/JHEP02(2012)010",
    journal = "JHEP",
    volume = "02",
    pages = "010",
    year = "2012"
}

@article{Landry:2015cva,
    author = "Landry, Philippe and Poisson, Eric",
    title = "{Gravitomagnetic response of an irrotational body to an applied tidal field}",
    eprint = "1504.06606",
    archivePrefix = "arXiv",
    primaryClass = "gr-qc",
    doi = "10.1103/PhysRevD.91.104026",
    journal = "Phys. Rev. D",
    volume = "91",
    number = "10",
    pages = "104026",
    year = "2015"
}

@article{Landry:2015zfa,
    author = "Landry, Philippe and Poisson, Eric",
    title = "{Tidal deformation of a slowly rotating material body. External metric}",
    eprint = "1503.07366",
    archivePrefix = "arXiv",
    primaryClass = "gr-qc",
    doi = "10.1103/PhysRevD.91.104018",
    journal = "Phys. Rev. D",
    volume = "91",
    pages = "104018",
    year = "2015"
}

@article{Porto:2016pyg,
    author = "Porto, Rafael A.",
    title = "{The effective field theorist\textquoteright{}s approach to gravitational dynamics}",
    eprint = "1601.04914",
    archivePrefix = "arXiv",
    primaryClass = "hep-th",
    doi = "10.1016/j.physrep.2016.04.003",
    journal = "Phys. Rept.",
    volume = "633",
    pages = "1--104",
    year = "2016"
}

@article{Poisson:2020mdi,
    author = "Poisson, Eric",
    title = "{Gravitomagnetic Love tensor of a slowly rotating body: post-Newtonian theory}",
    eprint = "2007.01678",
    archivePrefix = "arXiv",
    primaryClass = "gr-qc",
    doi = "10.1103/PhysRevD.102.064059",
    journal = "Phys. Rev. D",
    volume = "102",
    number = "6",
    pages = "064059",
    year = "2020"
}

@article{LeTiec:2020spy,
    author = "Le Tiec, Alexandre and Casals, Marc",
    title = "{Spinning Black Holes Fall in Love}",
    eprint = "2007.00214",
    archivePrefix = "arXiv",
    primaryClass = "gr-qc",
    doi = "10.1103/PhysRevLett.126.131102",
    journal = "Phys. Rev. Lett.",
    volume = "126",
    number = "13",
    pages = "131102",
    year = "2021"
}

@article{LeTiec:2020bos,
    author = "Le Tiec, Alexandre and Casals, Marc and Franzin, Edgardo",
    title = "{Tidal Love Numbers of Kerr Black Holes}",
    eprint = "2010.15795",
    archivePrefix = "arXiv",
    primaryClass = "gr-qc",
    doi = "10.1103/PhysRevD.103.084021",
    journal = "Phys. Rev. D",
    volume = "103",
    number = "8",
    pages = "084021",
    year = "2021"
}

@article{Chia:2020yla,
    author = "Chia, Horng Sheng",
    title = "{Tidal deformation and dissipation of rotating black holes}",
    eprint = "2010.07300",
    archivePrefix = "arXiv",
    primaryClass = "gr-qc",
    doi = "10.1103/PhysRevD.104.024013",
    journal = "Phys. Rev. D",
    volume = "104",
    number = "2",
    pages = "024013",
    year = "2021"
}

@article{Goldberger:2020fot,
    author = "Goldberger, Walter D. and Li, Jingping and Rothstein, Ira Z.",
    title = "{Non-conservative effects on spinning black holes from world-line effective field theory}",
    eprint = "2012.14869",
    archivePrefix = "arXiv",
    primaryClass = "hep-th",
    doi = "10.1007/JHEP06(2021)053",
    journal = "JHEP",
    volume = "06",
    pages = "053",
    year = "2021"
}

@article{Charalambous:2021mea,
    author = "Charalambous, Panagiotis and Dubovsky, Sergei and Ivanov, Mikhail M.",
    title = "{On the Vanishing of Love Numbers for Kerr Black Holes}",
    eprint = "2102.08917",
    archivePrefix = "arXiv",
    primaryClass = "hep-th",
    reportNumber = "INR-TH-2021-001",
    doi = "10.1007/JHEP05(2021)038",
    journal = "JHEP",
    volume = "05",
    pages = "038",
    year = "2021"
}

@article{Ivanov:2022qqt,
    author = "Ivanov, Mikhail M. and Zhou, Zihan",
    title = "{Vanishing of Black Hole Tidal Love Numbers from Scattering Amplitudes}",
    eprint = "2209.14324",
    archivePrefix = "arXiv",
    primaryClass = "hep-th",
    doi = "10.1103/PhysRevLett.130.091403",
    journal = "Phys. Rev. Lett.",
    volume = "130",
    number = "9",
    pages = "091403",
    year = "2023"
}

@article{Ivanov:2022hlo,
    author = "Ivanov, Mikhail M. and Zhou, Zihan",
    title = "{Revisiting the matching of black hole tidal responses: A systematic study of relativistic and logarithmic corrections}",
    eprint = "2208.08459",
    archivePrefix = "arXiv",
    primaryClass = "hep-th",
    doi = "10.1103/PhysRevD.107.084030",
    journal = "Phys. Rev. D",
    volume = "107",
    number = "8",
    pages = "084030",
    year = "2023"
}

@article{Pereniguez:2021xcj,
    author = "Pere\~niguez, David and Cardoso, Vitor",
    title = "{Love numbers and magnetic susceptibility of charged black holes}",
    eprint = "2112.08400",
    archivePrefix = "arXiv",
    primaryClass = "gr-qc",
    doi = "10.1103/PhysRevD.105.044026",
    journal = "Phys. Rev. D",
    volume = "105",
    number = "4",
    pages = "044026",
    year = "2022"
}

@article{Rai:2024lho,
    author = "Rai, Mudit and Santoni, Luca",
    title = {{Ladder symmetries and Love numbers of Reissner-Nordstr\"om black holes}},
    eprint = "2404.06544",
    archivePrefix = "arXiv",
    primaryClass = "gr-qc",
    doi = "10.1007/JHEP07(2024)098",
    journal = "JHEP",
    volume = "07",
    pages = "098",
    year = "2024"
}

@article{Landry:2014jka,
    author = "Landry, Philippe and Poisson, Eric",
    title = "{Relativistic theory of surficial Love numbers}",
    eprint = "1404.6798",
    archivePrefix = "arXiv",
    primaryClass = "gr-qc",
    doi = "10.1103/PhysRevD.89.124011",
    journal = "Phys. Rev. D",
    volume = "89",
    number = "12",
    pages = "124011",
    year = "2014"
}

@article{Poisson:2020vap,
    author = "Poisson, Eric",
    title = "{Compact body in a tidal environment: New types of relativistic Love numbers, and a post-Newtonian operational definition for tidally induced multipole moments}",
    eprint = "2012.10184",
    archivePrefix = "arXiv",
    primaryClass = "gr-qc",
    doi = "10.1103/PhysRevD.103.064023",
    journal = "Phys. Rev. D",
    volume = "103",
    number = "6",
    pages = "064023",
    year = "2021"
}

@article{Poisson:2021yau,
    author = "Poisson, Eric",
    title = "{Tidally induced multipole moments of a nonrotating black hole vanish to all post-Newtonian orders}",
    eprint = "2108.07328",
    archivePrefix = "arXiv",
    primaryClass = "gr-qc",
    doi = "10.1103/PhysRevD.104.104062",
    journal = "Phys. Rev. D",
    volume = "104",
    number = "10",
    pages = "104062",
    year = "2021"
}

@article{DeLuca:2023mio,
    author = "De Luca, Valerio and Khoury, Justin and Wong, Sam S. C.",
    title = "{Nonlinearities in the tidal Love numbers of black holes}",
    eprint = "2305.14444",
    archivePrefix = "arXiv",
    primaryClass = "gr-qc",
    doi = "10.1103/PhysRevD.108.024048",
    journal = "Phys. Rev. D",
    volume = "108",
    number = "2",
    pages = "024048",
    year = "2023"
}

@article{Riva:2023rcm,
    author = "Riva, Massimiliano Maria and Santoni, Luca and Savi\'c, Nikola and Vernizzi, Filippo",
    title = "{Vanishing of nonlinear tidal Love numbers of Schwarzschild black holes}",
    eprint = "2312.05065",
    archivePrefix = "arXiv",
    primaryClass = "gr-qc",
    doi = "10.1016/j.physletb.2024.138710",
    journal = "Phys. Lett. B",
    volume = "854",
    pages = "138710",
    year = "2024"
}

@article{Hadad:2024lsf,
    author = "Hadad, Tomer and Kol, Barak and Smolkin, Michael",
    title = "{Gravito-magnetic polarization of Schwarzschild black hole}",
    eprint = "2402.16172",
    archivePrefix = "arXiv",
    primaryClass = "hep-th",
    doi = "10.1007/JHEP06(2024)169",
    journal = "JHEP",
    volume = "06",
    pages = "169",
    year = "2024"
}

@article{Iteanu:2024dvx,
    author = "Iteanu, Simon and Riva, Massimiliano Maria and Santoni, Luca and Savi\'c, Nikola and Vernizzi, Filippo",
    title = "{Vanishing of Quadratic Love Numbers of Schwarzschild Black Holes}",
    eprint = "2410.03542",
    archivePrefix = "arXiv",
    primaryClass = "gr-qc",
    reportNumber = "DESY 24-141",
    month = "10",
    year = "2024"
}

@article{Katagiri:2024fpn,
    author = "Katagiri, Takuya and Cardoso, Vitor and Ikeda, Tact and Yagi, Kent",
    title = "{Tidal response beyond vacuum General Relativity with a canonical definition}",
    eprint = "2410.02531",
    archivePrefix = "arXiv",
    primaryClass = "gr-qc",
    reportNumber = "RUP-24-19",
    month = "10",
    year = "2024"
}

@article{Endlich:2017tqa,
    author = "Endlich, Solomon and Gorbenko, Victor and Huang, Junwu and Senatore, Leonardo",
    title = "{An effective formalism for testing extensions to General Relativity with gravitational waves}",
    eprint = "1704.01590",
    archivePrefix = "arXiv",
    primaryClass = "gr-qc",
    doi = "10.1007/JHEP09(2017)122",
    journal = "JHEP",
    volume = "09",
    pages = "122",
    year = "2017"
}

@article{Brandhuber:2019qpg,
    author = "Brandhuber, Andreas and Travaglini, Gabriele",
    title = "{On higher-derivative effects on the gravitational potential and particle bending}",
    eprint = "1905.05657",
    archivePrefix = "arXiv",
    primaryClass = "hep-th",
    reportNumber = "QMUL-PH-19-09, SAGEX-19-05",
    doi = "10.1007/JHEP01(2020)010",
    journal = "JHEP",
    volume = "01",
    pages = "010",
    year = "2020"
}

@article{AccettulliHuber:2020dal,
    author = "Accettulli Huber, Manuel and Brandhuber, Andreas and De Angelis, Stefano and Travaglini, Gabriele",
    title = "{From amplitudes to gravitational radiation with cubic interactions and tidal effects}",
    eprint = "2012.06548",
    archivePrefix = "arXiv",
    primaryClass = "hep-th",
    reportNumber = "QMUL-PH-20-19, SAGEX-20-19-E",
    doi = "10.1103/PhysRevD.103.045015",
    journal = "Phys. Rev. D",
    volume = "103",
    number = "4",
    pages = "045015",
    year = "2021"
}

@article{Cayuso:2023xbc,
    author = "Cayuso, Ramiro and Figueras, Pau and Fran\c{c}a, Tiago and Lehner, Luis",
    title = "{Self-Consistent Modeling of Gravitational Theories beyond General Relativity}",
    doi = "10.1103/PhysRevLett.131.111403",
    journal = "Phys. Rev. Lett.",
    volume = "131",
    number = "11",
    pages = "111403",
    year = "2023"
}

@article{Cardoso:2019mqo,
    author = "Cardoso, Vitor and Kimura, Masashi and Maselli, Andrea and Berti, Emanuele and Macedo, Caio F. B. and McManus, Ryan",
    title = "{Parametrized black hole quasinormal ringdown: Decoupled equations for nonrotating black holes}",
    eprint = "1901.01265",
    archivePrefix = "arXiv",
    primaryClass = "gr-qc",
    doi = "10.1103/PhysRevD.99.104077",
    journal = "Phys. Rev. D",
    volume = "99",
    number = "10",
    pages = "104077",
    year = "2019"
}

@article{McManus:2019ulj,
    author = "McManus, Ryan and Berti, Emanuele and Macedo, Caio F. B. and Kimura, Masashi and Maselli, Andrea and Cardoso, Vitor",
    title = "{Parametrized black hole quasinormal ringdown. II. Coupled equations and quadratic corrections for nonrotating black holes}",
    eprint = "1906.05155",
    archivePrefix = "arXiv",
    primaryClass = "gr-qc",
    doi = "10.1103/PhysRevD.100.044061",
    journal = "Phys. Rev. D",
    volume = "100",
    number = "4",
    pages = "044061",
    year = "2019"
}

@article{deRham:2020ejn,
    author = "de Rham, Claudia and Francfort, J\'er\'emie and Zhang, Jun",
    title = "{Black Hole Gravitational Waves in the Effective Field Theory of Gravity}",
    eprint = "2005.13923",
    archivePrefix = "arXiv",
    primaryClass = "hep-th",
    reportNumber = "Imperial/TP/2020/CdR/02",
    doi = "10.1103/PhysRevD.102.024079",
    journal = "Phys. Rev. D",
    volume = "102",
    number = "2",
    pages = "024079",
    year = "2020"
}

@article{Cano:2020cao,
    author = "Cano, Pablo A. and Fransen, Kwinten and Hertog, Thomas",
    title = "{Ringing of rotating black holes in higher-derivative gravity}",
    eprint = "2005.03671",
    archivePrefix = "arXiv",
    primaryClass = "gr-qc",
    doi = "10.1103/PhysRevD.102.044047",
    journal = "Phys. Rev. D",
    volume = "102",
    number = "4",
    pages = "044047",
    year = "2020"
}

@article{Maenaut:2024oci,
    author = "Maenaut, Simon and Carullo, Gregorio and Cano, Pablo A. and Liu, Anna and Cardoso, Vitor and Hertog, Thomas and Li, Tjonnie G. F.",
    title = "{Ringdown Analysis of Rotating Black Holes in Effective Field Theory Extensions of General Relativity}",
    eprint = "2411.17893",
    archivePrefix = "arXiv",
    primaryClass = "gr-qc",
    reportNumber = "LIGO-P2400551",
    month = "11",
    year = "2024"
}

@article{Sennett:2019bpc,
    author = "Sennett, Noah and Brito, Richard and Buonanno, Alessandra and Gorbenko, Victor and Senatore, Leonardo",
    title = "{Gravitational-Wave Constraints on an Effective Field-Theory Extension of General Relativity}",
    eprint = "1912.09917",
    archivePrefix = "arXiv",
    primaryClass = "gr-qc",
    doi = "10.1103/PhysRevD.102.044056",
    journal = "Phys. Rev. D",
    volume = "102",
    number = "4",
    pages = "044056",
    year = "2020"
}

@article{Silva:2022srr,
    author = "Silva, Hector O. and Ghosh, Abhirup and Buonanno, Alessandra",
    title = "{Black-hole ringdown as a probe of higher-curvature gravity theories}",
    eprint = "2205.05132",
    archivePrefix = "arXiv",
    primaryClass = "gr-qc",
    month = "5",
    year = "2022"
}

@article{deRham:2021bll,
    author = "de Rham, Claudia and Tolley, Andrew J. and Zhang, Jun",
    title = "{Causality Constraints on Gravitational Effective Field Theories}",
    eprint = "2112.05054",
    archivePrefix = "arXiv",
    primaryClass = "gr-qc",
    reportNumber = "Imperial/TP/2021/CdR/4",
    doi = "10.1103/PhysRevLett.128.131102",
    journal = "Phys. Rev. Lett.",
    volume = "128",
    number = "13",
    pages = "131102",
    year = "2022"
}

@article{Knorr:2024yiu,
    author = "Knorr, Benjamin and Platania, Alessia",
    title = "{Unearthing the intersections: positivity bounds, weak gravity conjecture, and asymptotic safety landscapes from photon-graviton flows}",
    eprint = "2405.08860",
    archivePrefix = "arXiv",
    primaryClass = "hep-th",
    reportNumber = "NORDITA 2024-014",
    month = "5",
    year = "2024"
}

@article{Camanho:2014apa,
    author = "Camanho, Xian O. and Edelstein, Jose D. and Maldacena, Juan and Zhiboedov, Alexander",
    title = "{Causality Constraints on Corrections to the Graviton Three-Point Coupling}",
    eprint = "1407.5597",
    archivePrefix = "arXiv",
    primaryClass = "hep-th",
    doi = "10.1007/JHEP02(2016)020",
    journal = "JHEP",
    volume = "02",
    pages = "020",
    year = "2016"
}

@article{Goon:2016une,
    author = "Goon, Garrett and Hinterbichler, Kurt",
    title = "{Superluminality, black holes and EFT}",
    eprint = "1609.00723",
    archivePrefix = "arXiv",
    primaryClass = "hep-th",
    doi = "10.1007/JHEP02(2017)134",
    journal = "JHEP",
    volume = "02",
    pages = "134",
    year = "2017"
}

@article{Brown:2024ajk,
    author = "Brown, Adam R. and Iliesiu, Luca V. and Penington, Geoff and Usatyuk, Mykhaylo",
    title = "{The evaporation of charged black holes}",
    eprint = "2411.03447",
    archivePrefix = "arXiv",
    primaryClass = "hep-th",
    month = "11",
    year = "2024"
}

@article{Fichet:2023dju,
    author = "Fichet, Sylvain and Megias, Eugenio and Quiros, Mariano",
    title = "{Holographic fluids from 5D dilaton gravity}",
    eprint = "2311.14233",
    archivePrefix = "arXiv",
    primaryClass = "hep-th",
    doi = "10.1007/JHEP08(2024)077",
    journal = "JHEP",
    volume = "08",
    pages = "077",
    year = "2024"
}

@article{Goldberger:2005cd,
    author = "Goldberger, Walter D. and Rothstein, Ira Z.",
    title = "{Dissipative effects in the worldline approach to black hole dynamics}",
    eprint = "hep-th/0511133",
    archivePrefix = "arXiv",
    doi = "10.1103/PhysRevD.73.104030",
    journal = "Phys. Rev. D",
    volume = "73",
    pages = "104030",
    year = "2006"
}

@article{Goldberger:2004jt,
    author = "Goldberger, Walter D. and Rothstein, Ira Z.",
    title = "{An Effective field theory of gravity for extended objects}",
    eprint = "hep-th/0409156",
    archivePrefix = "arXiv",
    reportNumber = "UCSD-PTH-04-17, CMU-HEP-04-06",
    doi = "10.1103/PhysRevD.73.104029",
    journal = "Phys. Rev. D",
    volume = "73",
    pages = "104029",
    year = "2006"
}

@article{Burgess:2017mhz,
    author = "Burgess, C. P. and Hayman, P. and Rummel, Markus and Zalavari, Laszlo",
    title = "{Reduced theoretical error for $^4He^+$ spectroscopy}",
    eprint = "1708.09768",
    archivePrefix = "arXiv",
    primaryClass = "hep-ph",
    doi = "10.1103/PhysRevA.98.052510",
    journal = "Phys. Rev. A",
    volume = "98",
    number = "5",
    pages = "052510",
    year = "2018"
}

@article{Endlich:2016jgc,
    author = "Endlich, Solomon and Penco, Riccardo",
    title = "{A Modern Approach to Superradiance}",
    eprint = "1609.06723",
    archivePrefix = "arXiv",
    primaryClass = "hep-th",
    doi = "10.1007/JHEP05(2017)052",
    journal = "JHEP",
    volume = "05",
    pages = "052",
    year = "2017"
}

@article{jess_BN,
    author = "Santiago, Jessica and Feng, Justin and Schuster, Sebastian and Visser, Matt",
    title = "{Immortality through the dark forces: Dark-charge primordial black holes as dark matter candidates}",
    eprint = "2503.20696",
    archivePrefix = "arXiv",
    primaryClass = "gr-qc",
    month = "3",
    year = "2025"
}

@article{Hiscock_Weems,
  title = {Evolution of charged evaporating black holes},
  author = {Hiscock, William A. and Weems, Lance D.},
  journal = {Phys. Rev. D},
  volume = {41},
  issue = {4},
  pages = {1142--1151},
  numpages = {0},
  year = {1990},
  month = {Feb},
  publisher = {American Physical Society},
  doi = {10.1103/PhysRevD.41.1142},
  url = {https://link.aps.org/doi/10.1103/PhysRevD.41.1142}
}

@book{gray2008linear,
  title={Linear Differential Equations and Group Theory from Riemann to Poincare},
  author={Gray, J.},
  isbn={9780817647728},
  series={Modern Birkh{\"a}user Classics},
  year={2008},
  publisher={Birkh{\"a}user Boston}
}

@article{Barura:2024uog,
    author = "Barura, Chams Gharib Ali and Kobayashi, Hajime and Mukohyama, Shinji and Oshita, Naritaka and Takahashi, Kazufumi and Yingcharoenrat, Vicharit",
    title = "{Tidal Love numbers from EFT of black hole perturbations with timelike scalar profile}",
    eprint = "2405.10813",
    archivePrefix = "arXiv",
    primaryClass = "gr-qc",
    reportNumber = "YITP-24-62, IPMU24-0018, RIKEN-iTHEMS-Report-24",
    doi = "10.1088/1475-7516/2024/09/001",
    journal = "JCAP",
    volume = "09",
    pages = "001",
    year = "2024"
}

@article{Mandal:2023hqa,
    author = "Mandal, Manoj K. and Mastrolia, Pierpaolo and Silva, Hector O. and Patil, Raj and Steinhoff, Jan",
    title = "{Renormalizing Love: tidal effects at the third post-Newtonian order}",
    eprint = "2308.01865",
    archivePrefix = "arXiv",
    primaryClass = "hep-th",
    reportNumber = "HU-EP-23/43-RTG",
    doi = "10.1007/JHEP02(2024)188",
    journal = "JHEP",
    volume = "02",
    pages = "188",
    year = "2024"
}

@article{Nair:2024mya,
    author = "Nair, Sreejith and Chakraborty, Sumanta and Sarkar, Sudipta",
    title = "{Asymptotically de Sitter black holes have nonzero tidal Love numbers}",
    eprint = "2401.06467",
    archivePrefix = "arXiv",
    primaryClass = "gr-qc",
    doi = "10.1103/PhysRevD.109.064025",
    journal = "Phys. Rev. D",
    volume = "109",
    number = "6",
    pages = "064025",
    year = "2024"
}

@article{Emparan:2017qxd,
    author = "Emparan, Roberto and Fernandez-Pique, Alejandro and Luna, Raimon",
    title = "{Geometric polarization of plasmas and Love numbers of AdS black branes}",
    eprint = "1707.02777",
    archivePrefix = "arXiv",
    primaryClass = "hep-th",
    doi = "10.1007/JHEP09(2017)150",
    journal = "JHEP",
    volume = "09",
    pages = "150",
    year = "2017"
}

@article{Sharma:2024hlz,
    author = "Sharma, Chanchal and Ghosh, Rajes and Sarkar, Sudipta",
    title = "{Exploring ladder symmetry and Love numbers for static and rotating black holes}",
    eprint = "2401.00703",
    archivePrefix = "arXiv",
    primaryClass = "gr-qc",
    doi = "10.1103/PhysRevD.109.L041505",
    journal = "Phys. Rev. D",
    volume = "109",
    number = "4",
    pages = "L041505",
    year = "2024"
}

@article{Cai:2019npx,
    author = "Cai, Shengsheng and Wang, Kai-Der",
    title = "{Non-vanishing of tidal Love numbers}",
    eprint = "1906.06850",
    archivePrefix = "arXiv",
    primaryClass = "hep-th",
    month = "6",
    year = "2019"
}

@article{Barbosa:2025uau,
    author = "Barbosa, Sergio and Brax, Philippe and Fichet, Sylvain and de Souza, Lucas",
    title = "{Running Love numbers and the Effective Field Theory of gravity}",
    eprint = "2501.18684",
    archivePrefix = "arXiv",
    primaryClass = "hep-th",
    doi = "10.1088/1475-7516/2025/07/071",
    journal = "JCAP",
    volume = "07",
    pages = "071",
    year = "2025"
}

@article{Barbosa:2025smt,
    author = "Barbosa, Sergio and Fichet, Sylvain and de Souza, Lucas",
    title = "{On the black hole weak gravity conjecture and extremality in the strong-field regime}",
    eprint = "2503.20910",
    archivePrefix = "arXiv",
    primaryClass = "hep-th",
    doi = "10.1007/JHEP10(2025)145",
    journal = "JHEP",
    volume = "10",
    pages = "145",
    year = "2025"
}

@article{Charalambous:2025ekl,
    author = "Charalambous, Panagiotis and Dubovsky, Sergei and Ivanov, Mikhail M.",
    title = "{Love numbers of black p-branes: fine tuning, Love symmetries, and their geometrization}",
    eprint = "2502.02694",
    archivePrefix = "arXiv",
    primaryClass = "hep-th",
    reportNumber = "MIT-CTP/5836",
    doi = "10.1007/JHEP06(2025)180",
    journal = "JHEP",
    volume = "06",
    pages = "180",
    year = "2025"
}

@article{Cano:2025zyk,
    author = "Cano, Pablo A.",
    title = "{Love numbers beyond GR from the modified Teukolsky equation}",
    eprint = "2502.20185",
    archivePrefix = "arXiv",
    primaryClass = "gr-qc",
    doi = "10.1007/JHEP07(2025)152",
    journal = "JHEP",
    volume = "07",
    pages = "152",
    year = "2025"
}

@article{Chakraborty:2025wvs,
    author = "Chakraborty, Sumanta and De Luca, Valerio and Gualtieri, Leonardo and Pani, Paolo",
    title = "{Dynamical Love numbers of black holes: Theory and gravitational waveforms}",
    eprint = "2507.22994",
    archivePrefix = "arXiv",
    primaryClass = "gr-qc",
    doi = "10.1103/fr3y-s1sz",
    journal = "Phys. Rev. D",
    volume = "112",
    number = "10",
    pages = "104015",
    year = "2025"
}

@article{Bhattacharyya:2025slf,
    author = "Bhattacharyya, Arpan and Ghosh, Saptaswa and Kumar, Naman and Kumar, Shailesh and Pal, Sounak",
    title = "{Love beyond Einstein: metric reconstruction and Love number in quadratic gravity using WEFT}",
    eprint = "2508.02785",
    archivePrefix = "arXiv",
    primaryClass = "hep-th",
    doi = "10.1007/JHEP11(2025)155",
    journal = "JHEP",
    volume = "11",
    pages = "155",
    year = "2025"
}

@article{Coviello:2025pla,
    author = "Coviello, Chiara and Vellucci, Vania and Lehner, Luis",
    title = "{Tidal response of regular black holes}",
    eprint = "2503.04287",
    archivePrefix = "arXiv",
    primaryClass = "gr-qc",
    doi = "10.1103/PhysRevD.111.104073",
    journal = "Phys. Rev. D",
    volume = "111",
    number = "10",
    pages = "104073",
    year = "2025"
}

@article{Parra-Martinez:2025bcu,
    author = "Parra-Martinez, Julio and Podo, Alessandro",
    title = "{Naturalness of vanishing black-hole tides}",
    eprint = "2510.20694",
    archivePrefix = "arXiv",
    primaryClass = "hep-th",
    month = "10",
    year = "2025"
}

@article{Berens:2025jfs,
    author = "Berens, Roman and Hui, Lam and McLoughlin, Daniel and Solomon, Adam R. and Staunton, John",
    title = "{Ladder Symmetries of Higher Dimensional Black Holes}",
    eprint = "2510.26748",
    archivePrefix = "arXiv",
    primaryClass = "hep-th",
    month = "10",
    year = "2025"
}

@article{Sharma:2025xii,
    author = "Sharma, Chanchal and Roy, Shuvayu and Sarkar, Sudipta",
    title = "{Ladder Symmetry: The Necessary and Sufficient Condition for Vanishing Love Numbers}",
    eprint = "2511.09670",
    archivePrefix = "arXiv",
    primaryClass = "gr-qc",
    month = "11",
    year = "2025"
}

@article{Garcia-Saenz:2025urd,
    author = "Garcia-Saenz, Sebastian and Lin, Hongbo",
    title = "{On the logarithmic Love number of black holes beyond general relativity}",
    eprint = "2512.19111",
    archivePrefix = "arXiv",
    primaryClass = "gr-qc",
    month = "12",
    year = "2025"
}

@article{Pani:2019cyc,
    author = "Pani, Paolo and Maselli, Andrea",
    title = "{Love in Extrema Ratio}",
    eprint = "1905.03947",
    archivePrefix = "arXiv",
    primaryClass = "gr-qc",
    doi = "10.1142/S0218271819440012",
    journal = "Int. J. Mod. Phys. D",
    volume = "28",
    number = "14",
    pages = "1944001",
    year = "2019"
}

@article{Datta:2019epe,
    author = "Datta, Sayak and Brito, Richard and Bose, Sukanta and Pani, Paolo and Hughes, Scott A.",
    title = "{Tidal heating as a discriminator for horizons in extreme mass ratio inspirals}",
    eprint = "1910.07841",
    archivePrefix = "arXiv",
    primaryClass = "gr-qc",
    doi = "10.1103/PhysRevD.101.044004",
    journal = "Phys. Rev. D",
    volume = "101",
    number = "4",
    pages = "044004",
    year = "2020"
}

@article{DeLuca:2021ite,
    author = "De Luca, Valerio and Pani, Paolo",
    title = "{Tidal deformability of dressed black holes and tests of ultralight bosons in extended mass ranges}",
    eprint = "2106.14428",
    archivePrefix = "arXiv",
    primaryClass = "gr-qc",
    doi = "10.1088/1475-7516/2021/08/032",
    journal = "JCAP",
    volume = "08",
    pages = "032",
    year = "2021"
}

@article{DeLuca:2022xlz,
    author = "De Luca, Valerio and Maselli, Andrea and Pani, Paolo",
    title = "{Modeling frequency-dependent tidal deformability for environmental black hole mergers}",
    eprint = "2212.03343",
    archivePrefix = "arXiv",
    primaryClass = "gr-qc",
    doi = "10.1103/PhysRevD.107.044058",
    journal = "Phys. Rev. D",
    volume = "107",
    number = "4",
    pages = "044058",
    year = "2023"
}

@article{Chakravarti:2018vlt,
    author = "Chakravarti, Kabir and Chakraborty, Sumanta and Bose, Sukanta and SenGupta, Soumitra",
    title = "{Tidal Love numbers of black holes and neutron stars in the presence of higher dimensions: Implications of GW170817}",
    eprint = "1811.11364",
    archivePrefix = "arXiv",
    primaryClass = "gr-qc",
    doi = "10.1103/PhysRevD.99.024036",
    journal = "Phys. Rev. D",
    volume = "99",
    number = "2",
    pages = "024036",
    year = "2019"
}

@article{Chakraborty:2024gcr,
    author = "Chakraborty, Sumanta and Comp{\`e}re, Geoffrey and Machet, Ludovico",
    title = "{Tidal Love numbers and quasinormal modes of the Schwarzschild-Hernquist black hole}",
    eprint = "2412.14831",
    archivePrefix = "arXiv",
    primaryClass = "gr-qc",
    doi = "10.1103/4p2c-rwdh",
    journal = "Phys. Rev. D",
    volume = "112",
    number = "2",
    pages = "024015",
    year = "2025"
}

@inproceedings{Gherghetta:2010cj,
    author = "Gherghetta, Tony",
    title = "{A Holographic View of Beyond the Standard Model Physics}",
    booktitle = "{Theoretical Advanced Study Institute in Elementary Particle Physics}: {Physics of the Large and the Small}",
    eprint = "1008.2570",
    archivePrefix = "arXiv",
    primaryClass = "hep-ph",
    doi = "10.1142/9789814327183_0004",
    pages = "165--232",
    year = "2011"
}

@article{Pomarol:2000hp,
    author = "Pomarol, Alex",
    title = "{Grand unified theories without the desert}",
    eprint = "hep-ph/0005293",
    archivePrefix = "arXiv",
    reportNumber = "CERN-TH-2000-154",
    doi = "10.1103/PhysRevLett.85.4004",
    journal = "Phys. Rev. Lett.",
    volume = "85",
    pages = "4004--4007",
    year = "2000"
}

@article{Friedland:2009zg,
    author = "Friedland, Alexander and Giannotti, Maurizio and Graesser, Michael L.",
    title = "{Vector Bosons in the Randall-Sundrum 2 and Lykken-Randall models and unparticles}",
    eprint = "0905.2607",
    archivePrefix = "arXiv",
    primaryClass = "hep-th",
    reportNumber = "LA-UR-08-08093",
    doi = "10.1088/1126-6708/2009/09/033",
    journal = "JHEP",
    volume = "09",
    pages = "033",
    year = "2009"
}

@article{Lozano:1995aq,
    author = "Lozano, Y.",
    title = "{S duality in gauge theories as a canonical transformation}",
    eprint = "hep-th/9508021",
    archivePrefix = "arXiv",
    reportNumber = "PUPT-1552",
    doi = "10.1016/0370-2693(95)01081-1",
    journal = "Phys. Lett. B",
    volume = "364",
    pages = "19--26",
    year = "1995"
}

@article{Kehagias:1995ic,
    author = "Kehagias, Alexandros A.",
    title = "{A canonical approach to s duality in Abelian gauge theory}",
    eprint = "hep-th/9508159",
    archivePrefix = "arXiv",
    reportNumber = "NTUA-95-51A",
    month = "8",
    year = "1995"
}

@article{Kodama:2003kk,
    author = "Kodama, Hideo and Ishibashi, Akihiro",
    title = "{Master equations for perturbations of generalized static black holes with charge in higher dimensions}",
    eprint = "hep-th/0308128",
    archivePrefix = "arXiv",
    doi = "10.1143/PTP.111.29",
    journal = "Prog. Theor. Phys.",
    volume = "111",
    pages = "29--73",
    year = "2004"
}

@article{Zhang:2023tfv,
    author = "Zhang, Chen and Zhang, Xin",
    title = "{Gravitational capture of magnetic monopoles by primordial black holes in the early universe}",
    eprint = "2302.07002",
    archivePrefix = "arXiv",
    primaryClass = "hep-ph",
    doi = "10.1007/JHEP10(2023)037",
    journal = "JHEP",
    volume = "10",
    pages = "037",
    year = "2023"
}

@article{Stojkovic:2004hz,
    author = "Stojkovic, Dejan and Freese, Katherine",
    title = "{A Black hole solution to the cosmological monopole problem}",
    eprint = "hep-ph/0403248",
    archivePrefix = "arXiv",
    doi = "10.1016/j.physletb.2004.12.019",
    journal = "Phys. Lett. B",
    volume = "606",
    pages = "251--257",
    year = "2005"
}

@article{Noumi:2026shc,
    author = "Noumi, Toshifumi and Wong, Sam S. C.",
    title = "{Extremal Love: tidal/electromagnetic deformability, logarithmic running and the weak gravity conjecture}",
    eprint = "2601.20962",
    archivePrefix = "arXiv",
    primaryClass = "hep-th",
    month = "1",
    year = "2026"
}

@article{Xia:2025zfp,
    author = "Xia, Minghao and Ma, Liang and Pang, Yi and Lu, H.",
    title = "{Full spectrum of Love numbers of Reissner-Nordstrom black hole in D-dimensions}",
    eprint = "2511.09642",
    archivePrefix = "arXiv",
    primaryClass = "hep-th",
    reportNumber = "USTC-ICTS/PCFT-25-56",
    month = "11",
    year = "2025"
}

\end{document}